\documentclass[10pt]{article}
\usepackage[utf8]{inputenc}

\usepackage{amsmath}
\usepackage{amsfonts}
\usepackage{amssymb}
\usepackage{makeidx}
\usepackage{graphicx}
\usepackage{caption}
\usepackage{subcaption}
\usepackage{listings}
\usepackage{apacite}
\usepackage{natbib}
\usepackage{indentfirst}
\usepackage{bbm}
\usepackage{xcolor}
\usepackage{siunitx}
\usepackage{multirow}
\usepackage{booktabs}
\usepackage{mathtools}
\usepackage{enumitem}
\usepackage{setspace}

\usepackage[margin=1in]{geometry}
 
\usepackage{authblk}
\usepackage[pdfborder={0 0 0}]{hyperref}

\title{Market Complete Option Valuation using a Jarrow-Rudd Pricing Tree with Skewness and Kurtosis }

\author[1]{Yuan Hu}
\author[1,3]{Abootaleb Shirvani}
\author[1,*]{W. Brent Lindquist}
\author[2]{Frank J. Fabozzi}
\author[1]{Svetlozar T. Rachev}

\affil[1]{\small Department of Mathematics \& Statistics, Texas Tech University, Lubbock, TX 79409-1042, USA\\
yuan.hu@ttu.edu;  abootaleb.shirvani@ttu.edu;  zari.rachev@ttu.edu}
\affil[2]{\small Finance Department, EDHEC Business School, 393/400 Promenade des Anglais-BP3116, CEDEX 3, 06202 Nice, France\\
frank.fabozzi@edhec.edu}
\affil[3]{\small Present address: Department of Actuarial Science \& Risk Management, Drake University, Des Moines, IA, 50311, USA}
\affil[*]{\small Corresponding author: brent.lindquist@ttu.edu}
\begin{document}
\thispagestyle{plain}
\begin{spacing}{1.0}
\maketitle
\noindent {\textbf {Abstract}}
 Applying the Cherny-Shiryaev-Yor invariance principle,
 we introduce a generalized Jarrow-Rudd (GJR) option pricing model with uncertainty driven by a skew random walk.
 The GJR pricing tree exhibits skewness and kurtosis in both the natural and risk-neutral world.
 We construct implied surfaces for the parameters determining the GJR tree.
 Motivated by Merton’s pricing tree incorporating transaction costs,
 we extend the GJR pricing model to include a hedging cost.
 We demonstrate ways to fit the GJR pricing model to a market driver that influences the price dynamics of the underlying asset.
 We supplement our findings with numerical examples. 
\\
\end{spacing}
\newpage
\section{Introduction}
\label{sec1}
\noindent
Pricing trees determined by a Markov chain have been studied in a number of academic papers.\footnote{
	See, for example, \cite{Duan2001}, \cite{Simonato2011}, \cite{Bhat2012},
	\cite{Fuh2012}, Van and Elliot (2012a,b), and \cite{Fan2016}.
	In \cite{Bhat2012}, the authors claim that the limiting return distribution is a mixture of normal distributions.
	A mixture of different normal distributions is not infinitely divisible \citep[Chapter VI] {Steutel2004},
	and  it is therefore not clear that their limiting continuous-time model is arbitrage-free.}
Since the Markov chain  pricing tree is constructed directly in the risk-neutral world in these papers,
it is not clear what discrete pricing model in the natural world would evolve to a risk-neutral Markov chain pricing
model in accordance with dynamic asset pricing theory.\footnote{
	\cite{Black1973}, \cite{Merton1973}, \cite{Delbaen1994}, \cite{Delbaen1998}, and
	\citet[Chapter 6]{Duffie2001}.}
In this work, we follow the classical binomial pricing model framework
(see \cite{Cox1979} and \citet[Chapters V and VI]{Jarrow1983}).
We begin with a Markov chain model in the natural world replicating a self-financing portfolio,
and then transform to risk-neutral option pricing preserving market completeness.
It is essential that our option pricing model is defined first in the natural world.
If the option pricing model is placed directly in the risk-neutral world, and calibrated with market option data,
no option market dislocation or option mispricing can be revealed.
Option pricing models which do not start with modeling the underlying assets in the natural world and then,
via risk-neutral valuation based on Black-Scholes-Merton dynamic pricing theory,
pass to the risk-neutral world are meaningless, if not dangerous, in practical applications.
Such pricing models are generally used to predict what the option traders {\it believe}
the correct option prices are and not what option prices are {\it actually fair}.
It is imperative that opinion prices be aligned with reliable spot market models.\footnote{
	\cite{Black1975}, \cite{Brenner1984}, \cite{Melick1997}, \cite{Hilber2009}, \cite{Pasquariello2014}, and
	\cite{Ross2015}.}
	
To illustrate this claim, suppose that a trader would like to determine if potential mispricing in the option market
exists due to a market bubble that the trader suspects will burst.\footnote{
	The trader can decide to ride the bubble as long as possible, as many other option traders will do,
	with the hope of unwinding option positions before the bubble bursts.
	In this case, the trader does not need to worry about what the true spot market model is that corresponds to the
	risk-neutral option pricing model.
	For an extensive study of market bubbles and related option markets,
	we refer to \cite{Heston2007}, \cite{Jarrow2010}, and \cite{Vogel2018}.}
The trader's option pricing model, when calibrated to option market data,
should recover, uniquely, the spot price dynamics \citep{Ross2015} of the underlying asset 
(\cite{Kim2016}, \cite{Hu2020a}).
If that recovered spot price process does not conform with market data on the asset spot price,
this could be a trading signal that there is potential option mispricing.\footnote{
	In this setting model risk does exist.
	That is why in real trading a suite of option pricing models is used.}

Our paper is close in spirit to \cite{Kijima1993}, where an option pricing model with Markov chain stochastic volatility
is introduced and the continuous-time limiting price process is determined as a subordinated geometric Brownian motion (GBM).\footnote{
	We employ the following abbreviations throughout this paper. Each abbreviation is also defined the first time it is referenced.
	BM: Brownian motion; CPM: continuous-time option pricing model; CSYIP: Cherny-Shiryaev-Yor Invariance Principle;
	DPM: discrete-time option pricing model; ECC: European contingency claim; GBM: geometric Brownian motion;
	GJR: generalized Jarrow-Rudd; HTC: hedging transaction cost; JR: Jarrow-Rudd; RMSE: relative mean-square error;
	SBM: skew Brownian motion; w.p.: with probability.}
While the discrete- and continuous-time market models are incomplete in the paper by Kijima and Yoshida,
in our paper we deal with complete market models, both in the discrete- and continuous-time settings.
We extend the Jarrow and Rudd (JR) binomial pricing model
(\cite{Jarrow1983}, \citet[p. 442]{Hull2012}, Kim et al. (2016, 2019))
with an additional parameter determining the skewness and excess kurtosis of the underlying asset return distribution.
We refer to our extended model as the {\it generalized Jarrow-Rudd} (GJR) {\it option pricing model}.

As in the original JR model (Kim et al., 2016, 2019; Hu et al., 2020a,b),
our GJR pricing model preserves market completeness.\footnote{
	As far the authors of this paper are aware,
	all discrete market models in the literature exhibiting skewness and kurtosis lead to incomplete market
	models.
	Thus, the problem of determining the unique spot price tree corresponding to the risk-neutral tree chosen
	in any of  those papers will be unsolved. }
In the GJR pricing model, the embedded Markov chain driving the discrete underlying price process is a skew
random walk which, in the limit, becomes a skew Brownian motion (SBM) after the necessary scalar
normalization.\footnote{
	See \cite{Harrison1981}, \cite{Cherny2003},  \citet[Chapters VII and X]{Revuz1994}, and \cite{Corns2007}.}
In discrete time, the distributional mapping between the GJR price dynamics of the underlying asset and its
risk-neutral dynamics is one-to-one.\footnote{
	That is, for every fixed trading frequency,
	 the probability law of the risk-neutral tree for the underlying asset uniquely determines the
	probability law of the asset's pricing tree in the natural world.}
The GJR pricing model allows us to study option pricing in a Markov chain market model with transaction costs.
Our market model has a relatively simple parametric form and is easily calibrated to option data, 
as illustrated by numerical examples.

Our paper proceeds as follows.
In section \ref{sec2} we introduce the Cherny-Shiryaev-Yor invariance principle (CSYIP) \citep{Cherny2003}
for SBM. 
As with the use of the Donsker-Prokhorov invariance principle\footnote{
	\cite{Donsker1951}, \cite{Prokhorov1956}, \citet[section 14]{Billingsley1999},
	\citet[Chapter IX]{Gikhman1969}, \citet[section 5.3.3]{Skorokhod2005}, \cite{Davydov2008}}
in the Cox-Ross-Rubinstein pricing tree \citep{Cox1979} and in the JR pricing tree,
we apply CSYIP to introduce the GJR pricing tree and obtain the limiting continuous-time price dynamics.
The mean return $\mu$ is retained as a parameter in the GJR model.
The GJR tree includes an additional parameter $\beta$ governing the skewness and kurtosis of the GJR tree in the natural world.
While the GJR tree ultimately converges to a GBM as in the classical JR pricing tree,
the GJR pre-limiting behavior is that of geometric SBM.
We investigate numerically the pre-limiting behavior of the GJR tree to obtain estimates of the smallest
number, $n$, of trading intervals to maturity $T = n\Delta t,\; n\uparrow \infty$, at which the skewness
and excess kurtosis vanish from the GJR tree distribution.
In section \ref{sec3}, we determine the risk-neutral probabilities in the GJR pricing model which,
together with the volatility $\sigma$ and the risk-free rate $r_f$, depend on $\mu$ and $\beta$.
Using daily closing price data for the SPDR S\&P 500 ETF Trust fund (SPY) and call option data for the same underlying asset,
we estimate the implied $\mu$, $\beta$ and $\sigma$ surfaces.
Motivated by Merton’s binomial tree model with transaction costs \citep[Chapter 14]{Merton1990},
in section \ref{sec4} we extend the GJR pricing tree model to include a hedging transaction cost term.
Using the SPY option data, we estimate option transaction costs, the implied transaction cost surface,
and the impact transaction costs have on the implied $\mu$, $\beta$ and $\sigma$ surfaces.
In section \ref{sec5}, we explore possibilities for fitting the GJR model to a market driver that affects the price dynamics of the
underlying asset.
We estimate the parameters for the new pricing tree model in a numerical example with the underlying asset being the stock
of Microsoft Corporation (MSFT).
We explore both an endogenous and exogenous approach.
The exogenous approach allows for greater generalization of the market driver;
in a numerical example we assume the asset returns are dependent on Fama-French five-factor loading values \citep{Fama2015}.
In section \ref{sec6}, we derive the risk-neutral probabilities for the extended pricing tree introduced in section \ref{sec5}
and estimate the corresponding implied volatility surface.
Chapter \ref{sec7} concludes of our work.
\section{The generalized Jarrow-Rudd pricing tree model}
\label{sec2}
\noindent
In this section, we introduce the GJR pricing tree model.
To do so, we describe SBM and then apply the CSYIP to formulate a new path-dependent pricing model
defined by a recombined tree that generalizes the JR binomial option pricing model.\footnote{
	See \cite{Jarrow1983}; \citet[p. 442]{Hull2012}; Kim et al. (2016, 2019).}
In contrast to the classical Cox-Ross-Rubinstein and JR binomial pricing models, in the GJR pricing model the pre-limiting pricing process is
determined by SBM.
That is, for a moderately small trading frequency $\Delta t$, the GJR dynamics exhibit the properties of geometric SBM.
This feature leads to a more flexible discrete pricing model for the underlying asset price behavior for any realistically
small trading frequency.
However, as $\Delta t \downarrow 0$ our pricing tree converges to a GBM, as in the traditional JR pricing model.
As shown in Hu et al. (2020a,b), discrete-time option pricing models contain considerably more information than
continuous-time pricing models.

Due to the option trader's  presumed ability  to trade continuously with no transaction costs,
in a continuous-time option pricing model the information about the underlying stock mean return and stock-price
direction at a given trading frequency is lost.
That is the main reason that we emphasize a discrete-time option pricing model,
rather than paying more attention to the limiting continuous-time option price dynamics.
In the GJR option pricing model, new features, such as skewness and excess kurtosis\footnote{
	Discrete-time option pricing models with underlying stock return distributions exhibiting skewness and
	excess kurtosis are known; see, for example, \cite{Yamada2004}.
	But the Yamada-Primbs' pricing model is based on mean-variance hedging in incomplete markets.
	Our GJR pricing model is based on no-arbitrage asset valuation arguments leading to a complete market
	model.},
of the underlying stock return distribution will be present.

\subsection{Definition and properties of SBM}
\label{sec21}
\begin{figure}[ht]
\begin{center}
    \begin{subfigure}[b]{0.29\textwidth} 
    	\includegraphics[width=\textwidth]{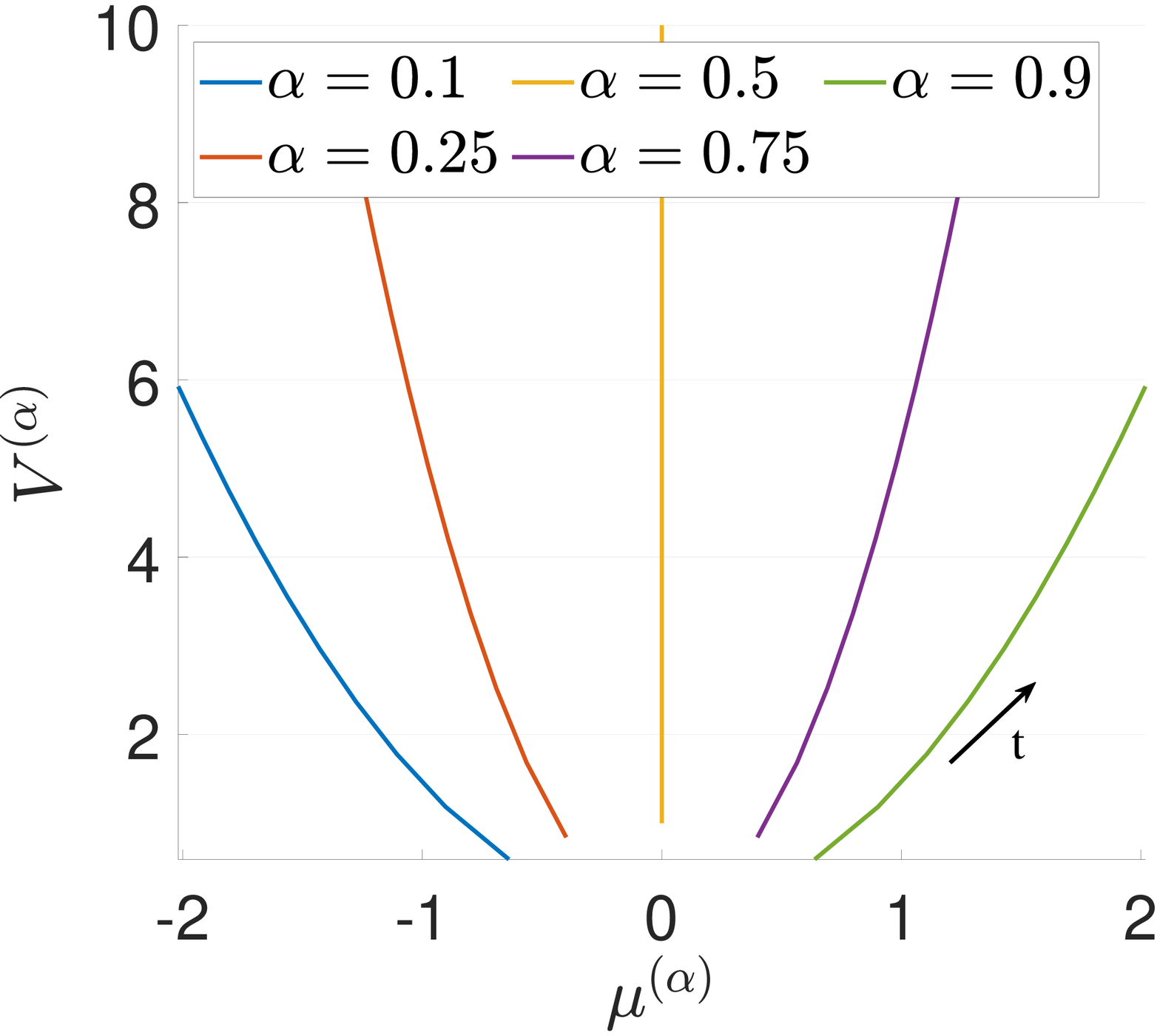}
    	\caption{}
    \end{subfigure}
    \begin{subfigure}[b]{0.29\textwidth} 
    	\includegraphics[width=\textwidth]{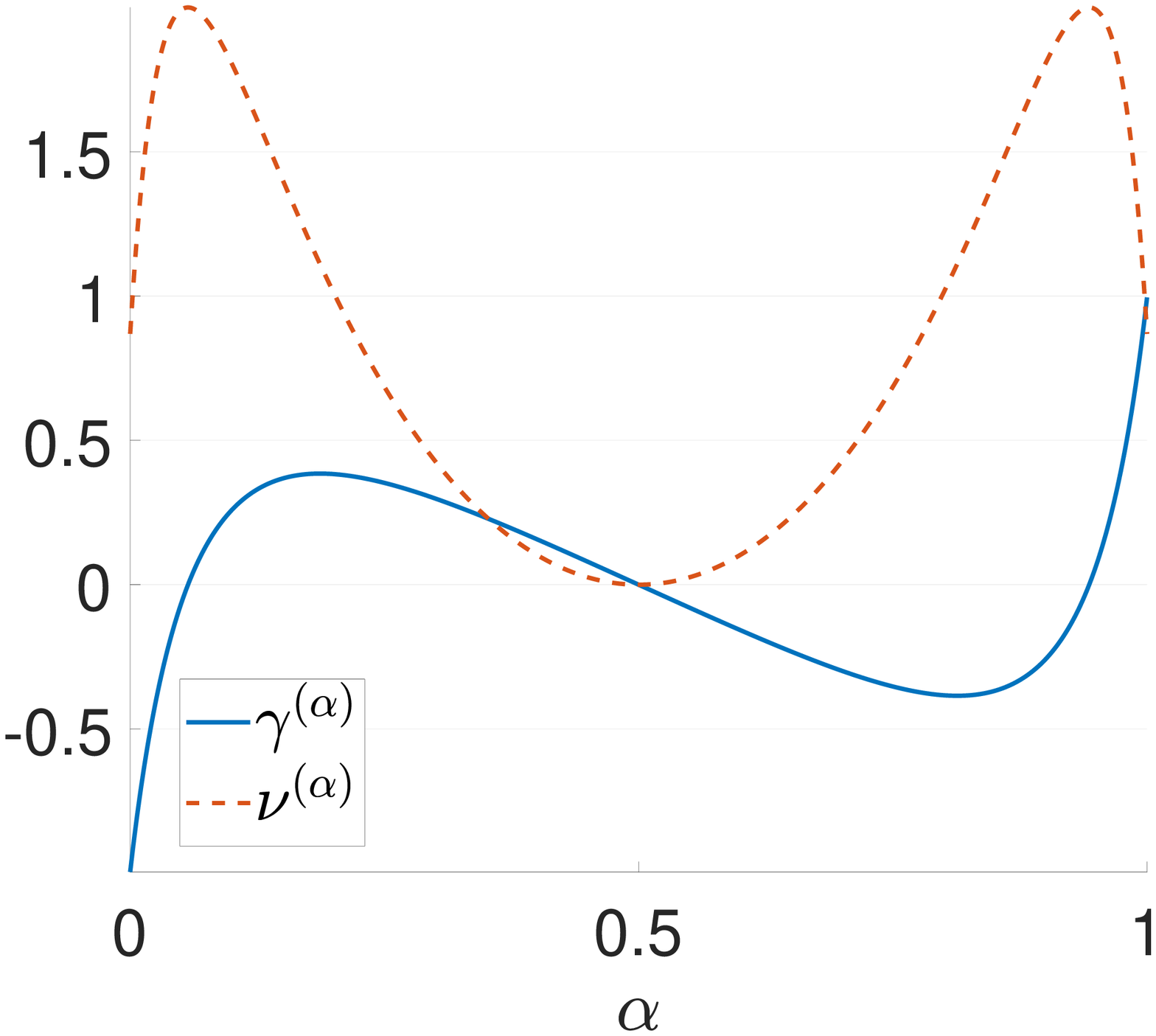}
    	\caption{}
    \end{subfigure}
    \caption{(a) Bivariate plot of mean $\mu_t^{(\alpha)}$ and variance  $V_t^{(\alpha)}$ of a SBM under different parameter
    values of $\alpha$ and $t\in[0,10]$.
    (b) Plots of skewness $\gamma^{(\alpha)}$ and excess kurtosis $\nu^{(\alpha)}$ of a SBM as a function of $\alpha\in [0,1]$.}
    \label{fig1}
\end{center}
\end{figure}

\noindent
We start with the definition of a SBM.\footnote{
	See \citet[section 4.2, problem 1, p. 115]{Ito1996}; \cite{Harrison1981};
	\citet[Chapters VII and X ]{Revuz1994}; \cite{Lang1995}; \cite{Lejay2006}; \cite{Corns2007};
	\cite{Cherny2003}; \cite{Ramirez2011}; \cite{Atar2015}; \cite{Trutnau2015}; and \cite{Li2019}.}
Let $\mathbb{B} = \left\{B_t,\;t\geq 0\right\}$ be a standard Brownian motion (BM)  generating a stochastic basis $\left(\Omega,\mathbb{F}=\{\mathcal{F}_t = \sigma(B_u,u\leq t)_{t\geq 0}\},\mathbb{P}\right)$.
Let $\alpha \in[0,1]$ and set\footnote{
	See \cite{Cherny2003}.}
\begin{align*}
\mathcal{A}_t^{(\alpha)} &= \int^t_0\left(\alpha^2I_{\left\{B_s \geq 0\right\}}+(1-\alpha)^2 I_{\left\{B_s<0\right\}}\right)ds,
\\
\tau_t^{(\alpha)} &= \inf \left\{s\geq 0: \mathcal{A}_s^{(\alpha)}\right\},
\\
B_t^{(\alpha)} &= \varphi_{\alpha}\left(B_{\tau_t^{(\alpha)}}\right),\; t\geq 0,
\end{align*}
where $\varphi_{\alpha}(x) = \alpha x I_{\left\{x \geq 0\right\}}+(1-\alpha)xI_{\left\{x < 0\right\}},\; x\in R$,
and $I_{\{\cdot\}}$ is the indicator function.
The process $\mathbb{B}^{(\alpha)} = \left\{B_t^{(\alpha)},t\geq 0\right\}$ with $B_0^{(\alpha)} = 0$
is a SBM with parameter $\alpha$ having the following properties.\footnote{
	See \citet[Chapter 4]{Cherny2003} and \cite{Corns2007}.}
\begin{itemize}
\item[(i)] $\mathbb{B}^{(1)}\overset{\mathrm{d}}{=}|\mathbb{B}| = \left\{|B_t|,t\geq 0 \right\}$.\footnote{
	Here, and in what follows, $\overset{\mathrm{d}}{=}$ stands for “equal in distribution”
	or “equal in probability law”.}
\item[(ii)]$\mathbb{B}^{(1/2)}\overset{\mathrm{d}}{=}  |\mathbb{B}| $.
\item[(iii)]$\mathbb{B}^{(0)}\overset{\mathrm{d}}{=} - |\mathbb{B}| $.
\item[(iv)]$|\mathbb{B}^{(\alpha)}|\overset{\mathrm{d}}{=}|\mathbb{B}|$.
\item[(v)]$\mathbb{B}^{(\alpha)}$ is a semimartingale satisfying the strong Markov property.
\item[(vi)] When $t \geq 0$, sample paths of $\mathbb{B}^{(\alpha)}$ can be generated using the
representation\footnote{
	See \citet[section 4.2, problem 1, p. 115]{Ito1996}, \cite{Lejay2006} and \cite{Corns2007}.
	We denote “with probability'' as “w.p.”}
\begin{equation*}
B_t^{(\alpha)} = \begin{cases}
\ \  |B_t| , &\textrm{w.p.}\;\; \alpha, \\
-|B_t| , &\textrm{w.p.}\;\; 1-\alpha.
\end{cases}
\end{equation*}
\item[(vii)] For $0 \leq s < t$,
\begin{equation}
\mathbb{P}\left(B_{s+t}^{(\alpha)}\in dx | B_s^{(\alpha)} = 0\right) = \begin{cases}
\alpha \sqrt{2/\pi t} \exp(-\frac{x^2}{2t})dx, \; \textrm{if}\; x\geq 0 ,\\
(1-\alpha) \sqrt{2/\pi t} \exp(-\frac{x^2}{2t})dx, \; \textrm{if}\; x< 0 ,
\end{cases}
\label{eq_pro_vii}
\end{equation}
is the conditional density $f_t^{(\alpha)}(x),\; x\in R$, with conditional cumulative distribution function
$F_t^{(\alpha)}(x) = \int^x_{-\infty}(y)dy,\;x\in R$, given by
\begin{equation*}
F_x^{(\alpha)}(x) = \begin{cases}
(1-\alpha)+ 2\alpha/\sqrt{\pi}\int^{x/\sqrt{2t}}_0 \exp(-z^2)dz, \; \textrm{if}\; x\geq 0 ,
\\
(1-\alpha) \left(1-2/\sqrt{\pi}\int_{x/\sqrt{2t}}^0 \exp(-z^2)dz\right), \; \textrm{if}\; x< 0 .
\end{cases}
\end{equation*}
\item[(viii)] \cite{Corns2007} developed a continuous-time option pricing model based on geometric Azzalini SBM.
The trajectories $A_t^{(\delta)}(\omega),\; t\geq 0,\; \omega\in\Omega$,
of an {\it Azzalini SBM}, $\mathbb{A}^{(\delta)} = \left\{A_t^{(\delta)},t\geq 0 \right\}$,
with parameter $\delta \in (-1,1)$ have the form
$A_t^{(\delta)}(\omega) = \sqrt{1-\delta^2}B_{1,t}(\omega)+\delta |B_{2,t}(\omega)|$,
where $B_{1,t}$ and $B_{2,t}$ are two independent BMs.
\cite{Corns2007} showed that $\mathbb{A}^{(\delta)}\overset{\mathrm{d}}{=}\mathbb{B}^{(\alpha)}$
with $\alpha = (1+\delta)/2$.  
\item[(ix)]  The moment-generating function of $B_t^{(\alpha)}$ has the form
\begin{equation*}
M_{B_t^{(\alpha)}}(u) = \mathbb{E}\left(\exp(uB_t^{(\alpha)})\right) = \exp(u^2t/2)\left(1+(2\alpha-1)\frac{2}{\sqrt{\pi}}\int^{u\sqrt{t/2}}_0\exp(-z^2)dz\right),\; u>0.
\end{equation*}
The moments $\mu_t^{(p,\alpha)} = \mathbb{E}\left( \left( B_t^{(\alpha)} \right)^p \right)$ are given by
\begin{equation}
\mu_t^{(p,\alpha)} = \sqrt{\frac{2^p}{\pi}}\Gamma\left(\frac{p+1}{2}\right)\left(\alpha+(-1)^p(1-\alpha)\right)t^{p/2},\;p>0,\;s\geq 0,\; t>0.
\label{eq_mu_p}
\end{equation}
Note that odd moments depend on $\alpha$ through the term $2 \alpha - 1$, while even moments are independent of $\alpha$.
\end{itemize}
From \eqref{eq_mu_p}, the mean, variance, skewness and excess kurtosis of $\mathbb{B}^{(\alpha)}$ are given by
\begin{align}
\mu_t^{(\alpha)} & = \mu_t^{(1,\alpha)} = (2\alpha-1)\sqrt{2t/\pi}, \label{eq_SBM_mean} \\
V_t^{(\alpha)} &= \mu_t^{(2,\alpha)}-\left(\mu_t^{(\alpha)}\right)^2 = \left(1-2(2\alpha-1)^2/\pi\right)t, \label{eq_SBM_var} \\
\gamma^{(\alpha)} & = \frac{\mathbb{E}\left( \left(B_t^{(\alpha)}- \mu_t^{(\alpha)}\right)^3 \right)}{\left(V_t^{(\alpha)}\right)^{3/2}}
				 = \frac{\sqrt{2}(2\alpha-1)\left(4(2\alpha-1)^2-\pi\right)}{\left(\pi-2(2\alpha-1)^2\right)^{3/2}}, \label{eq_SBM_skew} \\
\nu^{(\alpha)} & = \frac{\mathbb{E}\left( \left(B_t^{(\alpha)}- \mu_t^{(\alpha)}\right)^4 \right)}{\left(V_t^{(\alpha)}\right)^{2}}-3
			 = \frac{8\pi(2\alpha-1)^2-24(2\alpha-1)^{4}}{\left(\pi-2(2\alpha-1)^2\right)^{2}}. \label{eq_SBM_kurt}
\end{align}
Note that the skewness and excess kurtosis are time-independent quantities.
The behavior of these four quantities is demonstrated in Fig. \ref{fig1}.

\subsection{CSYIP for SBM}
\label{sec22}
\noindent
We start with the formulation of the CSYIP for SBM with the definition of a piecewise continuous function.
A function $h: R\rightarrow R$ is called a {\it piecewise continuous function} if there exists a collection
of disjoint intervals $J_n,\; n\in \mathcal{N}$, such that:\footnote{
	Each $J_n$ can be closed, open, semi-open, or a singleton.
	Define $\mathcal{N} = \left\{1,2,\ldots\right\}$ and $\mathcal{N}_0 = \left\{0,1,\ldots\right\}$.}
\begin{itemize}[leftmargin=45pt]
	\item[PSC(i)\;\;] $\cup^{\infty}_{n=1}J_n = R$;
	\item[PSC(ii)\;] for every compact interval $J$ there exists $n \in \mathcal{N}$
					such that $\cup^n_{k = 1}J_k \supseteq J$; and
	\item[PSC(iii)] on each $J_n,\; n\in \mathcal{N},\; h: J_n\rightarrow R$ is continuous and has finite limits
					at those endpoints of $J_n$ which do not belong to $J_n$. 
\end{itemize}

\begin{figure}[ht]
    \centering
    \subcaptionbox{$\alpha = 0.1$}{\includegraphics[width=0.27\textwidth]{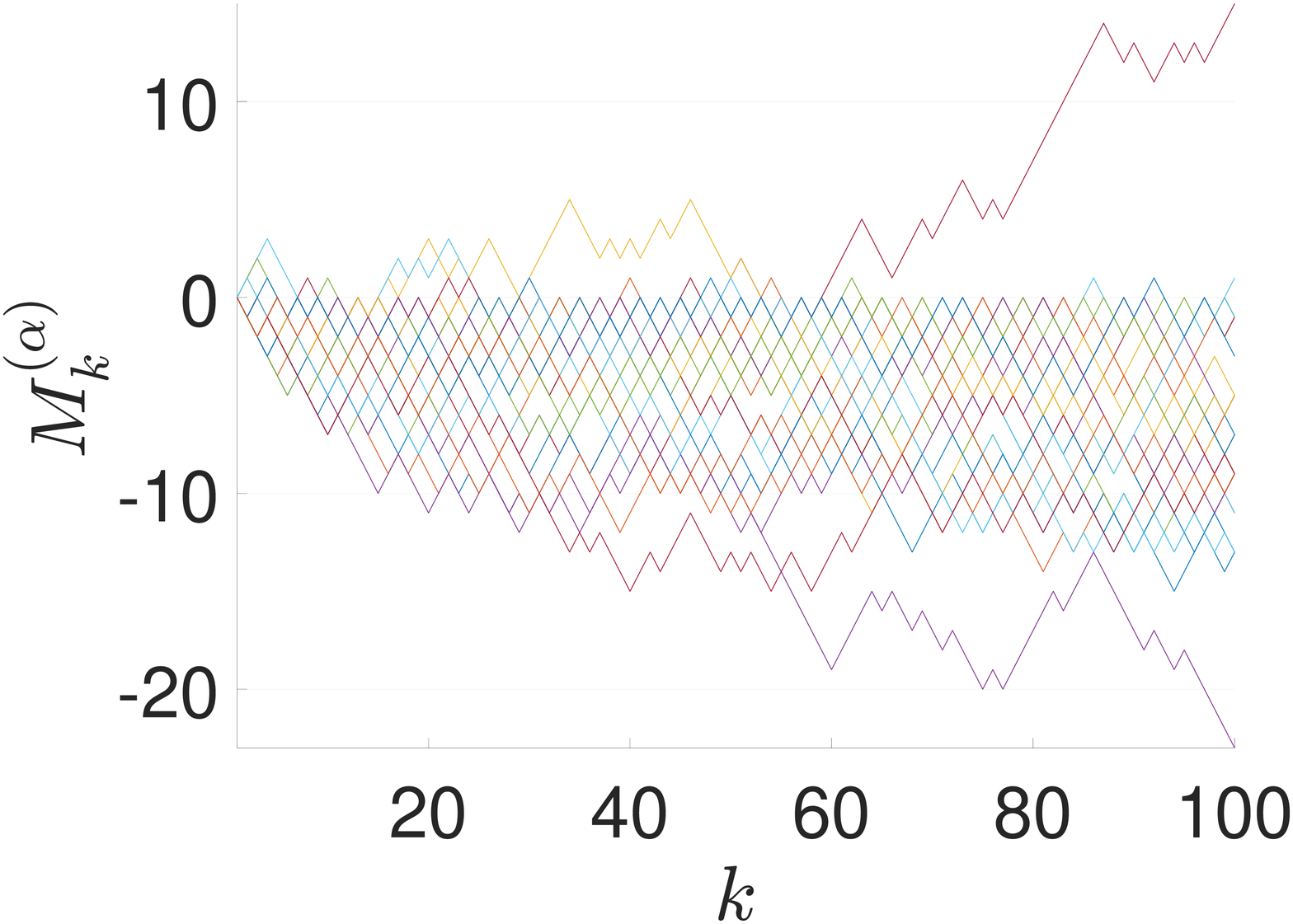}}\hspace{0em}\vspace{.5em}%
    \subcaptionbox{$\alpha = 0.25$}{\includegraphics[width=0.27\textwidth]{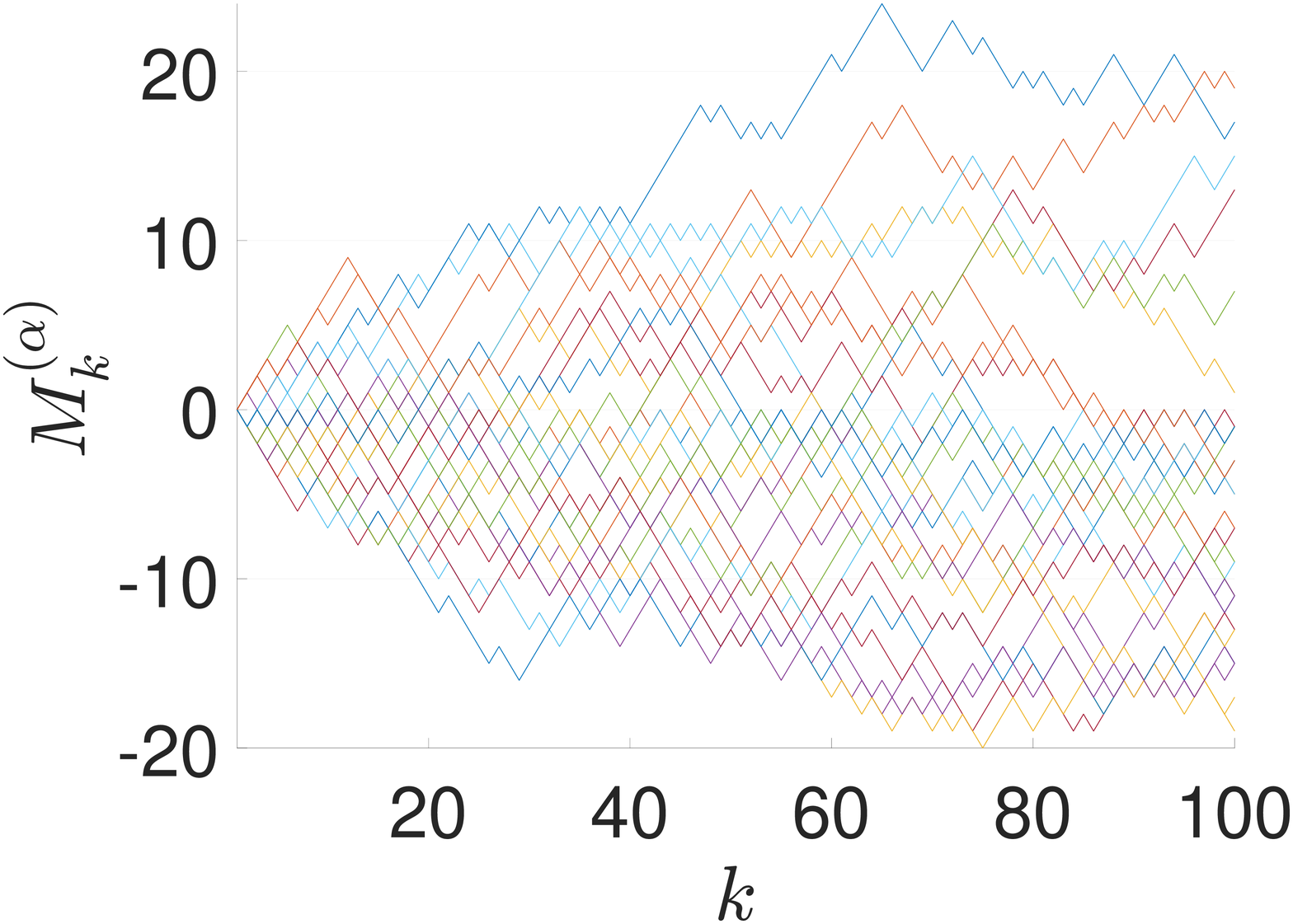}}\hspace{0em}\vspace{.5em}%
    \subcaptionbox{$\alpha = 0.5$}{\includegraphics[width=0.27\textwidth]{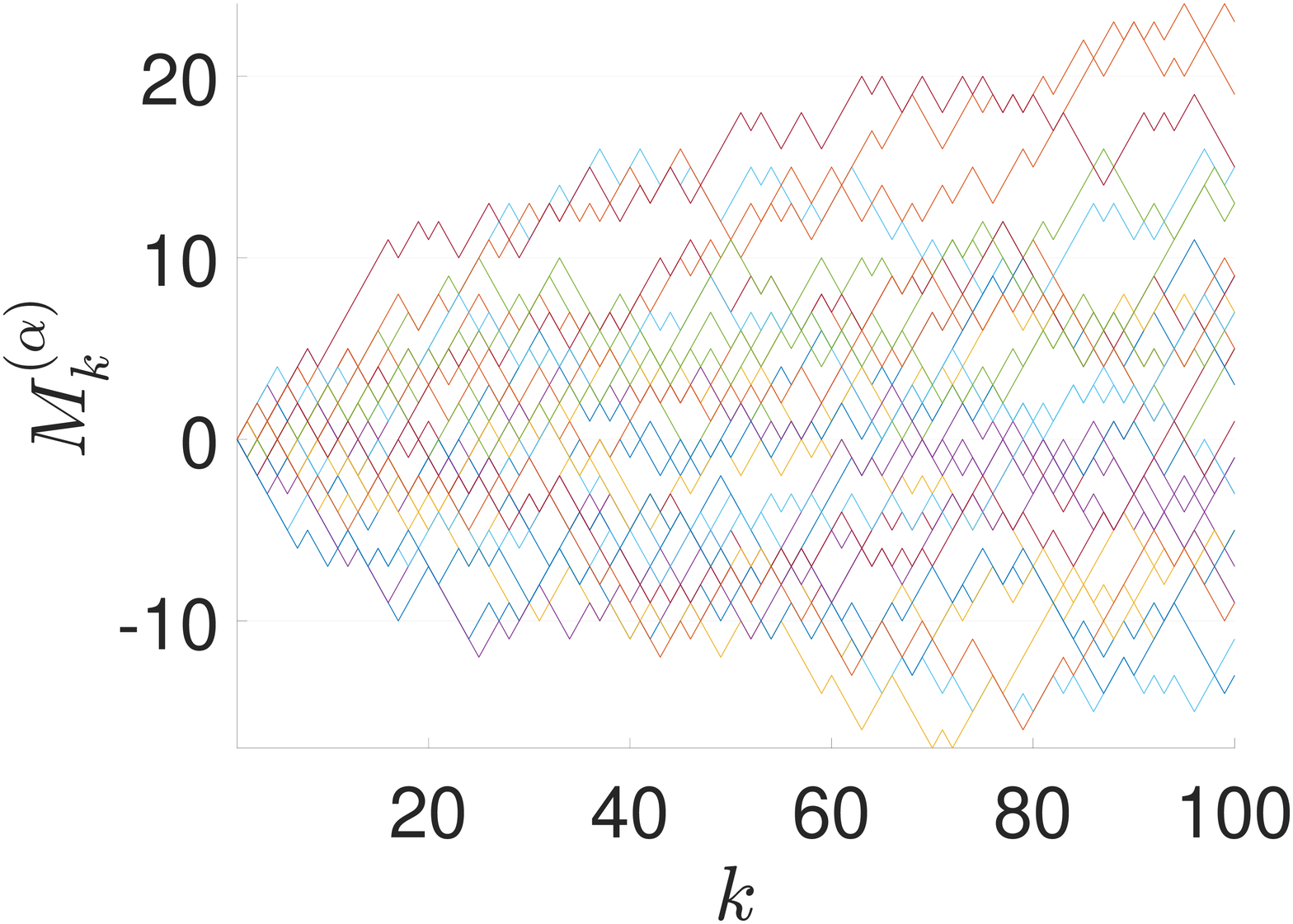}}\hspace{0em}\vspace{.5em}%
    \subcaptionbox{$\alpha = 0.75$}{\includegraphics[width=0.27\textwidth]{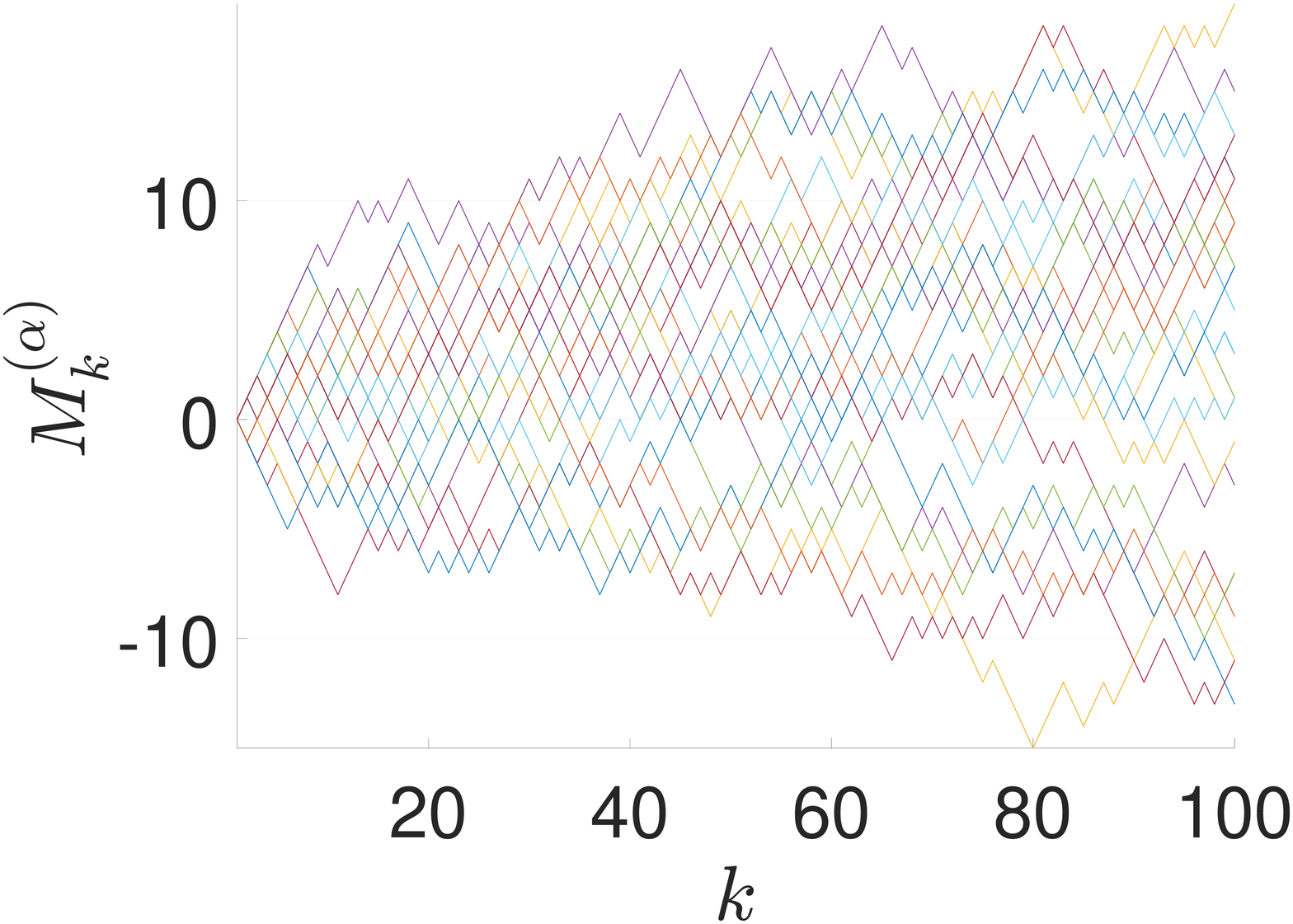}}\hspace{0em}%
    \subcaptionbox{$\alpha = 0.9$}{\includegraphics[width=0.27\textwidth]{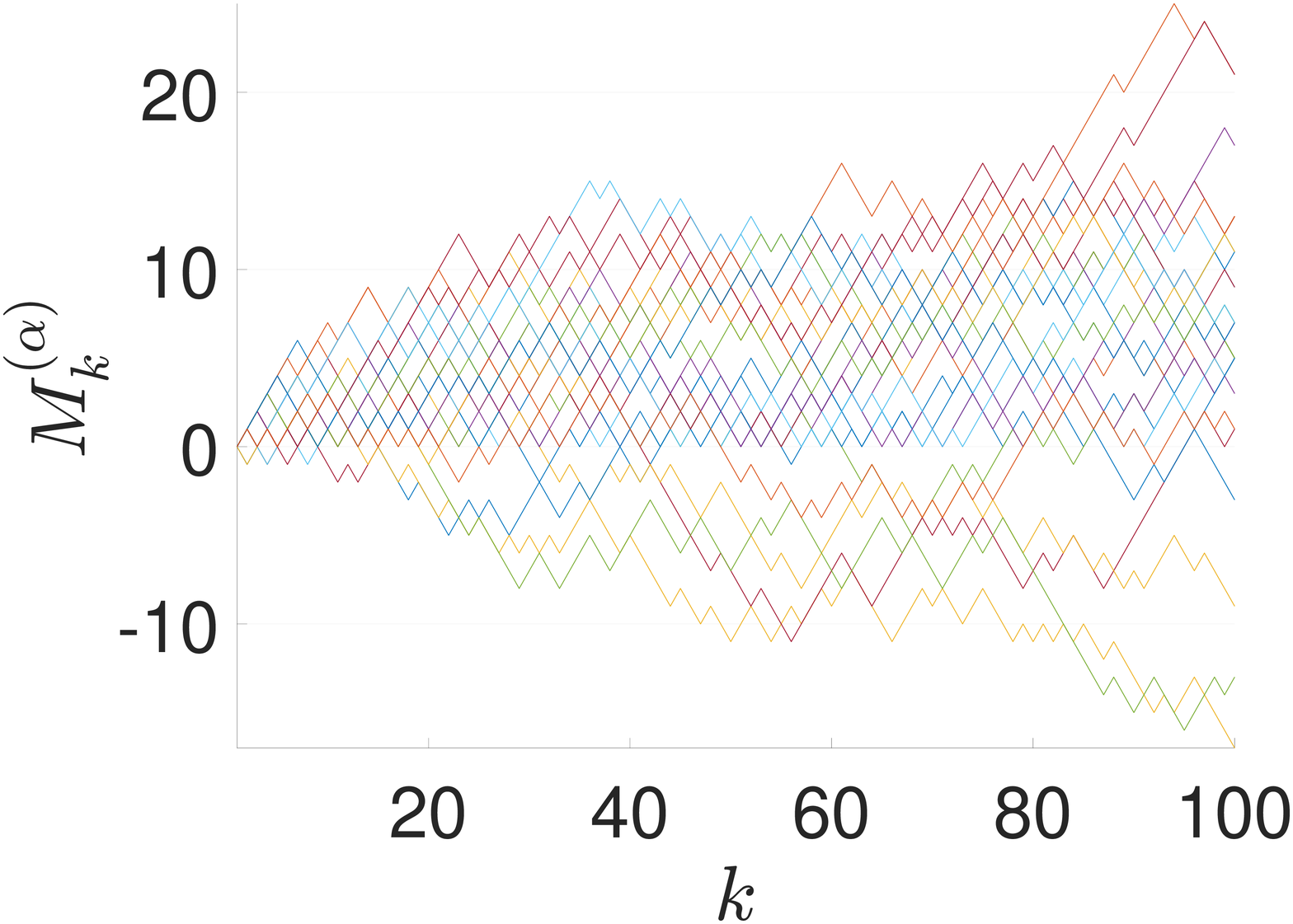}}\hspace{0em}%
    \caption{30 Monte Carlo scenarios for
    	$\mathbb{M}^{(\alpha)} = \left\{M_k^{(\alpha)}\in\mathbb{Z},\;k = 1,\ldots,100\right\}$
    	for select values of $\alpha \in (0,1)$.}
    \label{fig_MC}
\end{figure}

Next, for $\alpha \in (0,1)$, let $\mathbb{M}^{(\alpha)} = \left\{M_k^{(\alpha)}\in \mathbb{Z},\;k\in \mathcal{N}_0\right\},\;\mathbb{Z} = \left\{0,\pm 1,\pm 2,\ldots\right\}$
be a Markov chain with $M_0^{(\alpha)} = 0$ and transition probabilities\footnote{
	See \cite{Harrison1981} and \cite{Cherny2003}.}
\begin{equation}
\begin{aligned}
\mathbb{P}\left(M_{k+1}^{(\alpha)} = i+1 | M_{k}^{(\alpha)} = i\right) &= \begin{cases}
\frac{1}{2}, \;\textrm{if} \; i \neq 0,
\\
\alpha,\;\textrm{if} \; i = 0,
\end{cases}
\\
\mathbb{P}\left(M_{k+1}^{(\alpha)} = i-1 | M_{k}^{(\alpha)} = i\right) &= \begin{cases}
\frac{1}{2}, \;\textrm{if} \; i \neq 0,
\\
1-\alpha,\;\textrm{if} \; i = 0.
\end{cases}
\end{aligned}
\label{eq_trans_prob}
\end{equation}
We call $\mathbb{M}^{(\alpha)}$ a skew random walk with parameter $\alpha$.
Fig. \ref{fig_MC} shows the skew random walk for different values of $\alpha$.
The results exhibit a predominance of trajectories having $M_{k}^{(\alpha)}<0$ for $\alpha < 1/2$ and
a predominance of trajectories having $M_{k}^{(\alpha)}>0$ for $\alpha > 1/2$.

For $n\in \mathcal{N}$, set $M_0^{(\alpha,n)} = 0,\; M_k^{(\alpha,n)} = n^{-1/2}M_k^{(\alpha)}$ and
$X_{k/n}^{(\alpha,n)}=\sum^k_{i=1}M_i^{(\alpha,n)},\;k = 1, \dots, n$.
Fix $n \in \mathcal{N}$ and define $\mathbb{B}_t^{(\alpha,n)},\;t\geq 0$, to be the random process with
piecewise linear trajectories having  vertexes $\left(k/n,\mathbb{B}_{k/n}^{(\alpha,n)}\right)$,
where $\mathbb{B}_{k/n}^{(\alpha,n)} = X_{k/n}^{(\alpha,n)}$.
Let $h: R \rightarrow R$ be a piecewise continuous function and define
$Y_{k/n}^{(\alpha,n)} = \sum^k_{i = 1}h\left(X_{(i-1)/n}^{(\alpha,n)}\right)\left(X_{i/n}^{(\alpha,n)}-X_{(i-1)/n}^{(\alpha,n)}\right)$, $k=1, \dots, n$.
Define $\mathbb{C}_t^{(\alpha,n)},\; t\geq 0$, to be a random process with the piecewise linear trajectories
having vertexes  $\left(k/n,\mathbb{C}_{k/n}^{(\alpha,n)}\right)$,
where $\mathbb{C}_{k/n}^{(\alpha,n)} = Y_{k/n}^{(\alpha,n)}$.

\noindent
{\bf CSYIP for a SBM}\footnote{
	See Theorem 4.1 in \cite{Cherny2003}.}:
{\it If $h: R\rightarrow R$ is a piecewise continuous function,
then for $t \geq 0$, as $n \uparrow \infty$, the bivariate process $\left(\mathbb{B}_t^{(\alpha,n)},
\mathbb{C}_t^{(\alpha,n)}\right)$ converges in law to $\left(B_t^{(\alpha)},C_t^{(\alpha)}\right)$,
where $B_t^{(\alpha)}$ is a SBM and
$C_t^{(\alpha)} = \int_0^t h\left(B_s^{(\alpha)}\right)dB_s^{(\alpha)}$.}

\subsection{The GJR pricing model }
\label{sec23}
\noindent
We begin the construction of a binomial pricing tree by applying CSYIP
using only the lower moment SBM process $\mathbb{B}_t^{(\alpha,n)}$.
We will add the higher moment $\mathbb{C}_t^{(\alpha,n)}$ in section~\ref{sec52}.
We fix $n\in \mathcal{N}$,  the time interval  $\Delta t = T/n$,
and $\beta\in \left(-1/\sqrt{\Delta t},1/\sqrt{\Delta t}\right)$.
Set
\begin{equation}
\alpha_{\Delta t} =  (1+\beta\sqrt{\Delta t})/2\in (0,1).
\label{eq_alpha}
\end{equation}
If we consider the process $\mathbb{D}^{(\alpha_{\Delta t})} = \left\{D_t = B_{k\Delta t}^{(\alpha_{\Delta t})},\;k\Delta t\leq t <(k+1)\Delta t ,\; k = 1,\ldots,n-1,\; D_T = B_T^{(\alpha_{\Delta t})}\right\}$,
then $\mathbb{D}^{(\alpha_{\Delta t})}$ weakly converges in the Skorokhod space
$\mathcal{D}[0,T]$ to the BM $\mathbb{B}$ as $\Delta t \downarrow 0$.
However, for fixed $\Delta t$, $\mathbb{D}^{(\alpha_{\Delta t})}$ exhibits the properties of a SBM with
parameter $\alpha_{\Delta t}$.
From \eqref{eq_alpha}, \eqref{eq_trans_prob} and Fig.~\ref{fig_MC}, we see that
$\beta$ will affect the moments of the process $B_{k\Delta t}^{(\alpha_{\Delta t})}$.
This becomes clear by observing the form of the mean, variance, skewness and excess kurtosis of $B_{k\Delta t}^{(\alpha_{\Delta t})}$.
From \eqref{eq_SBM_mean} - \eqref{eq_SBM_kurt}, to leading order in $\Delta t$,
$\mu_{k \Delta t}^{(\alpha_{\Delta t})} = \beta\sqrt{2k/\pi}\Delta t$,
$V_{k \Delta t}^{(\alpha_{\Delta t})} = k\Delta t$,
$\gamma_{k \Delta t}^{(\alpha_{\Delta t})} = -\sqrt{2\Delta t / \pi}\beta$,
and
$\nu_{k \Delta t}^{(\alpha_{\Delta t})} = 8\beta^2\Delta t/\pi$.
Thus, for any fixed (arbitrarily small) time interval $\Delta t$, the increments $B_{k\Delta t}^{(\alpha_{\Delta t})}$
have skewed and heavy tailed distributions when $\beta \ne 0$. 
Consider the arithmetic SBM
\begin{equation}
B_t^{(\alpha;\mu^{(\alpha)},\sigma^{(\alpha)})} = \mu^{(\alpha)}t+\sigma^{(\alpha)}B_t^{(\alpha)},\;t\in[0,T],
\end{equation}
with mean $\mathbb{E}\left(B_t^{(\alpha;\mu^{(\alpha)},\sigma^{(\alpha)})}\right)$
and variance $\textrm{Var}\left(B_t^{(\alpha;\mu^{(\alpha)},\sigma^{(\alpha)})}\right)$.
Then, for $k = 1,...,n$, we have\footnote{
	As $\Delta t = 1/n$, we can assume that the value of
	$\left[\mathbb{E}\left(B_{k\Delta t}^{(\alpha;\mu^{(\alpha)},\sigma^{(\alpha)})}\right)\right]^2$
	is negligible.
	Then
	\begin{equation*}
	\textrm{Var}\left(B_{k\Delta t}^{(\alpha;\mu^{(\alpha)},\sigma^{(\alpha)})}\right)
	= \mathbb{E}\left[\left(B_{k\Delta t}^{(\alpha;\mu^{(\alpha)},\sigma^{(\alpha)})}\right)^2\right]
	- \left[\mathbb{E}\left(B_{k\Delta t}^{(\alpha;\mu^{(\alpha)},\sigma^{(\alpha)})}\right)\right]^2
	= {\sigma^{(\alpha)}}^2k\Delta t.
	\end{equation*}}
\begin{align*}
\mathbb{E}\left(B_{k\Delta t}^{(\alpha;\mu^{(\alpha)},\sigma^{(\alpha)})}\right)
	&= \mu^{(\alpha)}k\Delta t+\sigma^{(\alpha)}\beta\sqrt{2k/\pi}\Delta t, \\
\textrm{Var}\left(B_{k\Delta t}^{(\alpha;\mu^{(\alpha)},\sigma^{(\alpha)})}\right)
	&= {\sigma^{(\alpha)}}^2k\Delta t.
\end{align*}

We next define the GJR pricing tree, determined by the skew random walk $\mathbb{M}^{(\alpha_{\Delta t})}$.\footnote{
	If $\alpha_{\Delta t} = 1/2$ (i.e. $\beta = 0$), we obtain the JR binomial tree.
	\cite{Jarrow1983} and \citet[p. 442]{Hull2012} defined the JR binomial tree directly in the risk-neutral world,
	that is, when $\mu = r_f$, where $r_f$ is the risk-free rate.
	\cite{Kim2016} showed that JR binomial tree option pricing model provides the fastest rate of convergence
	to the corresponding GBM in the risk-neutral world.} 
\\
{\bf GJR Pricing Tree}:
Let $\mu>0$ and $\sigma>0$.
For $k=1,\ldots,n$, $n\Delta t = T$, $\alpha_{\Delta t} = (1+\beta\sqrt{\Delta t})/2$
and skew random walk $\mathbb{M}^{(\alpha_{\Delta t})}$, define the {\it GJR pricing recombined tree} by
\begin{equation}
S_{k\Delta t}^{(n)} = S_0 \exp\left(v_k\Delta t+ M_k^{(\alpha_{\Delta t})}\sigma\sqrt{\Delta t}\right),\; k = 1,\ldots,n,
\label{eq_GJR_pricing_tree}
\end{equation}
where $v_k = k\mu+\sigma\beta\left(\sqrt{2k/\pi}-1\right)$.
We study the limiting behavior of this tree as $\Delta t \downarrow 0$.
Note that, to leading order in $\Delta t$,
\begin{equation}
\mathbb{E}\left(M_k^{(\alpha_{\Delta t})}\right) = 2 \alpha - 1 = \beta\sqrt{\Delta t}, \qquad
\textrm{Var}\left(M_k^{(\alpha_{\Delta t})}\right) = k.
\label{eq_GJR_mean_var}
\end{equation}
Consider the cumulative log-return $R_{k\Delta t}^{(n)} = \ln\left(S_{k\Delta t}^{(n)}/S_0\right),\;k = 1,\ldots,n$.
From (\ref{eq_GJR_pricing_tree}) and (\ref{eq_GJR_mean_var}) we obtain
\begin{equation}
\begin{aligned}
\mathbb{E}\left(R_{k\Delta t}^{(n)}\right) &= \mu k \Delta t + \sigma\beta\sqrt{2k/\pi}\Delta t 
	= \mu k \Delta t+\sigma \mathbb{E}\left(B_{k\Delta t}^{(\alpha_{\Delta t})}\right),
\\
\textrm{Var}\left(R_{k\Delta t}^{(n)}\right) &= \sigma^2 k \Delta t
	= \sigma^2 \textrm{Var}\left(B_{k\Delta t}^{(\alpha_{\Delta t})}\right).
\end{aligned}
\label{eq_GJR_mean_var2}
\end{equation}
From (\ref{eq_GJR_mean_var2}) and the CSYIP it follows that, for a fixed but relatively small time-increment $\Delta t$, the pricing tree (\ref{eq_GJR_pricing_tree}) approximates
\begin{equation}
S_t = S_0  \exp\left(\mu t+ \sigma B_t^{(\alpha_{\Delta t})}\right),\;S_0>0,\; t\in[0,T],\; \mu>0,\; \sigma>0.
\end{equation}
However, if we let $\Delta t \downarrow 0$, then the $\mathcal{D}[0,T]$-process generated by the tree (\ref{eq_GJR_pricing_tree}) will ultimately converge to a GBM, that is, $S_t=S_0 \exp(\mu t+\sigma B_t ),\;t\in[0,T]$.

\subsection{An example}
\label{sec24}
\begin{figure}[ht]
 	\begin{center} 
    	\centering\includegraphics[width=0.25\textwidth]{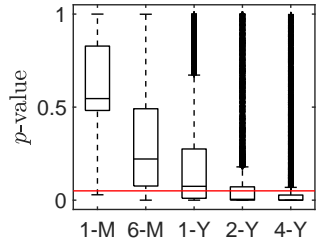}
    	\caption{Box-whisker plot of the $p$-values for two-sided z-tests for different window lengths $L$.
    	The horizontal line indicates the 0.05 significance level for rejecting the null hypothesis $H_0: \mathbbm{e}=0$.}
    	\label{fig3}
    \end{center}
\end{figure}

\noindent
As a numerical example to investigate the pre-limiting behavior of  the GJR pricing tree,
we use SPY daily closing prices.
Price data covering $N = 7136$ trading days over the period
1/29/1993 to 6/1/2021 was obtained from Bloomberg Professional Services.
We denote SPY daily closing prices as $S^{(\textrm{SPY})}_{k\Delta t}$ and cumulative log-returns as
$R_{k\Delta t}^{(\textrm{SPY})} = \ln\left(S^{(\textrm{SPY})}_{k\Delta t}/S^{(\textrm{SPY})}_0\right)$,
where $k = 1,...,N,$ and $\Delta t = 1/252$.
We first determine a length, $L$, for a moving-window estimator that provides relatively strong
explanatory power and acceptable precision for computing the parameters $\mu$, $\sigma$ and $\beta$,
by using the following procedure based on equations \eqref{eq_GJR_mean_var2}.
For each tested length $L$ and for each moving window $w_i$, $i = 1, \dots, N-L+1$, repeat the following steps.
\begin{itemize}[leftmargin=45pt]
\item[Step 1:] Fit a robust linear regression using a logistic weight function\footnote{
		See \cite{Holland2007} and \cite{Pregibon1981}.}
	to the SPY return data in $w_i$ using the model\footnote{
		Since the cumulative return is not stationary, we employ the model
		$\textrm{Var}\left ( R_{k\Delta t}^{(\textrm{SPY})} \right ) \propto \left ( R_{k\Delta t}^{(\textrm{SPY})}\right )^2$.}
	\begin{equation}
	\ln\left({R_{k\Delta t}^{(\textrm{SPY},w_i)}}^2\right)
		 = \ln\left({\sigma^{(\alpha,w_i)}}^2k\Delta t\right)+e^{(1,w_i)}_{k\Delta t},
	\label{eq_step1}
	\end{equation}
	to produce estimates for the value of $\hat{\sigma}^{(\alpha, w_i)}$ and the error terms
	$e^{(1,w_i)}_{k\Delta t},\;k = 1,...,L$.
\item[Step 2:] With  $\hat{\sigma}^{(\alpha,w_i)}$ estimated from Step 1,
	 apply the conditional least squares optimization,
	\begin{multline}
	\min_{\beta^{(\alpha,w_i)}\in \left(-1/\sqrt{\Delta t},1/\sqrt{\Delta t}\right)} \parallel e^{(2,w_i)}_{k\Delta t} 
	\parallel_2^2 \\
	= \min_{\beta^{(\alpha,w_i)}\in \left(-1/\sqrt{\Delta t},1/\sqrt{\Delta t}\right)}
	\parallel
	\mathbb{E}\left( R_{k\Delta t}^{(\textrm{SPY},w_i)} \right) - \mu^{(\alpha,w_i)} k \Delta t
		- \hat{\sigma}^{(\alpha,w_i)} \beta^{(\alpha,w_i)} \sqrt{2k/\pi} \Delta t
	\parallel_2^2,
	\label{eq_step2}
	\end{multline}	
	to determine $\hat{\beta}^{(\alpha,w_i)}$, $\hat{\mu}^{(\alpha,w_i)}$ and the second error sequence
	$e^{(2,w_i)}_{k\Delta t},\;k = 1,...,L$.
\item[Step 3:] Define  $\mathbbm{e}$ to be the random variable having the sample
	\begin{equation*}
	e_{k\Delta t}^{(w_i)} = \frac
		{e^{(1,w_i)}_{k\Delta t}+e^{(2,w_i)}_{k\Delta t}}
		{\sqrt{ \left ( e^{(1,w_i)}_{k\Delta t} \right )^2 + \left ( e^{(2,w_i)}_{k\Delta t} \right )^2} },\; k = 1,...,L,
	\end{equation*}
	and compute the $p$-value of the two-sided z-test with the hypothesis $H_0: \mathbbm{e}=0$.
\end{itemize}
The statistics of the $p$-values computed in Step 3 using tested values of $L$ ranging from one month to four years are shown
as box-whisker plots in Fig. \ref{fig3}.
The $p$-value ranges indicate stronger rejection of $H_0$ as $L$ increases.
As a compromise between satisfying the null hypothesis and retaining the accuracy provided by larger values 
of $L$, we chose $L=$ one year.

\begin{figure}[ht]
\begin{center}
    \begin{subfigure}[b]{0.32\textwidth} 
    	\includegraphics[width=\textwidth]{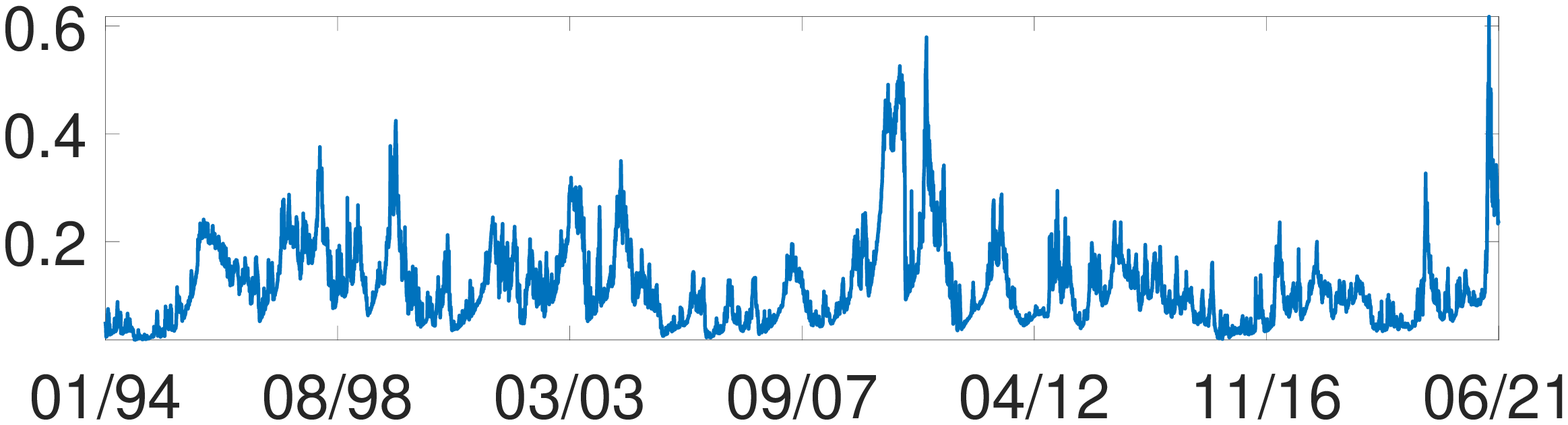}
    	\caption{$\hat{\sigma}^{(\alpha)}$}
    	\label{fig4_sigma}
    \end{subfigure}
    \begin{subfigure}[b]{0.32\textwidth} 
    	\includegraphics[width=\textwidth]{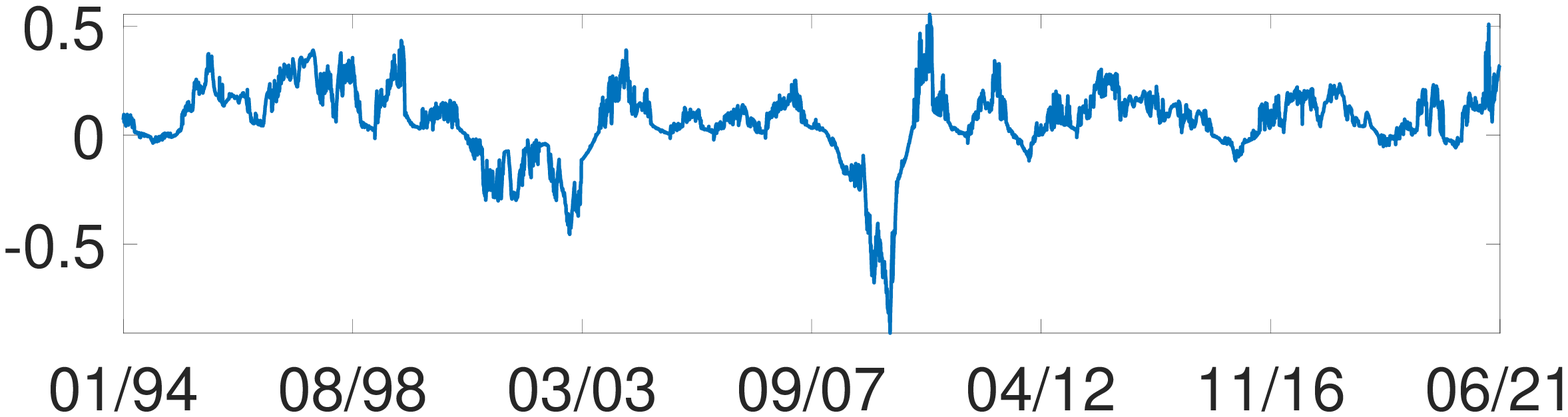}
    	\caption{$\hat{\mu}^{(\alpha)}$}
    	\label{fig4_mu}
    \end{subfigure}
    \begin{subfigure}[b]{0.32\textwidth} 
    	\includegraphics[width=\textwidth]{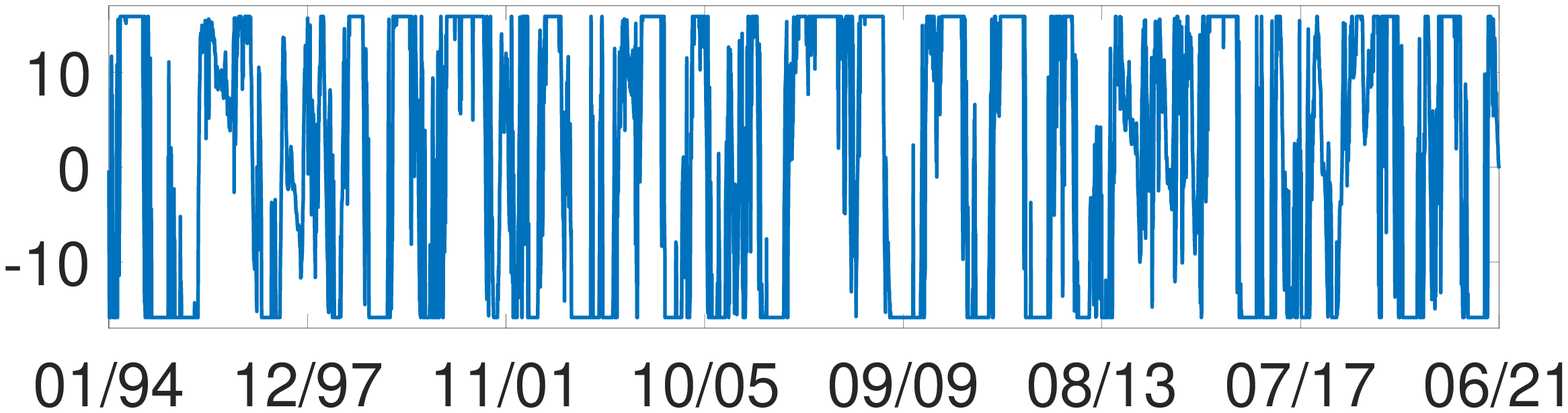}
    	\caption{$\hat{\beta}^{(\alpha)}$}
    	\label{fig4_beta}
    \end{subfigure}
        \begin{subfigure}[b]{0.32\textwidth} 
    	\includegraphics[width=\textwidth]{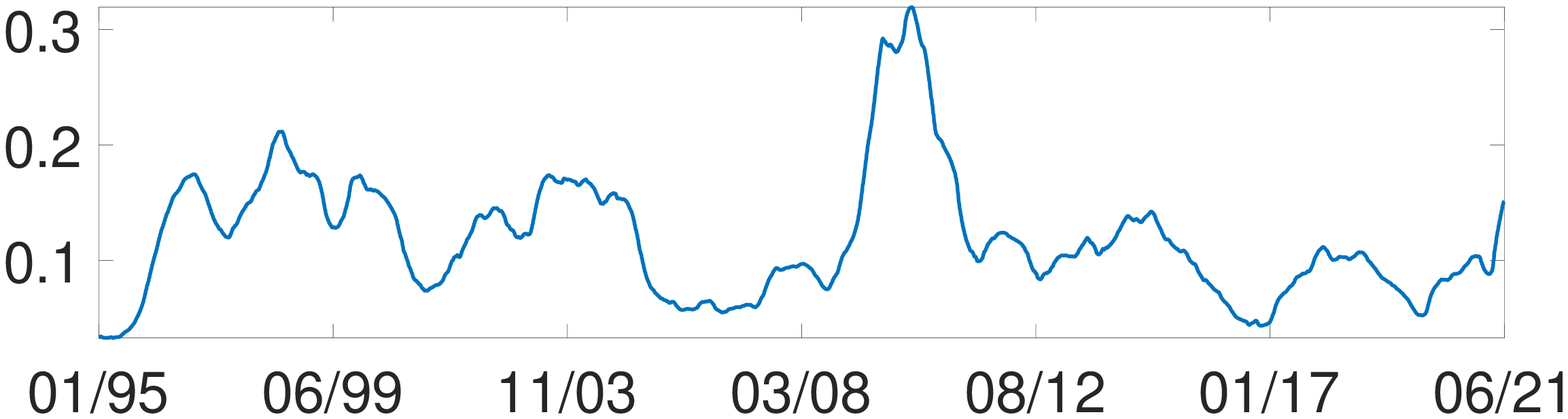}
    	\caption{$\bar{\sigma}^{(\alpha)}$}
    	\label{fig4_av_sigma}
    \end{subfigure}
    \begin{subfigure}[b]{0.32\textwidth} 
    	\includegraphics[width=\textwidth]{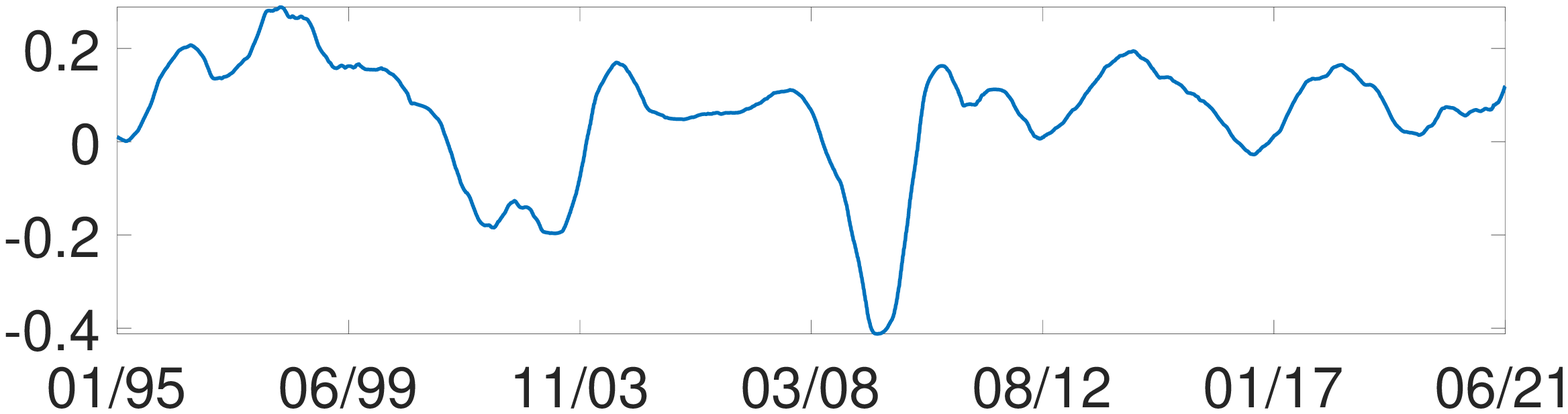}
    	\caption{$\bar{\mu}^{(\alpha)}$}
    	\label{fig4_av_mu}
    \end{subfigure}
    \begin{subfigure}[b]{0.32\textwidth} 
    	\includegraphics[width=\textwidth]{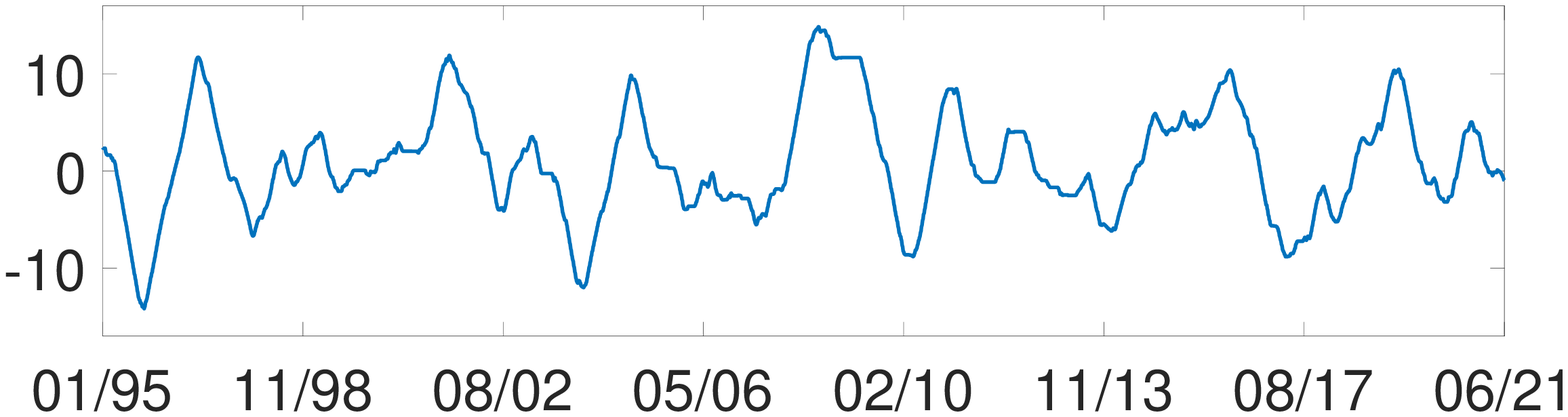}
    	\caption{$\bar{\beta}^{(\alpha)}$}
    	\label{fig4_av_beta}
    \end{subfigure}
    \caption{Time series of the values (a) $\hat{\sigma}^{(\alpha)}$ estimated from Step 1, and
			 (b) $\hat{\mu}^{(\alpha)}$ and (c) $\hat{\beta}^{(\alpha)}$ estimated from Step 2
		using the $L=$ one year moving window.
		Plots (d) to (f) show 1-year moving average values of (a) to (c), repectively.}
       \label{fig4}
\end{center}
\end{figure}
Fig. \ref{fig4} presents the results for $\hat{\mu}^{(\alpha)}$, $\hat{\sigma}^{(\alpha)}$
and $\hat{\beta}^{(\alpha)}$ computed from the one-year moving window.
The results clearly show a sharp drop in $\hat{\mu}^{(\alpha)}$
and increased volatility ($\hat{\sigma}^{(\alpha)}$) resulting from the 2008 global financial crisis.
The most recent peak in $\hat{\sigma}^{(\alpha)}$ in 2020 corresponds to the market reaction to the
Covid-19 pandemic.
In Fig.~\ref{fig4_beta}, we note that optimum values for $\hat{\beta}^{(\alpha)}$ change rapidly and frequently
hit the limits $\pm 1/\sqrt{\Delta t}$.
These results suggest that the least squares optimization in Step 2 may be improved with a smaller value of $\Delta t$,
which would require intra-day data.
We therefore smooth the time-series data in Figs. \ref{fig4_sigma} to \ref{fig4_beta} using one-year moving averages.
The time-series for the smoothed parameters, denoted $\bar{\sigma}^{(\alpha)}$, $\bar{\mu}^{(\alpha)}$ and $\bar{\beta}^{(\alpha)}$,
are presented in Figs. \ref{fig4_av_sigma} to \ref{fig4_av_beta}.
The impact of the global financial crises is retained in the smoothed series.
As the market disruption due to the Covid-19 pandemic was of shorter duration, the pandemic impact is lessened in the averaged data.
Most significantly, the averaged values $\bar{\beta}^{(\alpha)}$ are better behaved.

\begin{figure}[ht]
 	\begin{center} 
    	\centering\includegraphics[width=0.33\textwidth]{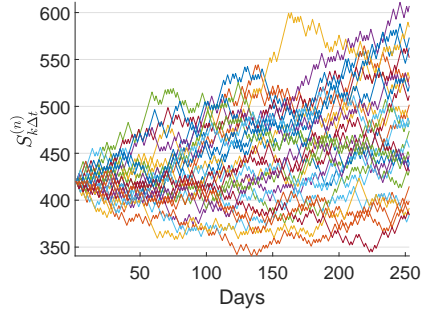}
    	\caption{A recombined,  GJR price tree (\ref{eq_GJR_pricing_tree}) computed from the smoothed parameter
		estimates $\bar{\sigma}^{(\alpha)} = 0.151$, $\bar{\mu}^{(\alpha)} = 0.119$, $\bar{\beta}^{(\alpha)} = -0.978$,
		and $\bar{\alpha} = 0.469$.
		30 price trajectories are displayed with initial capital \$$S_0 = 419.67$, which corresponds to the SPY closing price on 6/1/2021.}
    	\label{fig5_price_tree}
    \end{center}
\end{figure}

Using the last one-year estimation window (from 6/2/2020 to 6/1/2021),
we obtain the estimates $\bar{\sigma}^{(\alpha)} = 0.151$, $\bar{\mu}^{(\alpha)} = 0.119$,
$\bar{\beta}^{(\alpha)} = -0.978$,
and $\bar{\alpha} = (1+\bar{\beta}^{(\alpha)}\sqrt{\Delta t})/2 = 0.469$ for the date 6/1/2021.
Based on these estimates, Fig. \ref{fig5_price_tree} shows a constructed, recombined, GJR price tree
comprised of 30 price trajectories (\ref{eq_GJR_pricing_tree}).

\section{Option pricing using the GJR pricing tree}
\label{sec3}
\noindent
We now consider the following discrete version of Corns' and Satchell’s extension\footnote{
	See \citet[section 3]{Corns2007} where the asset dynamics was determined by an
	Azzalini SBM, $\mathbb{A}^{(\delta)}$.
	In what follows we will use $\mathbb{B}^{(\alpha)} = \left\{B_t^{(\alpha)},\; t\geq 0\right\}$
	instead of $\mathbb{A}^{(\delta)}$).
	Constructing a binomial model using $\mathbb{A}^{(\delta)}$ instead of $\mathbb{B}^{(\delta)}$
	would require two independent pricing trees to model the discrete price dynamics of the underlying asset,
	which is not desirable for derivative hedging.}
of the Black-Scholes-Merton market model.\footnote{
	For the Black-Scholes-Merton market model, see \cite{Black1973} and \citet[Chapter 6]{Duffie2001}.}
Suppose the dynamics\footnote{
	We call the risky asset a stock, and it will be denoted by $\mathcal{S}$.}
of the risky asset follows  the GJR pricing tree (\ref{eq_GJR_pricing_tree}).
The dynamics\footnote{
	We call the riskless asset a bond, and it will be denoted by $\mathcal{B}$.}
of the riskless asset is given by
\begin{equation}
\beta_{k\Delta t}^{(n)} = \beta_0 \exp(r_f k \Delta t),\; \beta_0>0,\; k = 0,1,\ldots,n,\; \mu > r_f > 0.
\label{eq_bond}
\end{equation}
Let
\begin{equation}
f_{k\Delta t}^{(n)} = f\left(S_{k\Delta t}^{(\mathcal{S})},k\Delta t\right),\ k = 1,\ldots,n,
\label{eq_f_price}
\end{equation}
where  $S_{k\Delta t}^{(\mathcal{S})}$ is the spot price of the underlying asset,
be the price dynamics of a European Contingent Claim (ECC)\footnote{
	We use the terms ECC, option, and derivative interchangeably.}
having terminal time $n \Delta t =T >0$ and final payoff $f_T^{(n)}$. 
Following the construction of the Cox-Ross-Rubinstein and JR binomial pricing models, our next goal is to use the pricing tree
(\ref{eq_GJR_pricing_tree}) to derive the price dynamics of  $f_{k\Delta t}^{(n)},\;k = 0,1,\ldots,n$.

We start by forming the replicating risk-neutral portfolio
$P_{k\Delta t}^{(n)} = D_{k\Delta t}^{(n)}S_{k\Delta t}^{(n)}-f_{k\Delta t}^{(n)},\; k = 0,1,\ldots,n-1$,
adapted to the filtration
$\mathbb{F}^{(n,\alpha_{\Delta t})} = \left\{\mathcal{F}_{k \Delta t}
= \sigma\left(M_1^{(\alpha_{\Delta t})},\ldots,M_k^{(\alpha_{\Delta t})}\right)\;k = 1,\ldots,n,\;
\mathcal{F}_0 = \left\{\varnothing,\Omega\right\}\right\}$.
Then, conditionally on $\mathcal{F}_{k\Delta t}$, the replicating portfolio should be riskless,
that is $P_{(k+1)\Delta t}^{(n)} = D_{k\Delta t}^{(n)}S_{(k+1)\Delta t}^{(n)}-f_{(k+1)\Delta t}^{(n)}$.
From (\ref{eq_trans_prob}) and (\ref{eq_GJR_pricing_tree}), and defining\footnote{
	Here, ``u" stands for upward movement and ``d" for downward movement in the binomial tree.} 
\begin{align*}
S_{(k+1)\Delta t}^{(n,u)} &= S_0\exp\left(v_{k+1}\Delta t+ \left(M_k^{(\alpha_{\Delta t})}+1\right)\sigma\sqrt{\Delta t}\right),
\\
S_{(k+1)\Delta t}^{(n,d)} &= S_0\exp\left(v_{k+1}\Delta t+\left(M_k^{(\alpha_{\Delta t})}-1\right)\sigma\sqrt{\Delta t}\right),
\end{align*}
conditionally on $\mathcal{F}_{k\Delta t}$, by (\ref{eq_GJR_mean_var2}) it follows that 
\begin{equation*}
\begin{aligned}
S_{(k+1)\Delta t}^{(n,u)} &= S_{k\Delta t}^{(n)}\exp\left(\mu\Delta t + \sqrt{2/\pi}\sigma\beta\left(\sqrt{k+1}-\sqrt{k}\right)\Delta t
							+ \sigma\sqrt{\Delta t}\right), \\
S_{(k+1)\Delta t}^{(n,d)} &= S_{k\Delta t}^{(n)}\exp\left(\mu\Delta t + \sqrt{2/\pi}\sigma\beta\left(\sqrt{k+1}-\sqrt{k}\right)\Delta t 
							- \sigma\sqrt{\Delta t}\right).
\label{eq_binomial_inP}
\end{aligned}
\end{equation*}
The risk-neutrality assumption implies that
$D_{k\Delta t}^{(n)}S_{(k+1)\Delta t}^{(n,u)}-f_{(k+1)\Delta t}^{(n,u)}
= D_{k\Delta t}^{(n)}S_{(k+1)\Delta t}^{(n,d)}-f_{(k+1)\Delta t}^{(n,d)}$, and thus
\begin{equation*}
 D_{k\Delta t}^{(n)} = \frac{ f_{(k+1)\Delta t}^{(n,u)} - f_{(k+1)\Delta t}^{(n,d)} }
 			{ S_{k\Delta t}^{(n)}\exp\left(\mu\Delta t + \sqrt{2/\pi}\sigma\beta\left(\sqrt{k+1}-\sqrt{k}\right)\Delta t\right)
 					   \left(e^{\sigma\sqrt{\Delta t}} - e^{-\sigma\sqrt{\Delta t}}\right) }.
\end{equation*}
Furthermore, given $\mathcal{F}_{k\Delta t}$, the portfolio $P_{(k+1)\Delta t}^{(n)} =P_{(k+1)\Delta t}^{(n,u)}$ is riskless, leading to
\begin{equation}
f_{k\Delta t}^{(n,u)} = D_{k\Delta t}^{(n)}S_{k\Delta t}^{(n,u)}-e^{-r_f \Delta t}P_{(k+1)\Delta t}^{(n,u)}
= e^{-r_f \Delta t}\left(q_{n,k+1}f_{(k+1)\Delta t}^{(n,u)}+(1-q_{n,k+1})f_{(k+1)\Delta t}^{(n,d)}\right),
\label{eq_f_price2}
\end{equation}
where the risk-neural probability $q_{n,k+1}$ for the time period $[k\Delta t, (k+1)\Delta t)$ is given by
\begin{equation*}
q_{n,k+1} = \frac{\exp\left(r_f \Delta t - \mu\Delta t - \sqrt{2/\pi}\sigma\beta( \sqrt{k+1} - \sqrt{k} ) \Delta t \right) - e^{\sigma\sqrt{\Delta t}}}{e^{\sigma\sqrt{\Delta t}} -e^{-\sigma\sqrt{\Delta t}}}.
\label{eq_riskneural_prob_q}
\end{equation*}
To leading term in $\Delta t$, $q_{n,k+1}$ has the form,
\begin{equation}
q_{n,k+1} = \left(1-\theta\sqrt{\Delta t}+\beta\sqrt{2\Delta t/\pi}(\sqrt{k+1}-\sqrt{k})\right)/2,
\label{eq_riskneural_prob_q2}
\end{equation}
where $\theta = \left(\mu - r_f + \sigma^2/2\right)/\sigma$ is the market price of risk.
For $\beta = 0$, we obtain the risk-neutral probabilities under the JR binomial model (see Kim et al. (2016, 2019)).

Now we consider the risk-neutral dynamics of the stock $\mathcal{S}$ under the GJR pricing tree model \eqref{eq_GJR_pricing_tree}
and the ECC dynamics \eqref{eq_f_price}.
Given $\mathcal{F}_{k \Delta t}$, to leading order in $\Delta t$, 
\begin{equation}
S_{(k+1)\Delta t}^{(n;q)} = \begin{cases}
S_{(k+1)\Delta t}^{(n,u;q)} &= S_{k\Delta t}^{(n)}\exp\left(\mu\Delta t + \sqrt{2/\pi}\sigma\beta\left(\sqrt{k+1}-\sqrt{k}\right)\Delta t
											+ \sigma\sqrt{\Delta t}\right),\; \textrm{w.p.}\;q_{n,k+1}, \\
S_{(k+1)\Delta t}^{(n,d;q)} &= S_{k\Delta t}^{(n)}\exp\left(\mu\Delta t + \sqrt{2/\pi}\sigma\beta\left(\sqrt{k+1}-\sqrt{k}\right)\Delta t
											 - \sigma\sqrt{\Delta t}\right),\; \textrm{w.p.}\;1-q_{n,k+1},
\end{cases}
\label{eq_binomialtree_32}
\end{equation}
where $k = 1,\ldots,n-1$.
Conditionally on $\mathcal{F}_{k\Delta t}$, the discrete risk-neutral return
$R_{(k+1)\Delta t}^{(n;q)} = \ln\left(S_{(k+1)\Delta t}^{(n;q)}/S_{k\Delta t}^{(n;q)}\right)$,
has mean $\mathbb{E}\left(R_{(k+1)\Delta t}^{(n;q)}\right) = \left(r_f - \sigma^2/2\right)\Delta t$
and variance $\textrm{Var}\left(R_{(k+1)\Delta t}^{(n;q)}\right) = \sigma^2\Delta t$,
and for $\gamma>2$, $\mathbb{E}\left(\left| R_{(k+1)\Delta t}^{(n;q)}\right| ^{\gamma}\right) = o(\Delta t) = 0$.
Let
\begin{equation*}
\begin{aligned}
\mathbb{S}_{[0,T]}^{(n;q)} &= \left\{S_t^{(n;q)} = S_{(k\Delta t)}^{(n;q)},\; t\in [k\Delta t,(k+1)\Delta t),\;
	k = 0,\ldots,n-1,\;S_T^{(n;q)} = S_{n\Delta t}^{(n;q)}\right\}, \\
\mathbb{S}_{[0,T]}^{(q)} &= \left\{S_0\exp\left((r_f - \sigma^2/2)t+\sigma B_t^{(q)}\right),\;t\in[0,T]\right\},
\end{aligned} 
\end{equation*}
where $\mathbb{B}_{[0,T]}^{(q)} = \left\{B_t^{(q)},\; t\in[0,T]\right\}$ is a standard BM.
Then, by the Donsker-Prokhorov invariance principle, it follows that $\mathbb{S}_{[0,T]}^{(n,q)}$ converges weakly
in $\left(\mathcal{D}[0,T],d^{(0)}\right)$ to $\mathbb{S}_{[0,T]}^{(q)}$.

\subsection{Implied  $\mathbf{\mu}$, $\mathbf{\beta}$, and $\mathbf{\sigma}$ surfaces}
\label{sec31}
\noindent Following this framework,
we use the data from section \ref{sec24} and the market option prices\footnote{
	The SPY call option data was collected from Bloomberg Professional Services on 6/1/2021, 19:32 EST.
	The data set includes call options,  with all strike values, having expiration date no later than 12/31/2021.
	In total the data involves $M = 1,913$ contracts with valid bid and ask quotes.
	The SPY spot price for this date and time was \$419.67.}
for the underlying SPY asset to compute implied $\mu$,  $\beta$, and $\sigma$ surfaces.
Let $C_i^{(\textrm{SPY,Market})}\left(S_0^{(\textrm{SPY})},K,T,t,r_f\right)$ denote the price of the $i^{\textrm{th}}$
market call option contract on day $t$
where: $S_0^{(\textrm{SPY})}$ is the spot price of the underlying SPY ETF;
$K$ is the strike price; $T$ is the terminal time; and $r_f$ is the risk-free rate.\footnote{
	We use 10-year Treasury yield curve rates for risk-free rate values;
	the annual rate was $r_f = 1.62\%$ on 6/1/2021.}
For brevity, we simply refer to $C_i^{(\textrm{SPY,Market})}$.

Applying equations \eqref{eq_f_price2}, \eqref{eq_riskneural_prob_q2} and \eqref{eq_binomialtree_32},
we construct the GJR pricing tree's theoretical call option price for day $t$,
denoted as $C_i^{(\textrm{SPY,GJR})}\left(S_0^{(\textrm{SPY})},K,T,t,r_f,\rho\right)$
For brevity, we refer to this as $C_i^{(\textrm{SPY,GJR})}(\rho)$,
where $\rho = \mu, \beta$ or $\sigma$ is the parameter indicating the surface to be computed.
For example, to obtain the implied $\mu$ surface, we use the estimated values for $\bar{\sigma}$ and $\bar{\beta}$
from section \ref{sec24} and designate
$\rho = \rho^{(\mu)} = (\mu \;|\; \bar{\sigma}^{(\alpha)} = 0.151,\ \bar{\beta}^{(\alpha)} = -0.978)$.
We estimate the value for $\rho^{(\mu)}_i $ for the  $i^{\textrm{th}}$ call option contract by
\begin{equation}
\hat{\rho}^{(\mu)}_i = \textrm{arg\;min}\left\{\left(\frac{C_i^{(\textrm{SPY,GJR})}\left(\rho\right)-C_i^{(\textrm{SPY,Market})}}{C_i^{(\textrm{SPY,Market})}}\right)^2\right\},\;i = 1,...,M.
\label{eq_implied_para}
\end{equation}
We proceed analogously using parameters $\rho^{(\beta)}$ and $\rho^{(\sigma)}$ to compute the $\beta$ and $\sigma$ surfaces. 

\begin{figure}[ht]
\begin{center}
    \begin{subfigure}[b]{0.32\textwidth} 
    	\includegraphics[width=\textwidth]{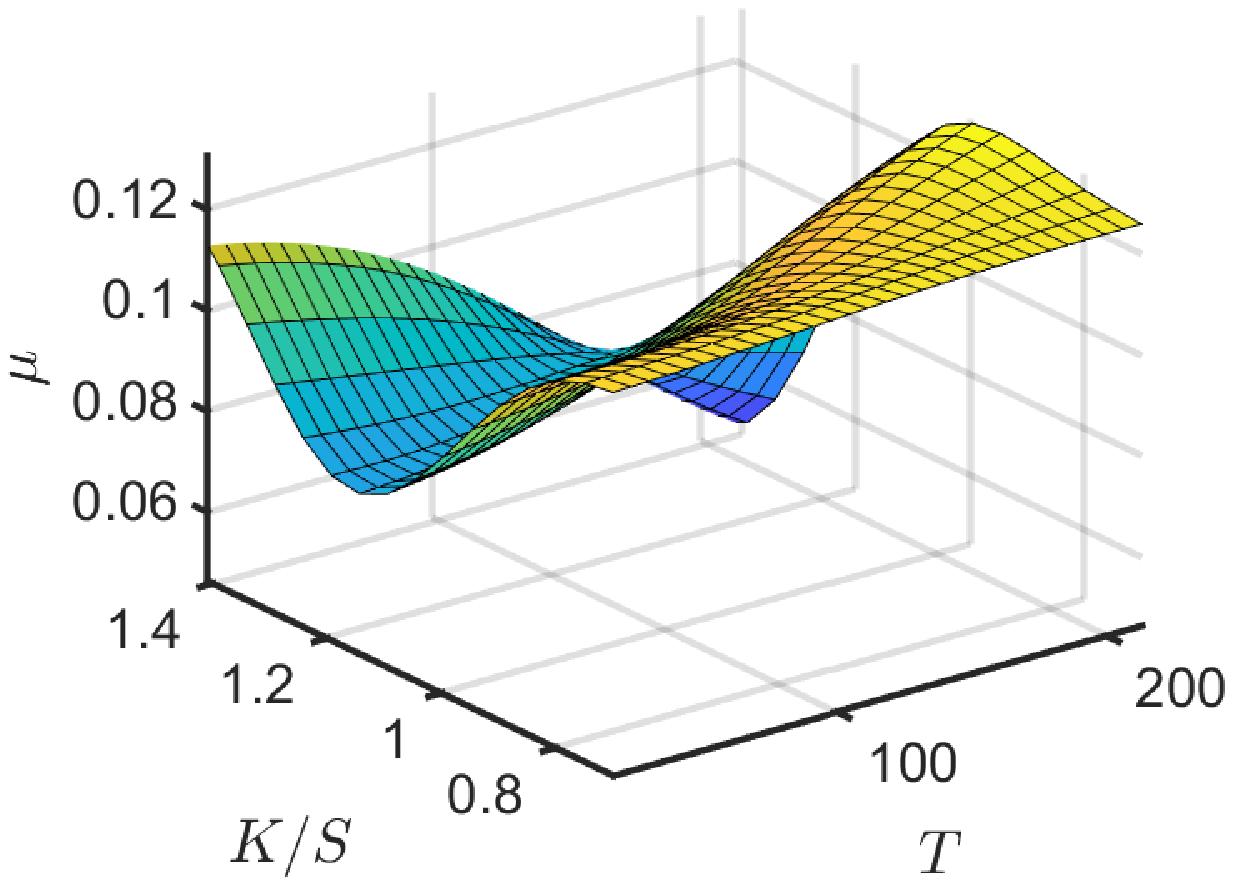}
    	\caption{}
    	\label{fig7_P8_mu}
    \end{subfigure}
    \begin{subfigure}[b]{0.32\textwidth} 
    	\includegraphics[width=\textwidth]{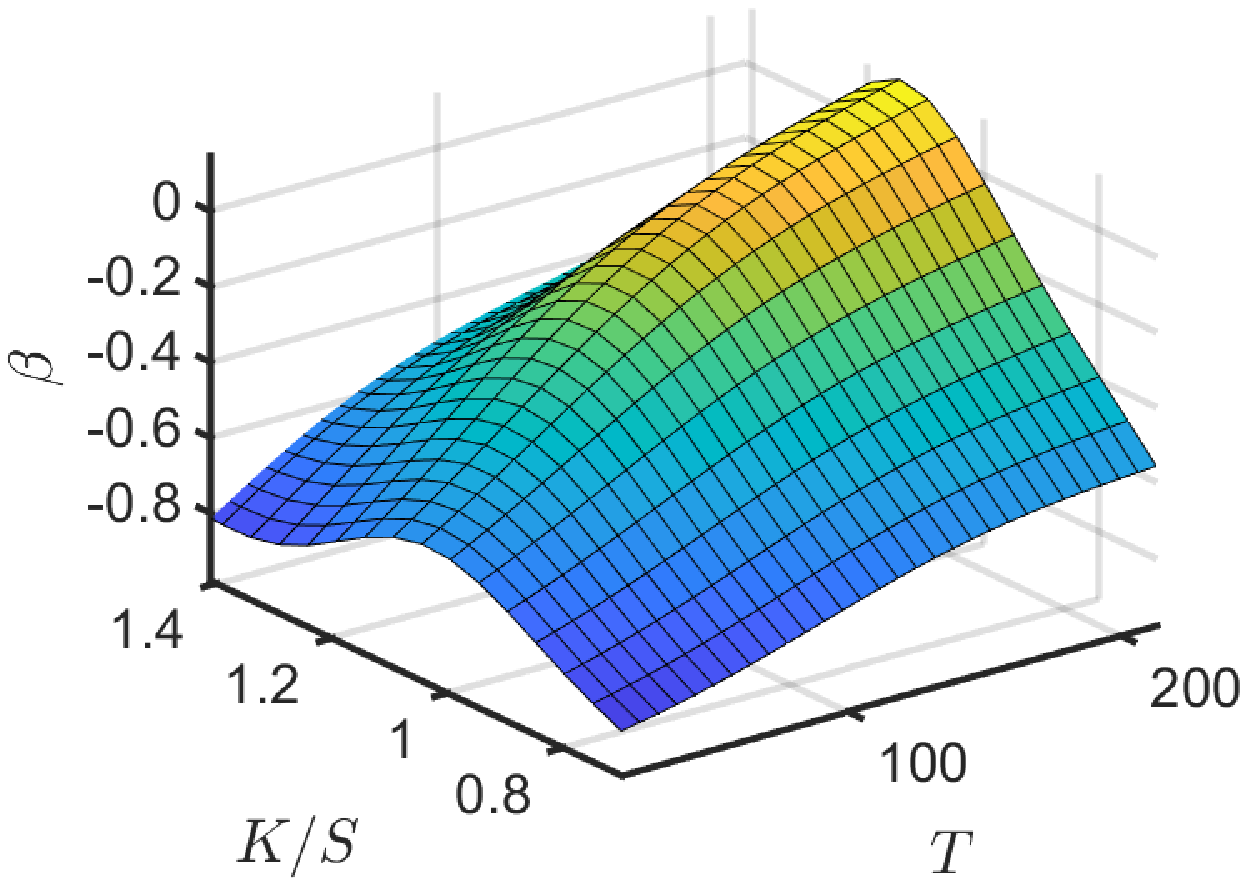}
    	\caption{}
    	\label{fig7_P7_beta}
    \end{subfigure}
    \caption{The implied (a) $\mu$  and (b) $\beta$ surfaces generated by the GJR pricing
     tree plotted as a function of moneyness $K/S$ and time to maturity $T$ (in days).}
    \label{fig6}
\end{center}
\end{figure}

Fig. \ref{fig6} shows the results for the implied $\mu$ and $\beta$ surfaces,
plotted in terms of moneyness $K/S$ and time to maturity $T$ (in days),
where $K$ is the strike price and $S$ is the current spot price of the underlying asset.
Of the two, the $\mu$ surface has the more complex behavior with different time-development of the surface as $K/S$
ranges from ``in the money'' to ``out of the money'' values.
In contrast, values of $\hat{\beta}$ decrease on both sides of a ridge of values that generally aligns with $K/S \sim 1.1$.
Values of $\beta$, ranging from $-0.978$ to $0.147$, are mostly negative indicating a negatively skewed pricing tree.
Values of $\beta$ increase with time at any constant value of $K/S$.

\begin{figure}[ht]
\begin{center}
    \begin{subfigure}[b]{0.32\textwidth} 
    	\includegraphics[width=\textwidth]{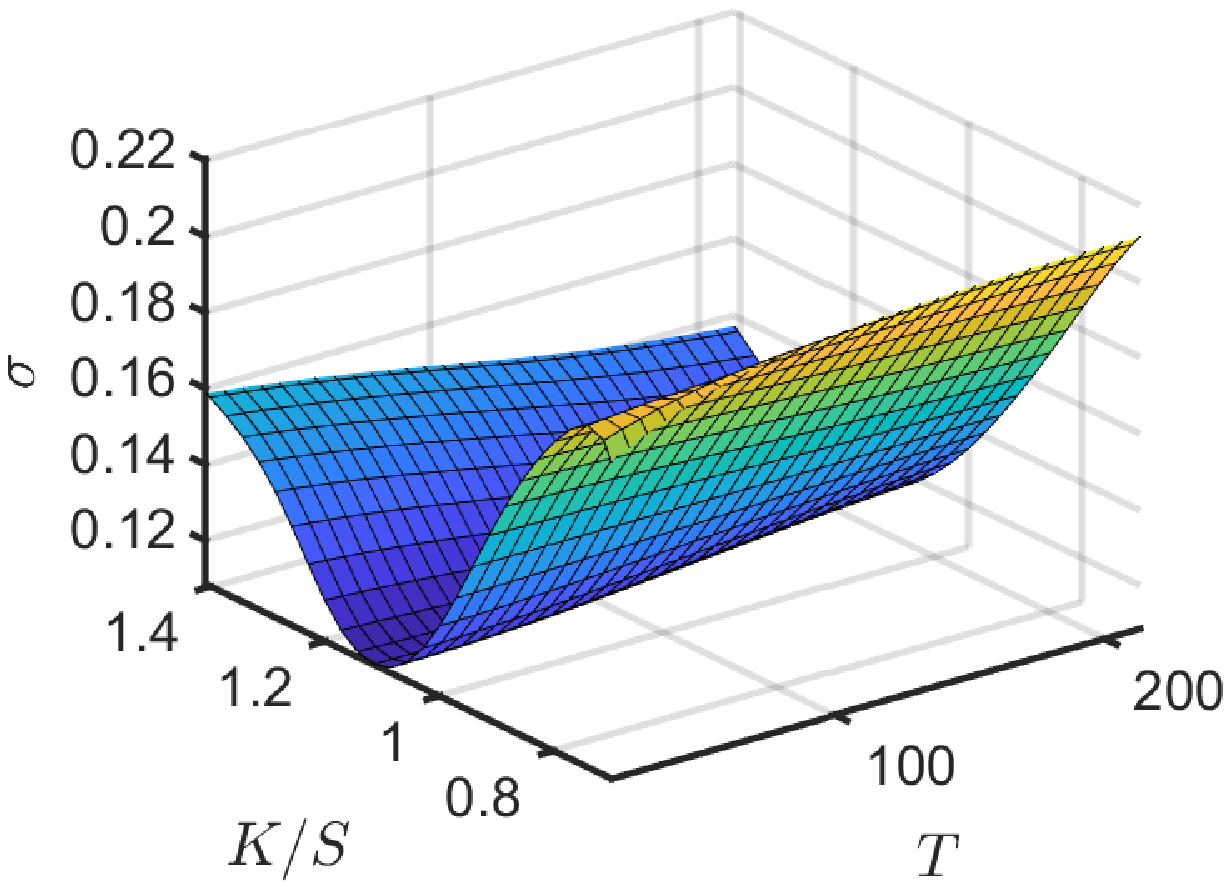}
    	\caption{}
    	\label{fig7_GJR}
    \end{subfigure}
    \begin{subfigure}[b]{0.32\textwidth} 
    	\includegraphics[width=\textwidth]{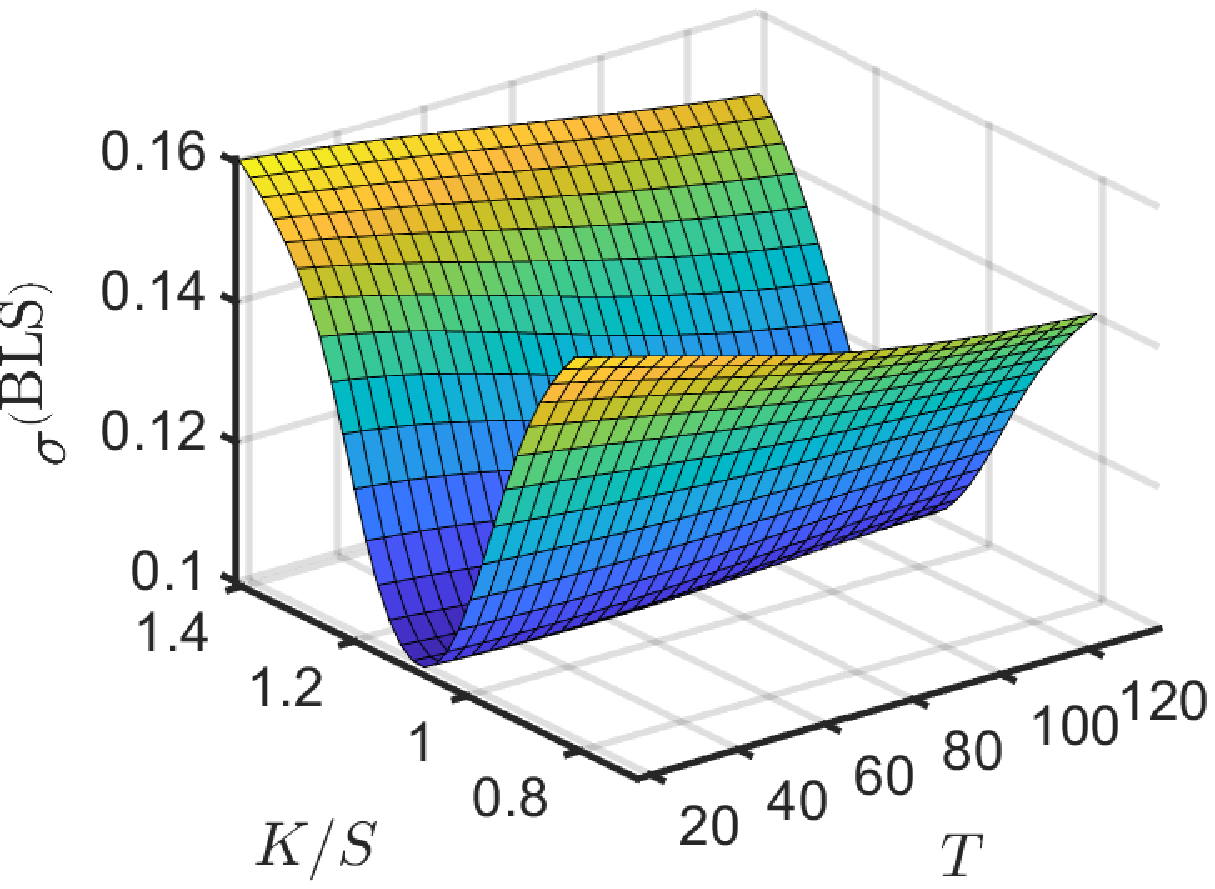}
    	\caption{}
    	\label{fig7_BLS}
    \end{subfigure}
    \begin{subfigure}[b]{0.32\textwidth} 
    	\includegraphics[width=\textwidth]{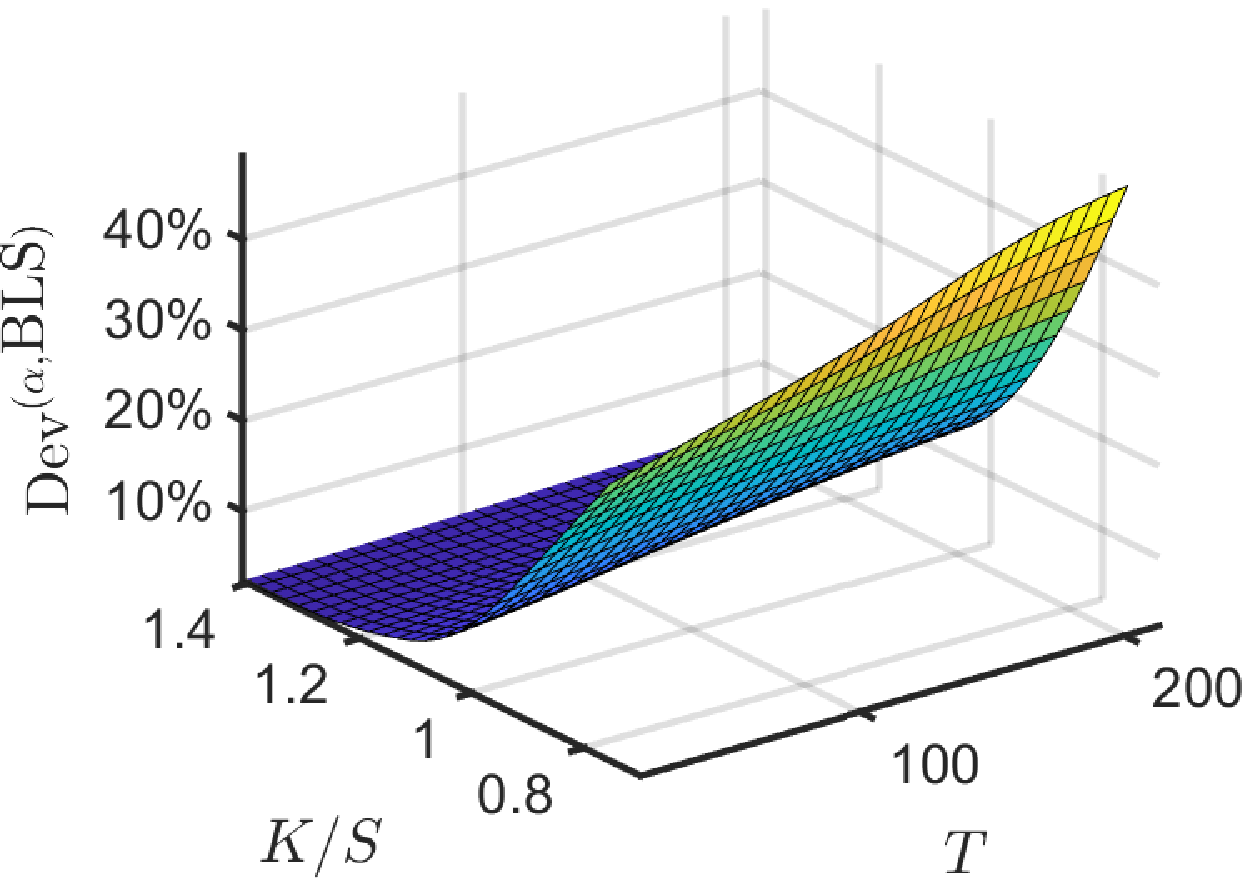}
    	\caption{}
    	\label{fig7_dev}
    \end{subfigure}
    \caption{(a) The implied $\sigma$ surface generated by the GJR pricing tree.
    		(b) The implied volatility surface, $\sigma^{(\textrm{BLS})}$, using the Black-Scholes formula.
    		(c) The deviation surface $\textrm{Dev}^{(\alpha,\textrm{BLS})}$.
    		 Each surface is plotted as a function of moneyness $K/S$ and time to maturity $T$ (in days).}
    \label{fig7}
\end{center}
\end{figure}

To evaluate the implied $\sigma$ surface,
we also calculated the Black-Scholes implied volatility surface\footnote{
	See \citet[Chapter 15]{Hull2012}.}
$\sigma^{\textrm{(BLS)}}$  for the same option prices, and considered their difference using the percent deviation,
$\textrm{Dev}^{(\alpha,\textrm{BLS})} = 100 \left( \sigma - \sigma^{\textrm{(BLS)}} \right) / \sigma^{\textrm{(BLS)}}$.
These surfaces are shown in Fig.~\ref{fig7}.
Both surfaces,  $\sigma$ and $\sigma^{\textrm{(BLS)}}$, show similar volatility smiles and roughly similar values when $K/S > 1$.
However, the $\sigma$ surface increases more rapidly than $\sigma^{\textrm{(BLS)}}$ as $K/S$ moves deeper into the money.

\section{GJR option price with hedging transaction costs}
\label{sec4}
\noindent
There is a vast literature on option pricing incorporating transaction costs.\footnote{
	Some basic references are: \cite{Leland1985}; \cite{Hodges1989}; \cite{Boyle1992}; \cite{Davis1993};
	\cite{Edirisinghe1993}; \cite{Kabanov1997}; \cite{Broadie1998}; \cite{Kabanov2001}; \cite{Palmer2001};
	\cite{Lai2009}; and \cite{Guasoni2012}.}
Discrete-time option pricing models have been studied under various assumptions
regarding the types of the transaction costs incurred in trading the replicating self-financing portfolios.\footnote{
	See \citet[Chapter 14]{Merton1990}, \cite{Stettner1997}, \cite{Palmer2001}, \cite{Delbaen2002},
	\cite{Melnikov2005}, and \cite{Chen2008}.}
Here, we extend the GJR option pricing model to the case when the hedging is subject to transaction costs.
Our approach in extending the GJR pricing tree model to include transaction costs is based on Merton’s binomial
option pricing model with transaction costs.\footnote{
	See \citet[Chapter 14]{Merton1990}.}

Consider a market with three securities $\left(\mathcal{S},\mathcal{B},\mathcal{C}\right)$ consisting of:
(a) a stock $\mathcal{S}$ with pricing tree $\{S_{k\Delta t}^{(n)},\;k = 1,\ldots,n,\;n\Delta t = T \}$
given by (\ref{eq_GJR_pricing_tree}),
(b) a bond $\mathcal{B}$ with price dynamics given by (\ref{eq_bond}), and
(c) an ECC $\mathcal{C}$ with price dynamics given by (\ref{eq_f_price}).
In contrast to the previous section,
we assume that the hedger (the ECC-contract seller) trades the replicating risk-neutral portfolio
$P_{k\Delta t}^{(n)} = D_{k\Delta t}^{(n)}S_{k\Delta t}^{(n)} - f_{k\Delta t}^{(n)}$ at a cost.
Namely, given
$\mathcal{F}_{k\Delta t} = \sigma\left(M_1^{(\alpha_{\Delta t})},\ldots,M_k^{(\alpha_{\Delta t})}\right)$,
the hedged portfolio at $(k+1)\Delta t$ is determined by 
\begin{equation}
\begin{aligned}
P_{(k+1)\Delta t}^{(n,u,\lambda)} &= D_{k\Delta t}^{(n)}S_{(k+1)\Delta t}^{(n,u)}+\lambda_{\Delta t}D_{k\Delta t}^{(n)}\left(S_{(k+1)\Delta t}^{(n,u)}-S_{k\Delta t}^{(n)}\right)-f_{(k+1)\Delta t}^{(n,u,\lambda)},
\\
P_{(k+1)\Delta t}^{(n,d,\lambda)} &= D_{k\Delta t}^{(n)}S_{(k+1)\Delta t}^{(n,d)}+\lambda_{\Delta t}D_{k\Delta t}^{(n)}\left(S_{(k+1)\Delta t}^{(n,d)}-S_{k\Delta t}^{(n)}\right)-f_{(k+1)\Delta t}^{(n,d,\lambda)}.
\end{aligned}
\label{eq_hedge_port}
\end{equation}
In (\ref{eq_hedge_port}),
$\lambda_{\Delta t} = \lambda^{(0)}+\lambda^{(1)}\sqrt{\Delta t},\;\lambda^{(0)} > 0$,
$\lambda^{(1)} \in R$,
is the hedging transaction cost (HTC).
From (\ref{eq_hedge_port}), to leading order in $\Delta t$, it follows that
\begin{equation}
D_{k\Delta t}^{(n)} = \frac{f_{(k+1)\Delta t}^{(n,u,\lambda)}  -f_{(k+1)\Delta t}^{(n,d,\lambda)}}{2S_{k\Delta t}^{(n)}\left(1+\lambda_{\Delta t}\right)\sigma\sqrt{\Delta t}}.
\label{eq_dkt}
\end{equation}
As in (\ref{eq_f_price2}) and (\ref{eq_riskneural_prob_q2}), we obtain the risk-neutral valuation of the ECC,
\begin{equation}
f_{k\Delta t}^{(n,\lambda)} = D_{k\Delta t}^{(n)}S_{k\Delta t}^{(n)} - e^{-r_f \Delta t}P_{(k+1)\Delta t}^{(n,u,\lambda)}
= e^{-r_f \Delta t}\left(q_{n,k+1}^{(\lambda)}f_{(k+1)\Delta t}^{(n,u,\lambda)}
+ (1-q_{n,k+1}^{(\lambda)})f_{(k+1)\Delta t}^{(n,d,\lambda)}\right),
\label{eq_fkt_41}
\end{equation}
where the risk-neutral probability $q_{n,k+1}^{(\lambda)}$ has the form
\begin{equation}
q_{n,k+1}^{(\lambda)} = \frac{1}{2}\left(1-\theta^{(\lambda^{(0)})}\sqrt{\Delta t}-\beta(\sqrt{1+k}- \sqrt{k})\sqrt{\frac{2\Delta t}{\pi}}\right)
-\frac{\lambda^{(1)} r_f \Delta t}{2\sigma(1+\lambda^{(0)})^2},
\label{eq_riskneutral_q_gimel}
\end{equation}
with $\theta^{(\lambda^{(0)})} = \left( \mu + \sigma^2/2 - r_f /\left( 1 + \lambda^{(0)} \right) \right)/\sigma$.\footnote{
	Formula (\ref{eq_dkt}) for the delta-position $D_{k\Delta t}^{(n)}$ is similar to formula (14.2a)
	in \citet[Chapter 14]{Merton1990}.}
In the special case, $\lambda^{(0)} = \lambda^{(1)} = 0$,
(\ref{eq_riskneutral_q_gimel}) coincides with (\ref{eq_riskneural_prob_q2}).

Now consider the risk-neutral dynamics of the stock S in the presence of HTC.
Conditional on $\mathcal{F}_{k\Delta t},\;k = 1,\ldots,n-1$, the risk-neutral value of the stock price at $(k+1)\Delta t$ is determined by
\begin{equation}
S_{(k+1)\Delta t}^{(n;q,\lambda)} =
\begin{cases}
S_{(k+1)\Delta t}^{(n,u;q,\lambda)} &= S_{k\Delta t}^{(n)}\exp\left(\mu\Delta t+\sigma\beta\left(\sqrt{k+1}-\sqrt{k}\right)\sqrt{2/\pi}\Delta t + \sigma\sqrt{\Delta t}\right),\; \textrm{w.p.}\;q_{n,k+1}^{(\lambda)},
\\
S_{(k+1)\Delta t}^{(n,d;q,\lambda)} &= S_{k\Delta t}^{(n)}\exp\left(\mu\Delta t+\sigma\beta\left(\sqrt{k+1}-\sqrt{k}\right)\sqrt{2/\pi}\Delta t - \sigma\sqrt{\Delta t}\right),\; \textrm{w.p.}\;1-q_{n,k+1}^{(\lambda)},
\end{cases}
\label{eq_sec42_tree}
\end{equation}
where $q_{n,k+1}^{(\lambda)}$ is given by (\ref{eq_riskneutral_q_gimel}).
Conditionally on $\mathcal{F}_{k\Delta t}$, the discrete risk-neutral log-return
$R_{(k+1)\Delta t}^{(n;q,\lambda)}=\ln\left(S_{(k+1)\Delta t}^{(n;q,\lambda)}/S_{k\Delta t}^{(n;q,\lambda)}\right)$ has mean and variance 
\begin{equation*}
\begin{aligned}
\mathbb{E}\left(R_{(k+1)\Delta t}^{(n;q,\lambda)}\mid \mathcal{F}_{k\Delta t}\right) &= \left(\frac{r_f}{1+\lambda^{(0)}}-\frac{\sigma^2}{2}\right)\Delta t,
\\
\textrm{Var}\left(R_{(k+1)\Delta t}^{(n;q,\lambda)}\mid \mathcal{F}_{k\Delta t}\right) &= \sigma^2\Delta t.
\end{aligned} 
\end{equation*}
Setting
\begin{equation}
\mathbb{S}_{[0,T]}^{(n;q,\lambda)} = \left\{S_t^{(n;q,\lambda)} = S_{k\Delta t}^{(n;q,\lambda)},\;t\in [k\Delta t,(k+1)\Delta t),\;k = 0,\ldots,n-1,\;S_T^{(n;q)} = S_{n\Delta t}^{(n)}\right\},
\label{eq_priceS_gimel}
\end{equation}
and
\begin{equation}
\mathbb{S}_{[0,T]}^{(q,\lambda)} = \left\{S_0\exp\left(\left(\frac{r_f}{1+\lambda^{(0)}}-\frac{\sigma^2}{2}\right)t+\sigma B_t^{(q)}\right),\;t\in[0,T]\right\},
\label{eq_priceB_gimel}
\end{equation}
where $\mathbb{B}_{[0,T]}^{(q)} = \left\{B_t^{(q)},\;t\in [0,T]\right\}$ is a standard BM,
we have that $\mathbb{S}_{[0,T]}^{(n,q)}$ converges weakly in $\left(\mathcal{D}[0,T],d^{(0)}\right)$ to $\mathbb{S}_{[0,T]}^{(q)}$.
From (\ref{eq_priceB_gimel}), it follows that the value of
$\lambda^{(1)}$ in the HTC is irrelevant in continuous-time trading.
However, for every fixed $\Delta t$, the risk-neutral tree (\ref{eq_binomialtree_32}) will exhibit the behavior of a geometric SBM
and the value of $\lambda^{(1)}$ becomes relevant, as evident from (\ref{eq_fkt_41}).

\noindent
{\bf Remark}: It appears natural to assume that HTC could be asymmetric;
that is, instead of (\ref{eq_hedge_port}), we could assume that
\begin{equation}
\begin{aligned}
P_{(k+1)\Delta t}^{(n,u,\lambda)} &= D_{k\Delta t}^{(n)}S_{(k+1)\Delta t}^{(n,u)}+\lambda^{(u)}_{\Delta t}D_{k\Delta t}^{(n)}\left(S_{(k+1)\Delta t}^{(n,u)}-S_{k\Delta t}^{(n)}\right)-f_{(k+1)\Delta t}^{(n,u,\lambda)},
\\
P_{(k+1)\Delta t}^{(n,d,\lambda)} &= D_{k\Delta t}^{(n)}S_{(k+1)\Delta t}^{(n,d)}+\lambda^{(d)}_{\Delta t}D_{k\Delta t}^{(n)}\left(S_{(k+1)\Delta t}^{(n,d)}-S_{k\Delta t}^{(n)}\right)-f_{(k+1)\Delta t}^{(n,d,\lambda)},
\end{aligned}
\label{eq_hegde_port2}
\end{equation}
with $\lambda_{\Delta t}^{(u)}\neq \lambda_{\Delta t}^{(d)}$, where
\begin{align*}
\lambda_{\Delta t}^{(u)} &= \lambda^{(0,u)}+\lambda^{(1,u)}\sqrt{\Delta t},\;\lambda^{(0,u)}>0,\;\lambda^{(1,u)}\in R, \\
\lambda_{\Delta t}^{(d)} &= \lambda^{(0,d)}+\lambda^{(1,d)}\sqrt{\Delta t},\;\lambda^{(0,d)}>0,\;\lambda^{(1,d)}\in R.
\end{align*}
A close inspection shows that, as $\Delta t \downarrow 0$, (\ref{eq_hegde_port2}) leads to a pricing model which is not arbitrage free.
However, for fixed $\Delta t$, the model (\ref{eq_hegde_port2}) is arbitrage free and an expression for the  risk-neutral probability
$q_{n,k+1}^{(\lambda)}$ similar to formula (\ref{eq_fkt_41}) can be readily obtained.\footnote{
	Discrete-time market models, which are free of arbitrages, but for which the corresponding continuous-time
	market model is not arbitrage-free, are known;
	see, for example, \cite{Karandikar1993} and \cite{Hurst1999}.
	In those models, the discrete equivalent martingale measure explodes in the limit.
	Similarly, if we assume that (\ref{eq_hegde_port2}) holds, the limiting martingale measure does not exist.}
	
\subsection{Implied surfaces with the inclusion of HTC}
\label{sec41}
\begin{figure}[ht]
\begin{center} 
	\begin{subfigure}[b]{0.32\textwidth} 
		\includegraphics[width=\textwidth]{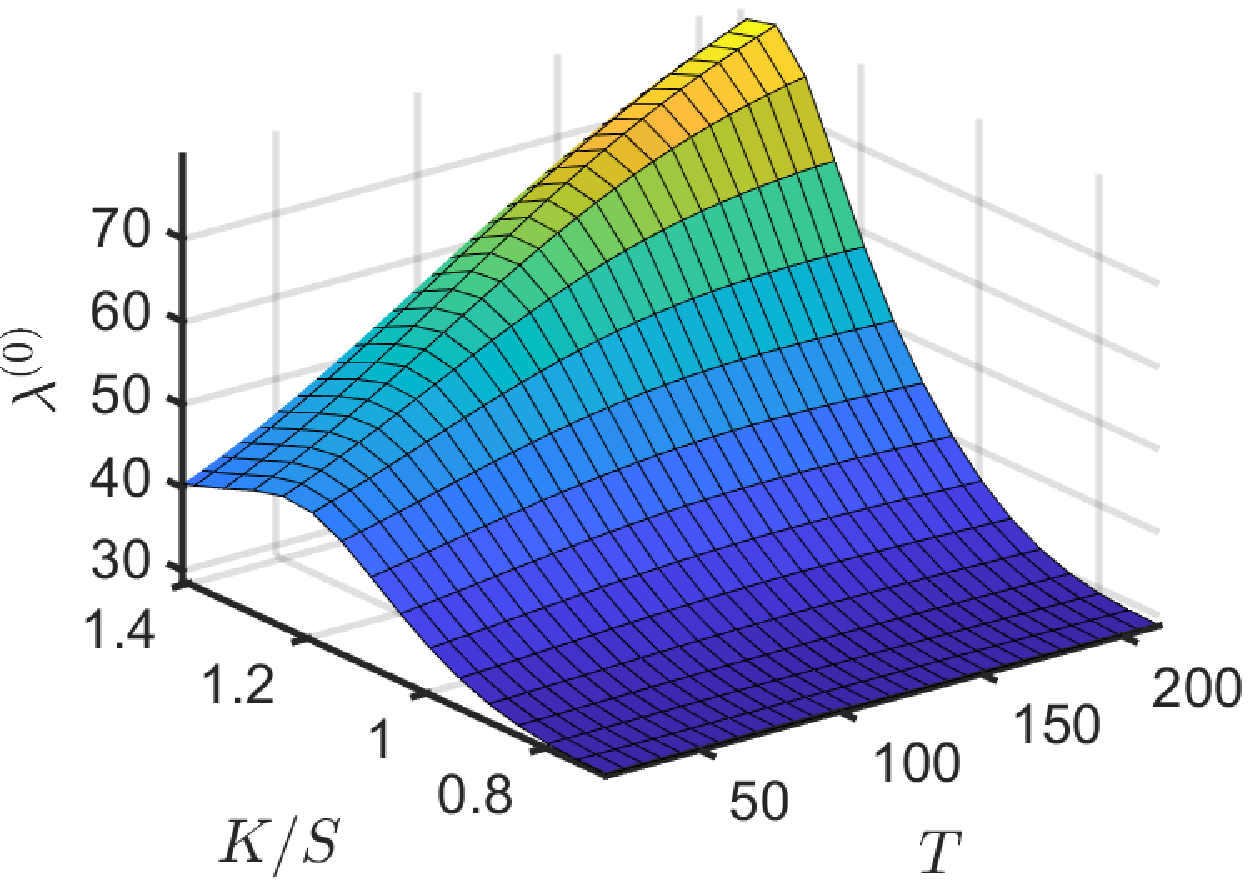}
	\end{subfigure}
	\begin{subfigure}[b]{0.32\textwidth}
		\includegraphics[width=\textwidth]{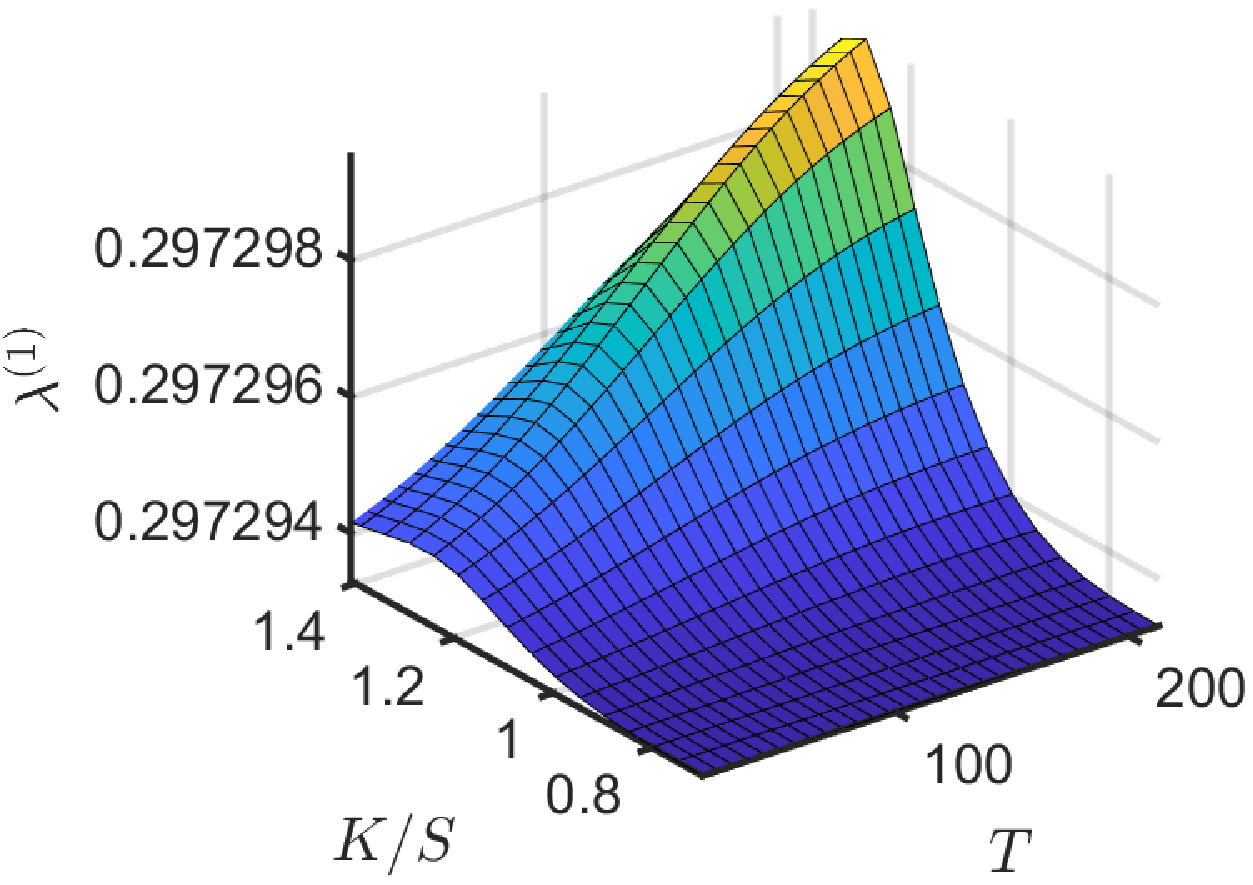}
	\end{subfigure}
	\caption{The implied $\lambda^{(0)}$ and $\lambda^{(1)}$ surfaces based on the GJR pricing tree
	 with HTC plotted as a function of moneyness $K/S$ and time to maturity $T$ (in days).}
    \label{fig8}
\end{center}
\end{figure}

\noindent
As in section \ref{sec31}, we numerically illustrate the HTC and its limiting behavior using the same data set,
including closing and call option prices, for the underlying asset SPY .
Following the notation in section \ref{sec31}, we set the parameter
$\rho = \rho^{(\lambda)}
= \left(\lambda^{(0)},\lambda^{(1)}\; |\; \bar{\mu}^{(\alpha)},\bar{\beta}^{(\alpha)},\bar{\sigma}^{(\alpha)}\right)$.
From \eqref{eq_fkt_41}, \eqref{eq_riskneutral_q_gimel} and \eqref{eq_sec42_tree},
we construct the GJR tree option price $C_i^{(\textrm{SPY,GJR})}(\rho^{(\lambda)})$.
We find the optimal solution for $\rho^{(\lambda)}$ by minimizating the relative mean-square error (relMSE)
\begin{equation}
\min_{\lambda^{(0)} > 0,\;\lambda^{(1)} \in R}\;\textrm{relMSE}
= \min_{\lambda^{(0)} > 0,\;\lambda^{(1)} \in R} \frac{1}{M}\sum_{i=1}^M
\left(\frac{C_i^{(\textrm{SPY,GJR})}\left(\rho\right)-C_i^{(\textrm{SPY,Market})}}{C_i^{(\textrm{SPY,Market})}}\right)^2.
\label{eq_implied_lambda}
\end{equation}
The resulting minimizing values are $\textrm{relMSE} = 0.433$ for $\lambda^{(0)} = 28.8$ and $\lambda^{(1)} = 0.297$.\footnote{
	The three terms on the right-hand side of \eqref{eq_hedge_port} have the form $DS + \lambda D \Delta S - f$,
	where the transaction cost term is $\lambda D \Delta S$.
	The estimated values of $\lambda^{(0)}$ and $\lambda^{(1)}$ yield $\lambda \approx \lambda^{(0)} = 28.8$.
	If daily stock price changes for the SPY fund are around 0.1\% of the fund price (i.e. $\Delta S \sim 10^{-3} S$),
	the transaction cost term would have the value $0.0288 DS$, indicating transactions costs contribute $\sim$3\% to
	the total price of the hedged portfolio.}
Using the parameters
$\rho^{(\lambda^{(0)})} = (\lambda^{(0)}\; |\;
\bar{\mu}^{(\alpha)}, \bar{\beta}^{(\alpha)}, \bar{\sigma}^{(\alpha)}, \lambda^{(1)} = 0.297)$
and $\rho^{(\lambda^{(1)})} = (\lambda^{(1)}\; |\;
\bar{\mu}^{(\alpha)}, \bar{\beta}^{(\alpha)}, \bar{\sigma}^{(\alpha)},\linebreak[1] \lambda^{(0)} = 28.8)$,
we reran the minimization problem \eqref{eq_implied_para} and computed the $\lambda^{(0)}$
and  $\lambda^{(1)}$ surfaces shown in Fig. \ref{fig8}.
Both surfaces have a similar ridge shape, but of  vastly different scales.
The value of $\lambda^{(1)}$ is constant through five computed digits, whereas $\lambda^{(0)}$ varies significantly in the
first computed digit.
Thus, for these values of $\bar{\mu}^{(\alpha)}$, $\bar{\beta}^{(\alpha)}$ and $\bar{\sigma}^{(\alpha)}$,
the constant value model, $\lambda_{\Delta t} = \lambda^{(0)}$, would suffice in \eqref{eq_hedge_port}.

\begin{figure}[ht]
\begin{center}
    \begin{subfigure}[b]{0.32\textwidth} 
    	\includegraphics[width=\textwidth]{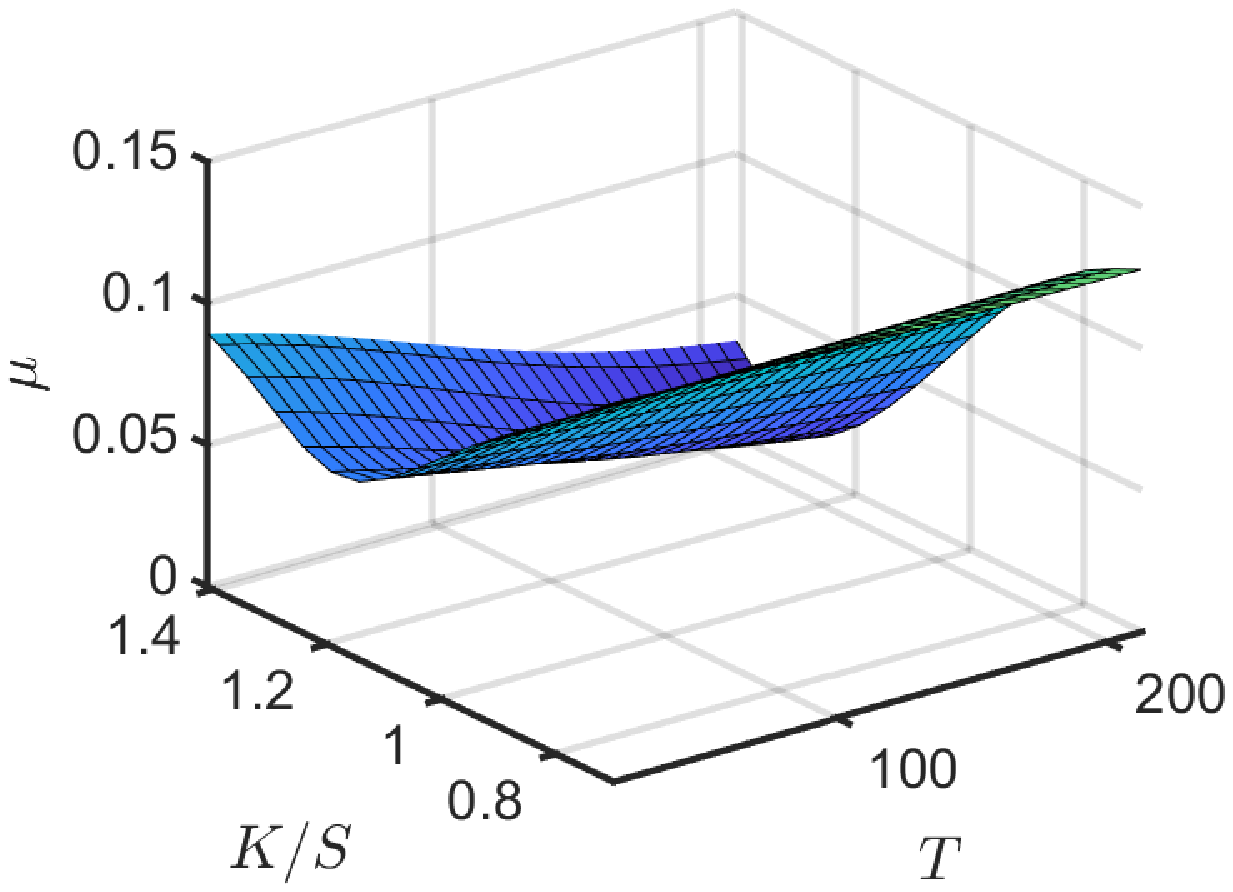}
    	\caption{}
    \end{subfigure}
    \begin{subfigure}[b]{0.32\textwidth} 
    	\includegraphics[width=\textwidth]{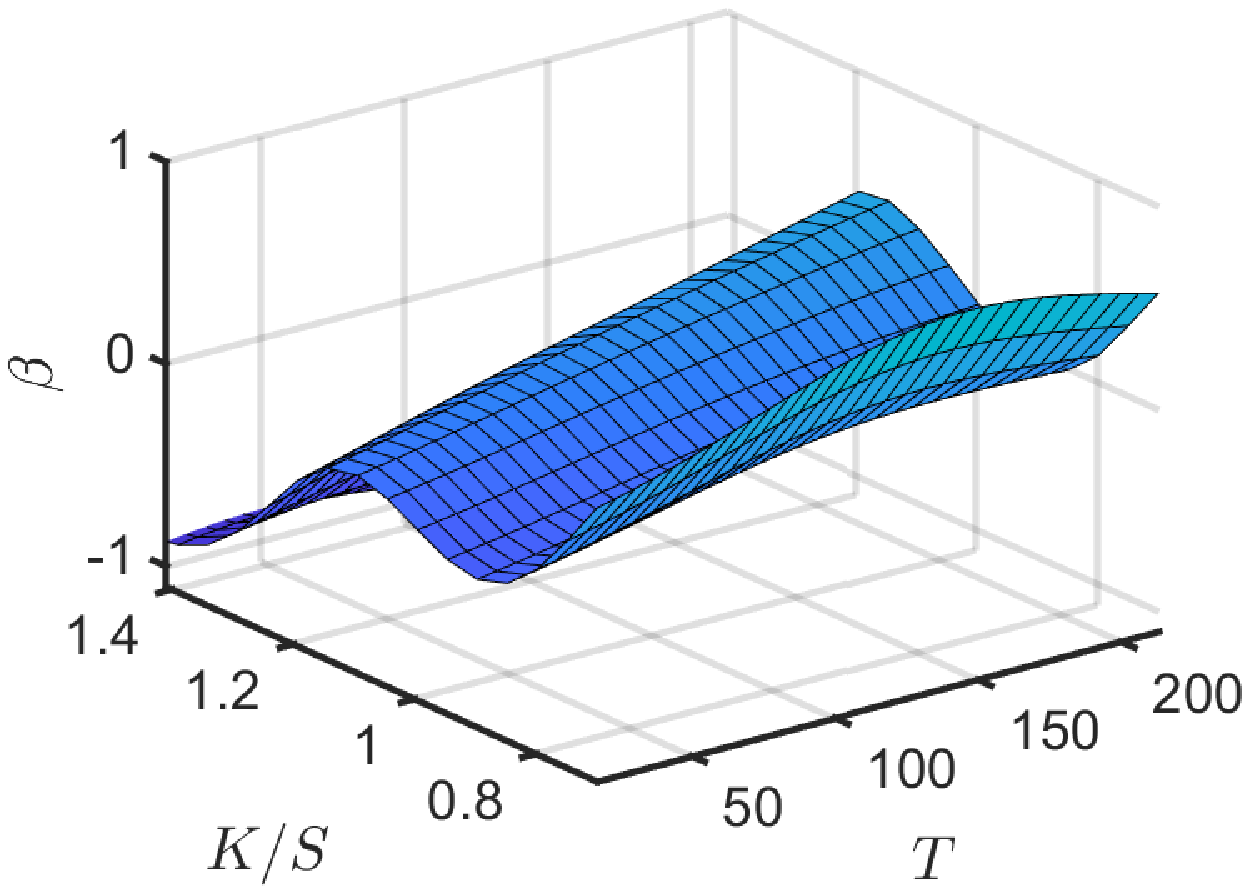}
    	\caption{}
    \end{subfigure}
    \begin{subfigure}[b]{0.32\textwidth} 
    	\includegraphics[width=\textwidth]{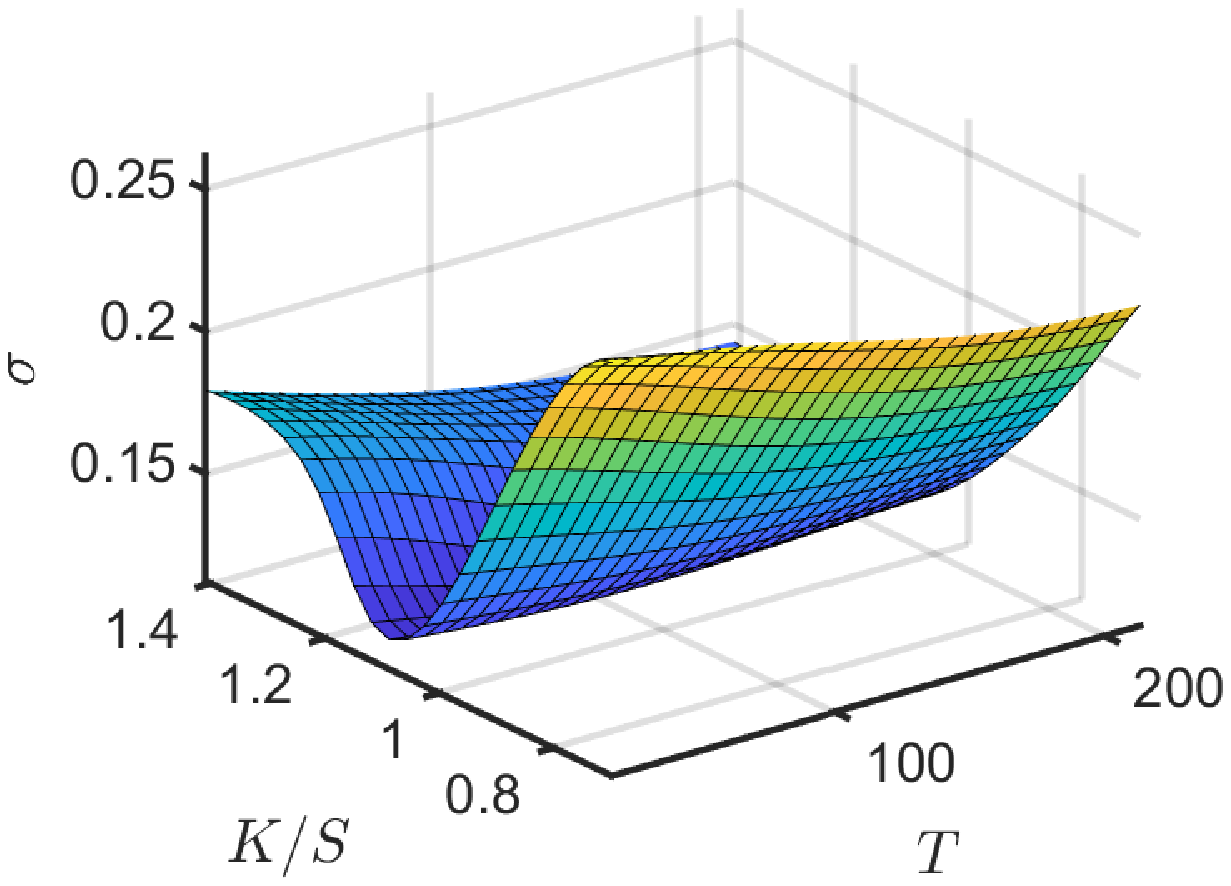}
    	\caption{}
    \end{subfigure}
        \begin{subfigure}[b]{0.32\textwidth} 
    	\includegraphics[width=\textwidth]{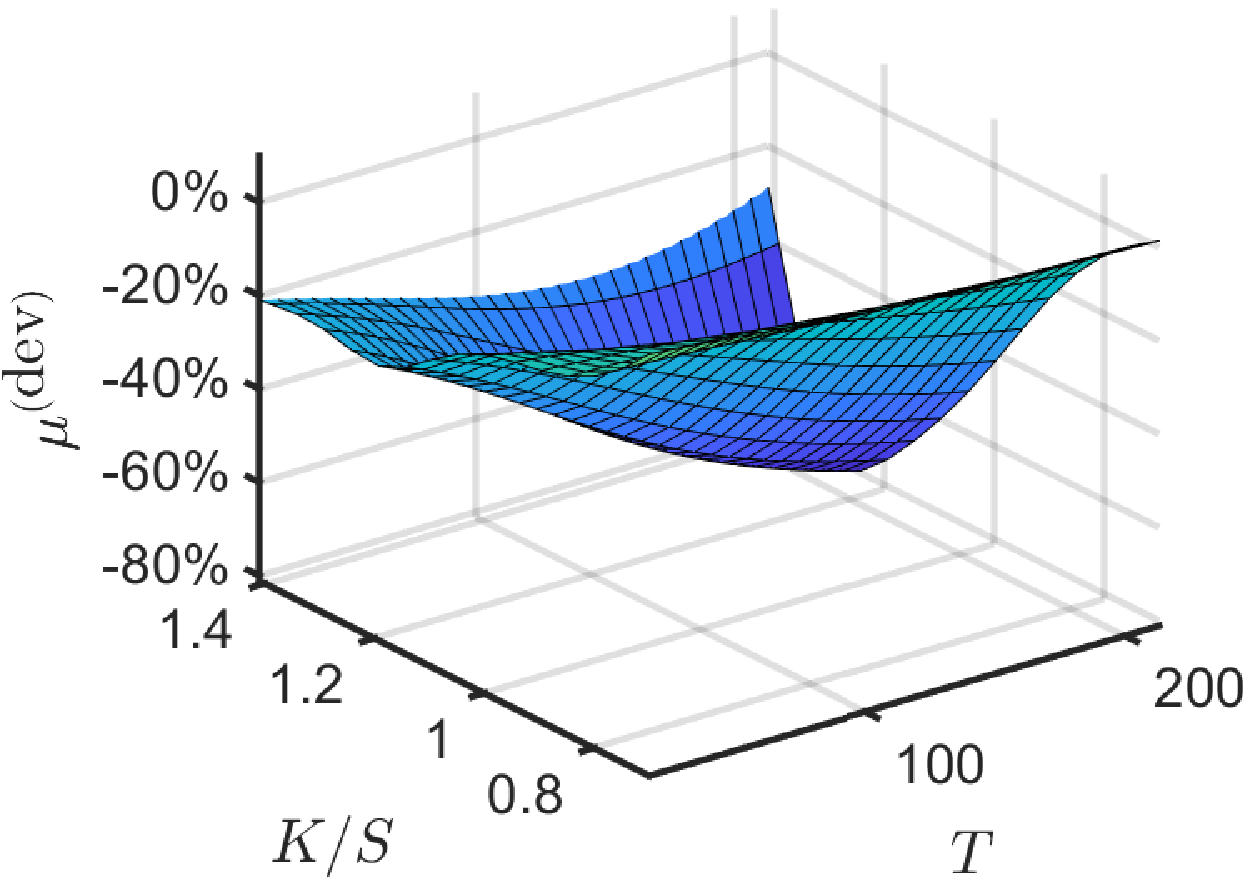}
    	\caption{}
    \end{subfigure}
    \begin{subfigure}[b]{0.32\textwidth} 
    	\includegraphics[width=\textwidth]{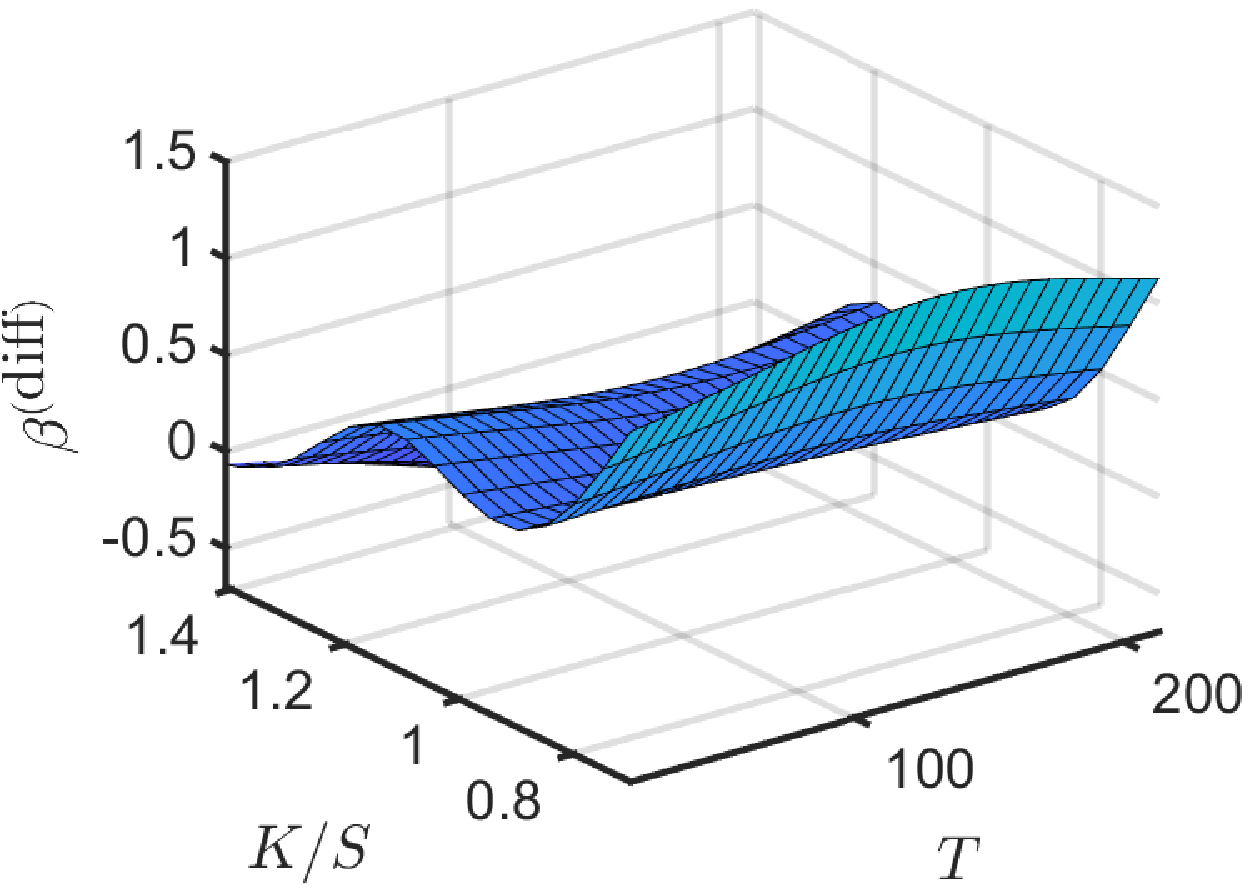}
    	\caption{}
    \end{subfigure}
    \begin{subfigure}[b]{0.32\textwidth} 
    	\includegraphics[width=\textwidth]{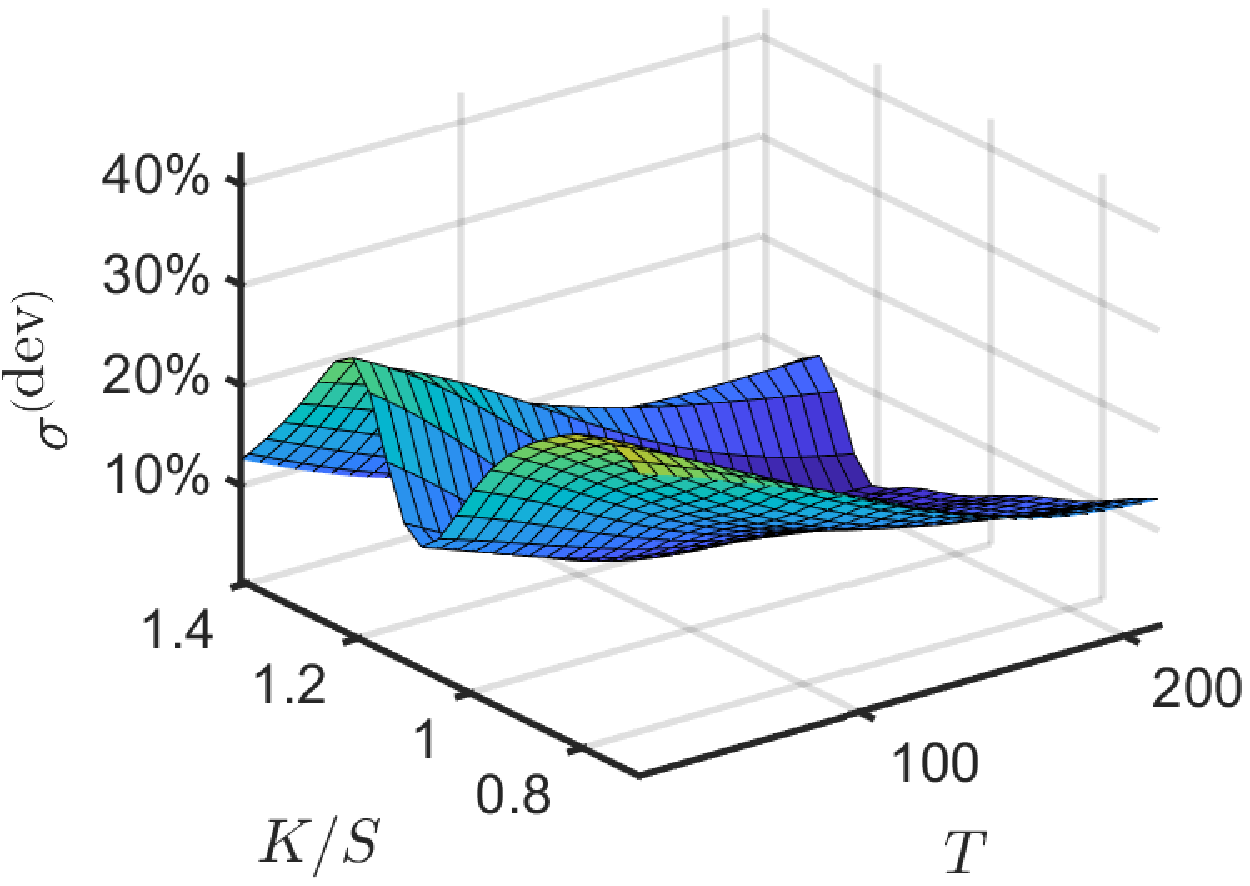}
    	\caption{}
    \end{subfigure}
    \caption{The implied (a) $\mu$, (b) $\beta$ and (c) $\sigma$ surfaces generated by the GJR pricing tree with HTC added.
	The percent deviation surfaces $\mu^{(\textrm{dev})}$ and $\sigma^{(\textrm{dev})}$ are shown in (d) and (f)
	respectively, while (e) displays the difference surface $\beta^{(\textrm{diff})}$.
	Each surface is plotted as a function of moneyness $K/S$ and time to maturity $T$ (in days).}
    \label{fig9}
\end{center}
\end{figure}

To ascertain the effects of adding HTC, 
we computed the implied $\mu$, $\beta$, and $\sigma$ surfaces using the minimizing values for $\lambda^{(0)}$ and $\lambda^{(1)}$.
(For example, for the $\mu$ surface we estimated values for
$\rho^{(\mu)} = (\mu \  | \  \bar{\sigma}^{(\alpha)}=0.151, \bar{\beta} ^{(\alpha)}=-0.978,
\lambda^{(0)} = 28.8, \lambda^{(1)} = 0.297)$ using \eqref{eq_implied_para}.)
To compare the results with those reported in section \ref{sec31}, we also computed the percent deviation
surfaces, $\rho^{(\textrm{dev})} = 100 \left( \rho^{(\textrm{HTC})} - \rho^{(\textrm{GJR})} \right) / \rho^{(\textrm{GJR})}$ for
$\rho = \{ \mu, \sigma \}$,
where $\rho^{(\textrm{GJR})}$ refers to the surface computed for the GJR model without HTC in section~\ref{sec31}
and $\rho^{(\textrm{HTC})}$ refers to the surface computed for the GJR model with HTC.
To avoid division by zero, for the $\beta$ surface comparison we plot the difference surface
$\beta^{(\textrm{diff})} = \beta^{(\textrm{HTC})} - \beta^{(\textrm{GJR})}$.
The results are shown in Fig.~\ref{fig9}.
Inclusion of HTC has the effect of differentially lowering $\mu$ values (as expected since the trader is expending money on fees),
most noticeably along the coordinate $K/S \sim 1.2$.
Similarly expected, inclusion of HTC differentially raises the volatility surface.
HTC differentially raises $\beta$ values, with greater increases arising as $K/S$ values move more ``into the money'',
resulting in larger regions of the phase space where the pricing tree is positively skewed.

\section{Fitting the GJR pricing tree to a market driver}
\label{sec5}
\noindent
In this section we explore possibilities for fitting the pricing tree model, $S_{k\Delta t}^{(n)}$, $k = 1,\ldots,n$,
to a market driver that affects the price dynamics, $S_{k \Delta t}^{(n,\mathcal{S})}$, of stock $\mathcal{S}$.
We consider both endogenous and exogenous approaches.
In the endogenous approach, the discrete dynamics for the price of $\mathcal{S}$ is modeled by
\begin{align}
	S_{k\Delta t}^{(n,\textrm{endo})} & = S_0 \exp\left(v_k^{(0,\mathcal{S})}\Delta t
	+ M_k^{(\textrm{endo})}\sigma^{(0,\mathcal{S})}\sqrt{\Delta t}\right), \label{eq_endoS} \\
	v_k^{(0,\mathcal{S})} & = k\mu^{(0,\mathcal{S})}+\sigma^{(0,\mathcal{S})}\beta^{(0,\mathcal{S})}\left(\sqrt{2k/\pi}-1\right),
	\; k=1, \ldots, n, \nonumber
\end{align}
with the path
$M_1^{(\textrm{endo})},\ldots,M_k^{(\textrm{endo})}$ of the Markov chain $\mathbb{M}^{(\textrm{endo})}$
determined by\footnote{
	Here $\textrm{sign}(x) = \begin{cases}1,\;\textrm{if}\; x\geq 0 \\ -1,\;\textrm{if}\;x<0\end{cases}$.}
\begin{equation*}
M_k^{(\textrm{endo})}
= \sum_{j=1}^k \textrm{sign}\left(r_{j\Delta t}^{(n,\mathcal{S})}\right),\ j = 1, \dots ,k;\  M_0^{(\textrm{endo})} = 0,
\end{equation*}
where $r_{j\Delta t}^{(n,\mathcal{S})} = \ln \left( S_{j \Delta t}^{(n,\mathcal{S})} / S_{(j-1)\Delta t}^{(n,\mathcal{S})} \right)$
is the market's daily log-return for stock $\mathcal{S}$.
In \eqref{eq_endoS}, the parameters $\sigma^{(0,\mathcal{S})}$, $\mu^{(0,\mathcal{S})}$ and $\beta^{(0,\mathcal{S})}$ are
determined for the initial time $k \Delta t = 0$ from equations \eqref{eq_step1} and \eqref{eq_step2} using the cumulative return data
for $\mathcal{S}$ applicable at the chosen start date $k \Delta t = 0$.
The values obtained from \eqref{eq_step1} and \eqref{eq_step2} are also smoothed using the moving-window averaging procedure
introduced in section \ref{sec24}.
This will be demonstrated explicitly by numerical example in section \ref{sec51}.

The endogenous approach is limited while
the exogenous approach assumes that the $M_k^{(\alpha_{\Delta t})}$ are determined by the upward
and downward movements of a more general market driver.
For concreteness, we initially assume the market driver is a factor model for the price dynamics of $\mathcal{S}$;
in particular we consider the Fama-French five-factor asset pricing model.

\noindent
{\bf Fama-French five-factor model.}\footnote{
	See Fama and French (2012, 2015, 2017).}
With $r_{j\Delta t}^{(n,\mathcal{S})}$ being the market log-return of stock $\mathcal{S}$ for the period $[j\Delta t, (j+1)\Delta t)$,
the coefficients of the terms in the Fama-French factor return,
\begin{equation}
	r_{j\Delta t}^{(n,F)} = r_{f,j\Delta t}
	+\mathbbm{a}
	+\mathbbm{b}\left(r_{j\Delta t}^{(n,M)} - r_{f,j\Delta t}\right)+\mathbbm{s}r_{j\Delta t}^{(n,SMB)}
	+\mathbbm{h}r_{j\Delta t}^{(n,HML)}+\mathbbm{r}r_{j\Delta t}^{(n,RMW)}
	+\mathbbm{c}r_{j\Delta t}^{(n,CMA)},
	\label{eq_FF}
\end{equation}
are determined from the regression
\begin{equation}
	r_{j\Delta t}^{(n,\mathcal{S})} = r_{j\Delta t}^{(n,F)} + \epsilon_{j\Delta t}, \quad j = 0,1,\ldots,m-1. \label{eq_FF_reg}
\end{equation}
In \eqref{eq_FF}, $r_{f,j\Delta t}$ is the risk-free rate in period $[j\Delta t, (j+1)\Delta t)$ and the remaining Fama-French factors are:
\begin{itemize}
\item[$\mathbb{F}_1$:]
	$r_{j\Delta t}^{(n,M)}$ (representing the market-factor effect) is the return over the period
	$[j\Delta t,(j+1)\Delta t)$ of the capitalization weighted stock market to which $\mathcal{S}$ belongs;
\item[$\mathbb{F}_2$:]
	$r_{j\Delta t}^{(n,SMB)}$ (representing the size-factor effect) is the difference in
	the return of a portfolio in $\mathcal{M}$ specializing in (a) small-cap and (b) large-cap stocks;
\item[$\mathbb{F}_3$:]
	$r_{j\Delta t}^{(n,HML)}$ (representing the value-factor effect) is the difference of the returns of a
	portfolio in $\mathcal{M}$ with (a) high and (b) small book-to-market ratios;
\item[$\mathbb{F}_4$:]
	$r_{j\Delta t}^{(n,RMW)}$ (representing the profitability-premium-factor effect) is the difference of
	the return of a portfolio in $\mathcal{M}$ with (a) high and (b) low profitability; and
\item[$\mathbb{F}_5$:]
	$r_{j\Delta t}^{(n,CMA)}$ (representing the investment-attitude-factor effect) is the difference of the
	return of a portfolio of stocks in $\mathcal{M}$ issued by firms that invest
	(a) conservatively  and (b) aggressively.
\end{itemize}

Let $j \Delta t$, $j = 1, \dots, m$ denote a time period $\tau_1$ and $j = m+1, \dots, m+n$ denote the following
time period $\tau_2$.
Let $\hat{r}_{j\Delta t}^{(n,F)}$, $j=1, \ldots, m$, denote sample return values (factor returns) for \eqref{eq_FF}
obtained from the regression \eqref{eq_FF_reg} over the period $\tau_1$,
and $\hat{R}_{j\Delta t}^{(n,F)}$ denote the resultant cumulative factor returns.
Using the cumulative factor returns on the left-hand side of \eqref{eq_step1} and \eqref{eq_step2},
and the moving-window averaging procedure introduced in section \ref{sec24}, we obtain estimates 
$\bar{\sigma}^{(F)}$, $\bar{\mu}^{(F)}$ and $\bar{\beta}^{(F)}$ for the values of the parameters
$\sigma$, $\mu$ and $\beta$ at timestep $m \Delta t$.
Let $k \Delta t$, $k = 1, \dots, n$, where $k = j-m$, label the trading days in time period $\tau_2$.
Then $\bar{\sigma}^{(F)}$, $\bar{\mu}^{(F)}$ and $\bar{\beta}^{(F)}$ correspond to parameter values for $k \Delta t = 0$.

We model the discrete dynamics for the factor-driven price over the period $\tau_2$ by 
\begin{align}
	S_{k\Delta t}^{(n,\textrm{exo})} & =
		S_0 \exp\left( v_k^{(F)} \Delta t
		+ M_k^{\left( \bar{\alpha}^{(F)},\textrm{exo}\right)} \bar{\sigma}^{(F)} \sqrt{\Delta t} \right),
		\label{eq_exoS} \\
	v_k^{(F)} & = k \bar{\mu}^{(F)} + \bar{\sigma}^{(F)} \bar{\beta}^{(F)}\left(\sqrt{2k/\pi}-1\right), \nonumber
\end{align} 
where $\bar{\alpha}^{(F)}_{\Delta t} = (1+\bar{\beta}^{(F)}\sqrt{\Delta t})/2$.
To estimate the path $M_k^{\left( \bar{\alpha}^{(F)},\textrm{exo} \right)}$,
we generate a large ensemble of sample paths $\left \{ M_k^{\left( \bar{\alpha}^{(F)} \right)}, \  k = 1, \dots, n \right \}$.
$M_k^{\left( \bar{\alpha}^{(F)},\textrm{exo} \right)}$ is then chosen as that path that minimizes the relMSE between the two sides
of \eqref{eq_exoS} when the stock prices $S_{k\Delta t}^{(n,\mathcal{S})}$ are used
on the left-hand side of \eqref{eq_exoS}.
This exogenous method is also demonstrated in section \ref{sec51}.

\subsection{Numerical example of the endogenous and exogenous approaches}
\label{sec51}
\noindent
We demonstrate the endogenous and exogenous approaches using the daily return series for the stock $\mathcal{S} = \textrm{MSFT}$.
The price data set\footnote{
	MSFT market closing prices from Bloomberg Professional Services.
	Return values for the Fama-French factors from the U.S. Research Returns Data Center
	(\url{https://mba.tuck.dartmouth.edu/pages/faculty/ken.french/Data_Library.html}).}
covers the period 4/30/2015 tthrough 4/30/2021.
Since the estimation (equations \eqref{eq_step1} and \eqref{eq_step2}) and averaging procedures of section \ref{sec24}
for $\sigma$, $\mu$ and $\beta$ require one-year moving windows,
we divide the data set into two time periods, $\tau_1 = $ 4/31/2015 through 4/28/2017
and $\tau_2 = $ 5/1/2017 through 4/30/2021.
MSFT data on the close of 4/28/2017 in $\tau_1$ serves as $t=0$ data for $\tau_2$;
the closing price of MSFT on  4/28/2017 serves as $S_0$ for the period $\tau_2$
and cumulative returns over $\tau_2$ are then computed relative to this $S_0$.
And applying the estimation and averaging procedures of section \ref{sec24} to the cumulative log-return data for MSFT
over $\tau_1$, we obtain estimates $\bar{\sigma}^{(0,\textrm{MSFT})}$, $\bar{\mu}^{(0,\textrm{MSFT})}$
and $\bar{\beta}^{(0,\textrm{MSFT})}$ determined for the close of trading on 4/28/2017.
These values, along with $\bar{\alpha}^{(0,\textrm{MSFT})}$ and the value of the relMSE, are given in Table~\ref{tab1}.
Using $\bar{\sigma}^{(0,\textrm{MSFT})}$, $\bar{\mu}^{(0,\textrm{MSFT})}$ and $\bar{\beta}^{(0,\textrm{MSFT})}$
in \eqref{eq_endoS} we obtain an endogenous estimate $S_{k\Delta t}^{(n,\textrm{endo})}$ for the cumulative price dynamics
of MSFT over the period $\tau_2$.
\begin{table}[ht]
	\begin{center}
	\begin{tabular}{c c c c c c}
		\toprule
		\multirow{2}{*}{endogenous}
			     & $\bar{\mu}^{(0,\textrm{MSFT})}$ & $\bar{\sigma}^{(0,\textrm{MSFT})}$ & $\bar{\beta}^{(0,\textrm{MSFT})}$
			     & $\bar{\alpha}^{(0,\textrm{MSFT})}$ & relMSE \\
			     & $7.38\cdot 10^{-4} $ & $8.91\cdot 10^{-2}$      & $-4.08$                   & 0.371  & $1.78\cdot 10^{-3}$ \\
		\midrule
		\multirow{2}{*}{exogenous}
			    & $\bar{\mu}^{(F)}$ & $\bar{\sigma}^{(F)}$ & $\bar{\beta}^{(F)}$ & $\bar{\alpha}^{(F)}$ & relMSE \\
			    & $2.31\cdot 10^{-4} $ & $7.01\cdot 10^{-3}$      & $-11.2 $                  & 0.147  & $3.22\cdot 10^{-3} $  \\
		\bottomrule
	\end{tabular}
	\caption{Parameters estimates from the endogenous and exogenous estimation procedures}
	\label{tab1}
	\end{center}
\end{table}

\begin{figure}[ht]
\begin{center}
    \begin{subfigure}[b]{\textwidth} 
    	\includegraphics[width=0.49 \textwidth]{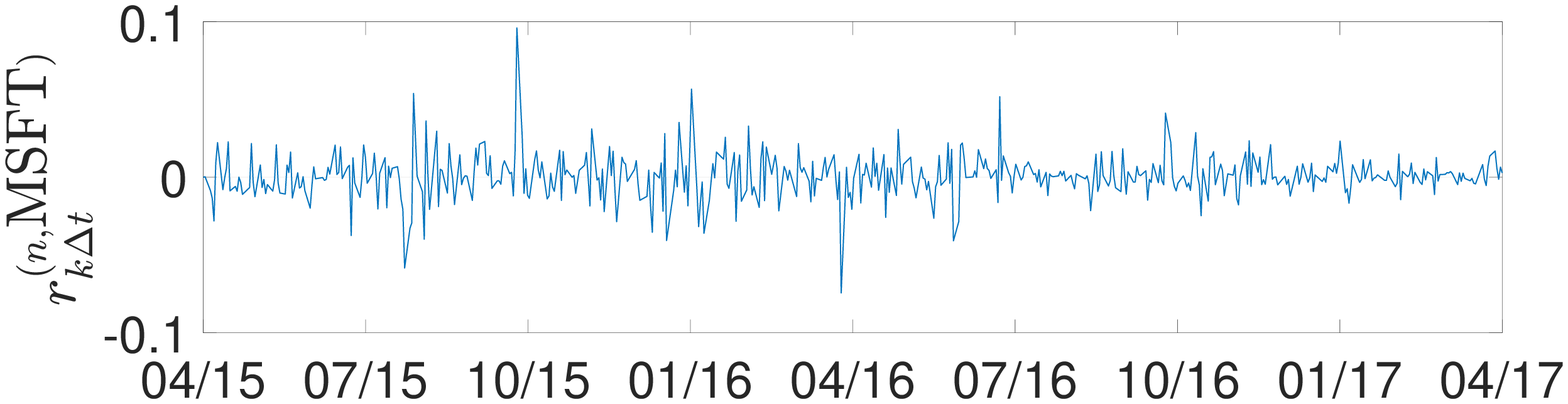}
    	\includegraphics[width=0.49 \textwidth]{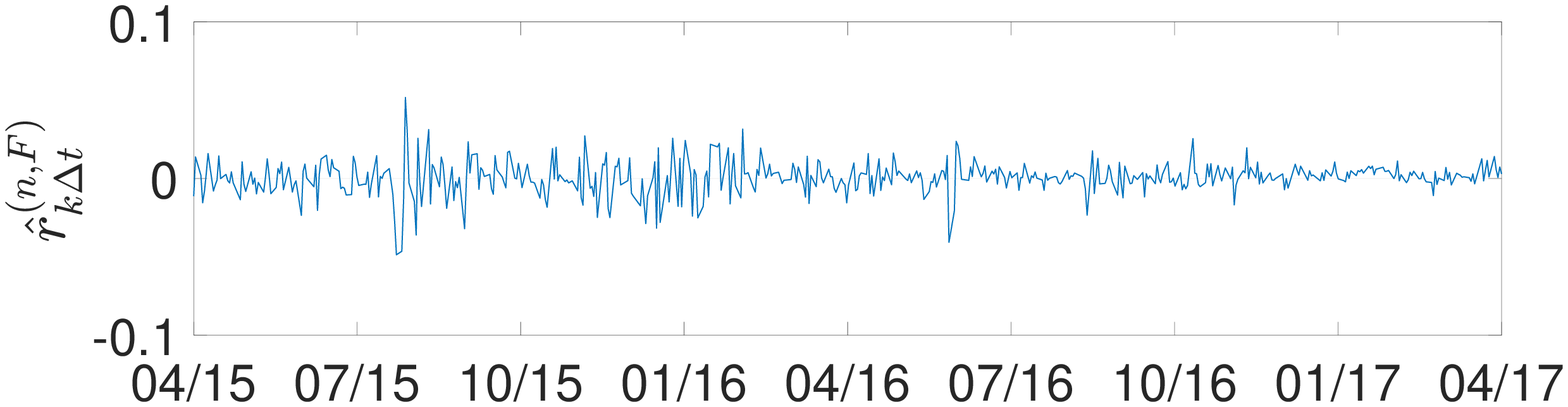}
	\caption{}
	\label{fig10_rtn}
    \end{subfigure}
    \begin{subfigure}[b]{0.75\textwidth}
        \vspace{.7em}
    	\includegraphics[width=\textwidth]{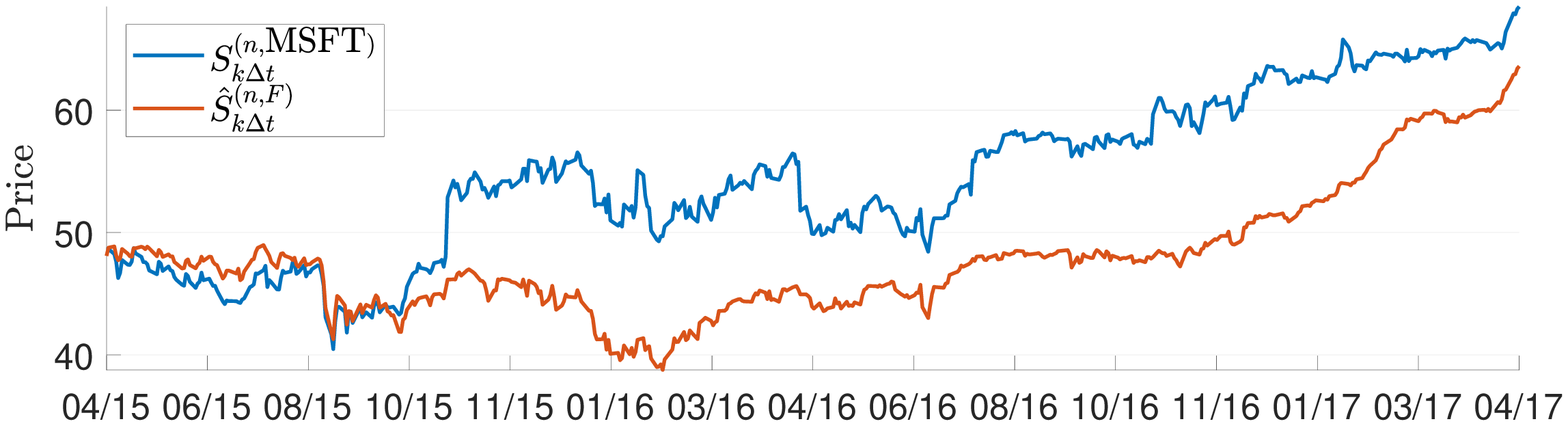}
    	\caption{}
    	\label{fig10_price}
    \end{subfigure}
    \caption{(a) Comparison of the MSFT market (a) return series $r^{(n,\textrm{MSFT})}_{k\Delta t}$ and
		 (b) price series $S_{k\Delta t}^{(n,\textrm{MSFT})}$ with with the return series $\hat{r}^{(n,F)}_{k\Delta t}$
		 and corresponding price series $\hat{S}^{(n,F)}_{k\Delta t}$
		 produced by a regression fit to the Fama-French five-factor model over the period $\tau_1$.}
\label{fig10}
\end{center}
\end{figure}

For the exogenous approach, we use the Fama-French five-factor model outlined in the previous section as the market driver.
The Fama-French factor returns, $r_{k\Delta t}^{(n,F)}$in \eqref{eq_FF}, were computed over the period $\tau_1$
using a robust form for the regression \eqref{eq_FF_reg}.\footnote{
	Motivated by results in \cite{Knez1997}, we choose robust regression in estimating the parameters
	of the Fama-French five-factor model.}
The values of the coefficients, as well as the R-square and root mean square-error (RMSE) values, are given in Table \ref{tab2}.
The sample return $\hat{r}^{(n,F)}_{k\Delta t}$ and price $\hat{S}^{(n,F)}_{k\Delta t}$ series computed from
these coefficients are presented in Fig. \ref{fig10} where they are compared with the market return $r^{(n,\textrm{MSFT})}_{k\Delta t}$
and price $S^{(n,\textrm{MSFT})}_{k\Delta t}$ series over the period $\tau_1$.
The significant differences between the series $\hat{S}^{(n,F)}_{k\Delta t}$ and $S^{(n,\textrm{MSFT})}_{k\Delta t}$ are
indicative of issues with applying market driver analyses to stock returns.
Since the Fama-French model is based upon a regression fit to return values,
errors produced in return values have large-time decay structure and lead ultimately to significant cumulative differences in price performance.

\begin{table}[th]
	\begin{center}	
	\begin{tabular}{c c c c c c c c}
		\toprule
		$\mathbbm{a}$ & $\mathbbm{b}$ & $\mathbbm{s}$ & $\mathbbm{h}$
				& $\mathbbm{r}$ & $\mathbbm{c}$ & R-square & RMSE  \\
		\midrule
		$-6.5\cdot 10^{-4}$ & $1.2\cdot 10^{-2}$ & $-3.9\cdot 10^{-3}$ & $-4.0\cdot 10^{-4}$
		& $3.0\cdot 10^{-3}$ & $-5.8\cdot 10^{-3}$ & 0.58 & $9.1\cdot 10^{-3}$ \\
		\bottomrule
	\end{tabular}
	\caption{Parameter estimates from the regression \eqref{eq_FF_reg}}
	\label{tab2}
	\end{center}
\end{table}

The parameter estimates $\bar{\mu}^{(F)}$, $\bar{\sigma}^{(F)}$, $\bar{\beta}^{(F)}$, and $\bar{\alpha}^{(F)}$
obtained for 4/28/2017 using the cumulative Fama-French returns over $\tau_1$ are given in Table \ref{tab1}.
These values generally differ by a factor of $\sim3$ compared to the corresponding endogenous variables,
with the exception of a factor of $\sim13$ difference between $\bar{\sigma}^{(F)}$ and $\bar{\sigma}^{(0,\textrm{MSFT})}$.
Using the estimated value $\bar{\alpha}^{(F)}$, we generate $10^6$ scenarios of the sample path
$M_k^{\left( \bar{\alpha}^{(F)} \right)}$ covering the time period $\tau_2$.
Using the price data $S_{k\Delta t}^{(n,\textrm{MSFT})}$ for $\tau_2$ on the left-hand side of \eqref{eq_exoS},
the optimal path $M_k^{\left( \bar{\alpha}^{(F)},\textrm{exo} \right)}$ was selected as the path scenario that minimized
the relMSE between the left- and right-hand sides.
The exogenous price series estimate $S_{k\Delta t}^{(n,\textrm{exo})}$ for $\tau_2$ is then computed from \eqref{eq_exoS}
using $M_k^{\left( \bar{\alpha}^{(F)},\textrm{exo} \right)}$ on the right-hand side.

\begin{figure}[ht]
	\begin{center}
    	\includegraphics[width=0.75\textwidth]{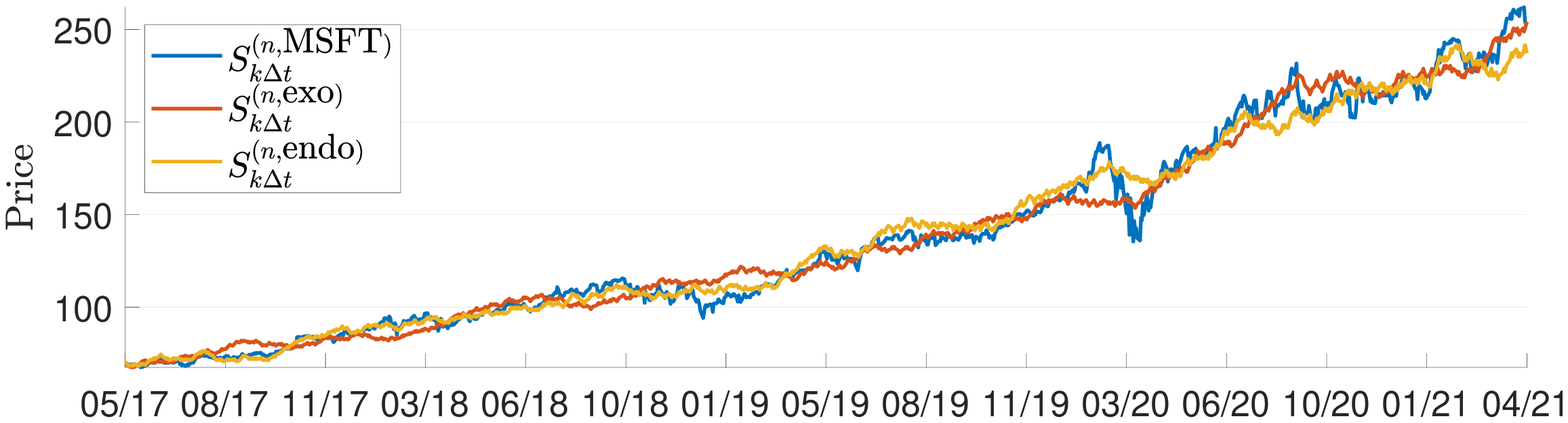}
	\caption{Comparisons of the MSFT market price series $S_{k\Delta t}^{(n,\textrm{MSFT})}$ to
	the estimated endogenous, $S_{k\Delta t}^{(n,\textrm{endo})}$, and exogenous, $S_{k\Delta t}^{(n,\textrm{exo})}$,
	price dynamics over the time period $\tau_2$.}
    	\label{fig11}
\end{center}
\end{figure}
The cumulative price $S_{k\Delta t}^{(n,\textrm{MSFT})}$ over $\tau_2$ is compared with the estimates
$S_{k\Delta t}^{(n,\textrm{endo})}$ and $S_{k\Delta t}^{(n,\textrm{exo})}$ in Fig.~\ref{fig11}.
Both estimates produce reasonable agreement to the actual price data; however both significantly smooth and decrease
the impact of the Covid-19 pandemic; the exogenous method showing less impact from the pandemic than the endogenous method.

\subsection{Inclusion of the higher moment SBM process $\mathbb{C}_t^{(\alpha,n)}$}
\label{sec52}
\noindent
For a general market driver, we extend to the full functionality of the CYSIP by adding the higher moment
SBM process $\mathbb{C}_t^{(\alpha,n)}$ into our model.\footnote{
	See \cite{Hu2020b} for an application of the CSYIP using BM.}
According to (\ref{eq_exoS}), $S_{k\Delta t}^{(n,\textrm{exo})}$ is adapted to the filtration
\begin{equation*}
\mathbb{F}^{(n;\mathbb{M}^{(\alpha_{\Delta t})})}
	=\left\{ \mathcal{F}_k^{(n;\mathbb{M}^{(\alpha_{\Delta t})})}
	=\sigma\left(M_j^{(\alpha_{\Delta t})};\;j=1,\ldots,k\right),\;k=1,\ldots,n,\;
	\mathcal{F}_k^{(n;\mathbb{M}^{(\alpha_{\Delta t})})}=\left\{\varnothing,\Omega\right\}\right\}.
\end{equation*}
Paralleling the development for the lower moment SBM process in section \ref{sec22},
with $n\Delta t=T$ we set $M_0^{(\alpha_{\Delta t},n)}=0$,
$M_k^{(\alpha_{\Delta t},n)}=n^{-1/2} M_k^{(\alpha_{\Delta t})}$, and
$X_{k/n}^{(\alpha_{\Delta t},n)}= \sum_{i=1}^{k} M_i^{(\alpha_{\Delta t},n)},\;k=1,\ldots,n$.
Let $B_t^{(\alpha_{\Delta t},n)},\; t\geq 0$ be a random process with piecewise linear trajectories
having  vertexes $\left(k/n,\mathbb{B}_{k/n}^{(\alpha_{\Delta t},n)}\right),\;k=1,\ldots,n$,
where $\mathbb{B}_{k/n}^{(\alpha_{\Delta t},n)}=X_{k/n}^{(\alpha_{\Delta t},n)}$.
Let  $h:R \rightarrow R$  be a piecewise continuous function and define 
$Y_{k/n}^{(\alpha_{\Delta t},n)}= \sum_{i=1}^{k} h\left( X_{(i-1)/n}^{(\alpha_{\Delta t},n)}\right)\left(X_{i/n}^{(\alpha_{\Delta  t},n)}-X_{(i-1)/n}^{(\alpha_{\Delta t},n)}\right),\;k\in \mathcal{N},\;n\in \mathcal{N}$.
Define $\mathbb{C}_t^{(\alpha_{\Delta t},n;h)},\;t \geq 0$ to be a random process with the piecewise linear
trajectories having vertexes 
$\left(k/n,\mathbb{C}_{k/n}^{(\alpha_{\Delta t},n;h)}\right),\;k\in\mathcal{N},\;n \in \mathcal{N}$,
where $\mathbb{C}_{k/n}^{(\alpha_{\Delta t},n;h)}=Y_{k/n}^{(\alpha_{\Delta t},n;h)}$.
Then, for a fixed, relatively small value of $\Delta t$, the bivariate process
$\left(\mathbb{B}_t^{(\alpha_{\Delta t},n)},\mathbb{C}_t^{(\alpha_{\Delta t},n;h)}\right),\;t\geq 0$,
approaches $\left(B_t^{(\alpha_{\Delta t})},C_t^{(\alpha_{\Delta t};h)}\right)$,
where $B_t^{(\alpha_{\Delta t})}$ is a SBM and
$C_t^{(\alpha_{\Delta t};h)}=\int_0^t h\left(B_s^{(\alpha_{\Delta t})}\right)d B_s^{(\alpha_{\Delta t})}$. 
However, as $n\uparrow \infty$ and
$\Delta t \downarrow 0,\;\left(\mathbb{B}_t^{(\alpha_{\Delta t},n)},\mathbb{C}_t^{(\alpha_{\Delta t},n)}\right)$
converges in law to $\left(B_t,C_t\right)$ for $t\geq 0$, where $B_t$ is a BM and
$C_t=\int_0^t h\left(B_s \right)dB_s$.\footnote{
	See \cite{Cherny2003}.}

Next, consider the following two processes in $\mathcal{D}[0,T]$:
\begin{align*}
\mathbb{B}_{[0,T]}^{(\alpha_{\Delta t},n)} &= \left\{\mathbb{B}_t^{(\alpha_{\Delta t},n)} = \sqrt{T}X_{k/n}^{(\alpha_{\Delta t},n)},\; t\in [k\Delta t,(k+1)\Delta t),\;\mathbb{B}_1^{(\alpha_{\Delta t},n)} = X_1^{(\alpha_{\Delta t},n)}\right\},
\\
\mathbb{C}_{[0,T]}^{(\alpha_{\Delta t},n,h)} &= \left\{
\begin{aligned}
\mathbb{C}_t^{(\alpha_{\Delta t},n;h)} &= \sum_{i=1}^k h\left(\sqrt{T}X_{(i-1)/n}^{(\alpha_{\Delta t},n)}\right)\left(\sqrt{T}X_{i/n}^{(\alpha_{\Delta t},n)}-\sqrt{T}X_{(i-1)/n}^{(\alpha_{\Delta t},n)}\right),\; t\in [k\Delta t,(k+1)\Delta t),
\\
\mathbb{C}_1^{(\alpha_{\Delta t},n)} &= Y_1^{(\alpha_{\Delta t},n)}
\end{aligned}
\right\}.
\end{align*}
Then, for large $n$, the bivariate process
$\left(\mathbb{B}_{[0,T]}^{(\alpha_{\Delta t},n)},\mathbb{C}_{[0,T]}^{(\alpha_{\Delta t},n;h)}\right)$ approaches
$\left(\mathbb{B}_{[0,T]}^{(\alpha_{\Delta t})},\mathbb{C}_{[0,T]}^{(\alpha_{\Delta t};h)}\right)$ in
$\mathcal{D}[0,T]\cdot \mathcal{D}[0,T]$,
where $\mathbb{B}_{[0,T]}^{(\alpha_{\Delta t})} = \left\{B_t^{(\alpha_{\Delta t})},\; t\in[0,T]\right\}$ is a SBM on
$[0,T]$ and $\mathbb{C}_{[0,T]}^{(\alpha_{\Delta t};h)} = \left\{C_t^{(\alpha_{\Delta t};h)},\;t\in [0,T]\right\}$.
Ultimately, as $n \uparrow \infty$,
$\left(\mathbb{B}_{[0,T]}^{(\alpha_{\Delta t},n)},\mathbb{C}_{[0,T]}^{(\alpha_{\Delta t},n;h)}\right)$ converges
weakly in $\mathcal{D}[0,T]\cdot \mathcal{D}[0,T]$ to $\left(\mathbb{B}_{[0,T]},\mathbb{C}_{[0,T]}^{(h)}\right)$,
where $\mathbb{B}_{[0,T]} = \left\{B_t,\; t\in [0,T]\right\}$ is a BM on $[0,T]$
and $\mathbb{C}_{[0,T]}^{(h)} = \left\{C_t^{(h)} = \int^t_0 h\left(B_s\right) dB_s,\; t\in [0,T]\right\}$.

Now define the stock price discrete dynamics as a functional of $\mathbb{B}_{[0,T]}^{(\alpha_{\Delta t},n)}$
and $\mathbb{C}_{[0,T]}^{(\alpha_{\Delta t},n;h)}$.
Let
\begin{equation}
\mathbb{S}_{[0,T]}^{(\alpha_{\Delta t},n;h)} = \left\{
    \begin{aligned}
	S_t^{(\alpha_{\Delta t},n;h)} &= S_0 \exp\left(vk\Delta t+\sigma \mathbb{B}_t^{(\alpha_{\Delta t},n)}
	+ \gamma \mathbb{C}_t^{(\alpha_{\Delta t},n;h)}\right),\; t\in [k\Delta t,(k+1)\Delta t), k=1,\ldots,n, \\
	S_T^{(\alpha_{\Delta t},n;h)} & =
		S_0 \exp\left( vT+\sigma \mathbb{B}_T^{(\alpha_{\Delta t},n)} + \gamma \mathbb{C}_T^{(\alpha_{\Delta t},n;h)} \right)
    \end{aligned}
    \right\},
    \label{eq_BCS}
\end{equation}
where $v\in R$, $\sigma \in R\backslash\{0\}$ and $\gamma \in R$ are parameters determining the dynamics of
the stock price as a function of the index dynamics.
The discrete dynamics of the stock log-return is given by
\begin{equation}
r_{k\Delta t}^{(\alpha_{\Delta t},n;h)}
	= \ln \left(\frac{S_{k\Delta t}^{(\alpha_{\Delta t},n;h)}}{S_{(k-1)\Delta t}^{(\alpha_{\Delta t},n;h)}}\right)
	= v\Delta t + \sigma\sqrt{\Delta t}M_k^{(\alpha_{\Delta t})} 
	+ \gamma\sqrt{\Delta t}M_k^{(\alpha_{\Delta t})} h\left(\sum_{i=1}^{k-1}\sqrt{\Delta t}M_i^{(\alpha_{\Delta t})}\right),
	k = 1,\ldots,n.
\label{eq_BCr}
\end{equation}
Thus, if $\gamma\neq 0$, the stock log-return $r_{k\Delta t}^{(\alpha_{\Delta t},n;h)}$ depends on the entire path
$\{M_i^{(\alpha_{\Delta t})},\;i=1,\ldots,k \}$.
In the pre-limiting case, $\mathbb{S}_{[0,T]}^{(\alpha_{\Delta t},n;h)}$ approaches
$\mathbb{S}_{[0,T]}^{(\alpha_{\Delta t};h)}
= \{S_0 \exp\left(vt+\sigma B_t^{(\alpha_{\Delta t})}+\gamma C_t^{(\alpha_{\Delta t};h)}\right),\;t\in[0,T]\}$.
In the limit $n \uparrow \infty$, $\mathbb{S}_{[0,T]}^{(\alpha_{\Delta t},n;h)}$ converges weakly in $\mathcal{D}[0,T]$
to $\mathbb{S}_{[0,T]}^{(h)} = \{S_0\exp\left(vt+\sigma B_t+\gamma C_t^{(h)}\right),\; t\in [0,T]\}$.

To demonstrate the application of the model \eqref{eq_BCS}, \eqref{eq_BCr}, we use \eqref{eq_BCr} to estimate the MSFT log-return
$r_{k\Delta t}^{(n,\textrm{MSFT})}$ over the period $\tau_2$.
To capture likely heavy-tailed behavior, we identify $h(\cdot)$ as the Student's $t$
probability density function with $\kappa$ degrees of freedom.
For convenience, we utilize the optimal path $M_k^{\left( \bar{\alpha}^{(F)},\textrm{exo} \right)}$ determined from
the exogenous procedure in section~\ref{sec51} for the required path in \eqref{eq_BCr}.
To establish the model parameters, $v$, $\sigma$, $\gamma$, and $\kappa$,
we construct the conditional least squares minimization problem,
\begin{equation}
\min_{\nu \in R, \sigma \in R, \gamma \in R, \kappa \in[5,30]}
\parallel r_{k\Delta t}^{(n,\textrm{MSFT})} - r_{k\Delta t}^{(\alpha_{\Delta t},n;h)} \parallel_2^2.
\label{eq_53_mini}
\end{equation}
Table \ref{tab3} provides the resulting parameter estimates.
The model \eqref{eq_BCS}, \eqref{eq_BCr} with these parameter estimates will be used further in section~\ref{sec61}.
\begin{table}[ht]
	\begin{center}	
	\begin{tabular} {c c c c c}
		\toprule
		 $v$ & $\sigma$ & $\gamma$ & $\kappa$ & RMSE \\
		0.272 & $3.44\cdot 10^{-5}$ & $4.22\cdot 10^{-2}$ & 6.24  & $1.86\cdot 10^{-2}$\\
		\bottomrule
	\end{tabular}
	\caption{Parameters estimates obtained from \eqref{eq_53_mini} }
	\label{tab3}
	\end{center}
\end{table}

\section{GJR option pricing}
\label{sec6}
\noindent
Given the filtration $\mathbb{F}^{(n;\mathbb{M}^{(\alpha_{\Delta t})})}
= \{\mathcal{F}_k^{(n;\mathbb{M}^{(\alpha_{\Delta t})}}
= \sigma \left(M_j^{(\alpha_{\Delta t})};j=1,\ldots,k\right),
\;\mathcal{F}_k^{(n;\mathbb{M}^{(\alpha_{\Delta t})})} 
= \{\varnothing,\Omega\},\;k=1,\ldots,n\}$,
consider the $\mathbb{F}^{(n;\mathbb{M}^{(\alpha_{\Delta t})})}$-adapted path-dependent
GJR pricing tree determined by (\ref{eq_BCr}),
\begin{multline}
S_{k\Delta t}^{(\alpha_{\Delta t},n;h)} = S_{(k-1)\Delta t}^{(\alpha_{\Delta t},n;h)}
\exp\left[
	v \Delta t + \sigma \sqrt{\Delta t} \left( M_k^{(\alpha_{\Delta t})} - M_{k-1}^{(\alpha_{\Delta t})} \right)
\right .\\
\left .
+ \gamma \sqrt{\Delta t} \left( M_k^{(\alpha_{\Delta t})}-M_{k-1}^{(\alpha_{\Delta t})} \right)
	h \left( \sqrt{\Delta t}M_{k-1}^{(\alpha_{\Delta t})} \right)
\right),
\label{eq_GJR_price60}
\end{multline}
where $k=1,\ldots,n,\;n\Delta t = T$. Let $h(x)\geq 0,\; x\in R$, be a chosen probability density function.\footnote{
	For example, $h$ is the Student's $t$ density function with $\kappa$-degrees of freedom.}
Define
\begin{equation}
	\eta_{k\Delta t} =\sigma + \gamma h\left(\sqrt{\Delta t}M_{k-1}^{(\alpha_{\Delta t})}\right),\; k = 1, \ldots, n; \qquad
	\eta_{0} = \sigma.
	\label{eq_eta}
\end{equation}
Note that $\eta_{k\Delta t}^{(\alpha_{\Delta t})}$ represents the stock's time-varying factor-volatility at
$k\Delta t$ as a function of the path of factor up and down movements $M_{k-1}^{(\alpha_{\Delta t})}$.
From (\ref{eq_GJR_price60}), conditionally on
$\mathcal{F}_k^{(n;\mathbb{M}^{(\alpha_{\Delta t})})},\;k = 0,\ldots,n-1,$
\begin{equation}
S_{(k+1)\Delta t}^{(\alpha_{\Delta t},n;h)} = \begin{cases}
S_{(k+1)\Delta t}^{(\alpha_{\Delta t},n,u;h)} &= S_{k\Delta t}^{(\alpha_{\Delta t},n;h)}\exp\left(v\Delta t+\eta_{k\Delta t}^{(\alpha_{\Delta t})}\sqrt{\Delta t}\right),\; \textrm{w.p.}\;\mathbb{P}\left(M_{k+1}^{(\alpha)}=M_k^{(\alpha)}+1\mid M_k^{(\alpha)}\right),
\\
S_{(k+1)\Delta t}^{(\alpha_{\Delta t},n,d;h)} &= S_{k\Delta t}^{(\alpha_{\Delta t},n;h)}\exp\left(v\Delta t-\eta_{k\Delta t}^{(\alpha_{\Delta t})}\sqrt{\Delta t}\right),\; \textrm{w.p.}\;\mathbb{P}\left(M_{k+1}^{(\alpha)}=M_k^{(\alpha)}-1\mid M_k^{(\alpha)}\right).
\end{cases}
\label{eq_sec6_last}
\end{equation}

For $k=0,1,\ldots,n-1$, consider the replicating risk-neutral portfolio
$P_{k\Delta t}^{(\alpha_{\Delta t},n;h)}
= S_{k\Delta t}^{(\alpha_{\Delta t},n;h)}D_{k\Delta t}^{(\alpha_{\Delta t},n;h)}
- f_{k\Delta t}^{(\alpha_{\Delta t},n;h)}$ with $P_{(k+1)\Delta t}^{(\alpha_{\Delta t},n;h)}
= S_{(k+1)\Delta t}^{(\alpha_{\Delta t},n;h)}D_{k\Delta t}^{(\alpha_{\Delta t},n;h)}
- f_{(k+1)\Delta t}^{(\alpha_{\Delta t},n;h)}$,
given $\mathcal{F}_k^{(n;\mathbb{M}^{(\alpha_{\Delta t})})}$.
Thus, conditionally on $\mathcal{F}_k^{(n;\mathbb{M}^{(\alpha_{\Delta t})})}$,
\begin{equation*}
D_{k\Delta t}^{(\alpha_{\Delta t},n;h)}
= \frac{f_{(k+1)\Delta t}^{(\alpha_{\Delta t},n,u;h)}
- f_{(k+1)\Delta t}^{(\alpha_{\Delta t},n,d;h)}}{e^{v\Delta t}S_{k\Delta t}^{(\alpha_{\Delta t},n;h)}
\left(\exp\left(\eta_{k\Delta t}^{(\alpha_{\Delta t})}\sqrt{\Delta t}\right)
- \exp\left(-\eta_{k\Delta t}^{(\alpha_{\Delta t})}\sqrt{\Delta t}\right)\right)}.
\end{equation*}
As portfolio
$P_{(k+1)\Delta t}^{(\alpha_{\Delta t},n;h)}$
is now riskless, it follows that
$P_{k\Delta t}^{(\alpha_{\Delta t},n;h)} = e^{-r_f\Delta t}P_{(k+1)\Delta t}^{(\alpha_{\Delta t},n,u;h)}$
with
$P_{(k+1)\Delta t}^{(\alpha_{\Delta t},n,u;h)}
= D_{k\Delta t}^{(\alpha_{\Delta t},n;h)}S_{k\Delta t}^{(\alpha_{\Delta t},n;h)}
\exp\left(v\Delta t+\eta_{k\Delta t}^{(\alpha_{\Delta t})}\sqrt{\Delta t}\right)
- f_{(k+1)\Delta t}^{(n,u)}$.
Thus, conditionally on $\mathcal{F}_k^{(n;\mathbb{M}^{(\alpha_{\Delta t})})}$,
\begin{equation*}
f_{k\Delta t}^{(\alpha_{\Delta t},n;h)}
= e^{-r_f \Delta t}\left(q_{(k+1)\Delta t}^{(\alpha_{\Delta t},n;h)}f_{(k+1)\Delta t}^{(\alpha_{\Delta t},n,u;h)}
+ (1-q_{(k+1)\Delta t}^{(\alpha_{\Delta t},n;h)})f_{(k+1)\Delta t}^{(\alpha_{\Delta t},n,d;h)}\right),
\label{eq_fkt_61}
\end{equation*}
where the risk-neutral probability, $q_{(k+1)\Delta t}^{(\alpha_{\Delta t},n;h)}$, for the time period
$[k\Delta t,(k+1)\Delta t)$ is given by
\begin{equation}
q_{(k+1)\Delta t}^{(\alpha_{\Delta t},n;h)}
= \frac{\exp\left(r_f \Delta t-v\Delta t\right)-\exp\left(-\eta_{k\Delta t}^{(\alpha_{\Delta t})}\sqrt{\Delta t}\right)}
{\exp\left(\eta_{k\Delta t}^{(\alpha_{\Delta t})}\sqrt{\Delta t}\right)
- \exp\left(-\eta_{k\Delta t}^{(\alpha_{\Delta t})}\sqrt{\Delta t}\right)}.
\label{eq_30}
\end{equation} 
To leading order in $\Delta t$ we have
\begin{equation}
q_{(k+1)\Delta t}^{(\alpha_{\Delta t},n;h)}
= \frac{1}{2}+\frac{r_f - v - (\eta_{k\Delta t}^{(\alpha_{\Delta t})})^2/2}{2\eta_{k\Delta t}^{(\alpha_{\Delta t})}}\sqrt{\Delta t}.
\label{eq_sec6_q2}
\end{equation} 
Note that, conditionally on $\mathcal{F}_k^{(n;\mathbb{M}^{(\alpha_{\Delta t})})}$,
the risk-neutral probability $q_{(k+1)\Delta t}^{(\alpha_{\Delta t},n;h)}$
depends on the entire path $M_k^{(\alpha_{\Delta t})}$.

\subsection{Implied volatility}
\label{sec61}
\noindent
Using the daily closing prices and the corresponding log-returns for MSFT,
we apply the full GJR model to generate a  binomial tree to construct call option
prices using \eqref{eq_sec6_last} and \eqref{eq_30}.
CBOE option price data\footnote{
	See \url{https://datashop.cboe.com/options-intervals}.},
with all available strike values and expiration dates prior to 1/17/2020,
were collected for the date 7/1/2019 for the underlying stock MSFT.
The time series $\left\{\eta_{k \Delta t}^{(\alpha_{\Delta t})},k = 1,...,n\right\}$ \eqref{eq_eta}
is based on the optimal trajectory
$\left\{ M_k^{\left( \bar{\alpha}^{(F)},\textrm{exo} \right)},\ k = 1, \dots, n \right\}$ obtained in section \ref{sec51}.
The coefficients $v$, $\eta_0 = \sigma$, $\gamma$, and the degrees of freedom $\kappa$ for the Student-t
distribution used as the model for $h(\cdot)$, are taken from Table~\ref{tab3}.

\begin{figure}[ht]
\begin{center}
    \begin{subfigure}[b]{0.32\textwidth} 
    	\includegraphics[width=\textwidth]{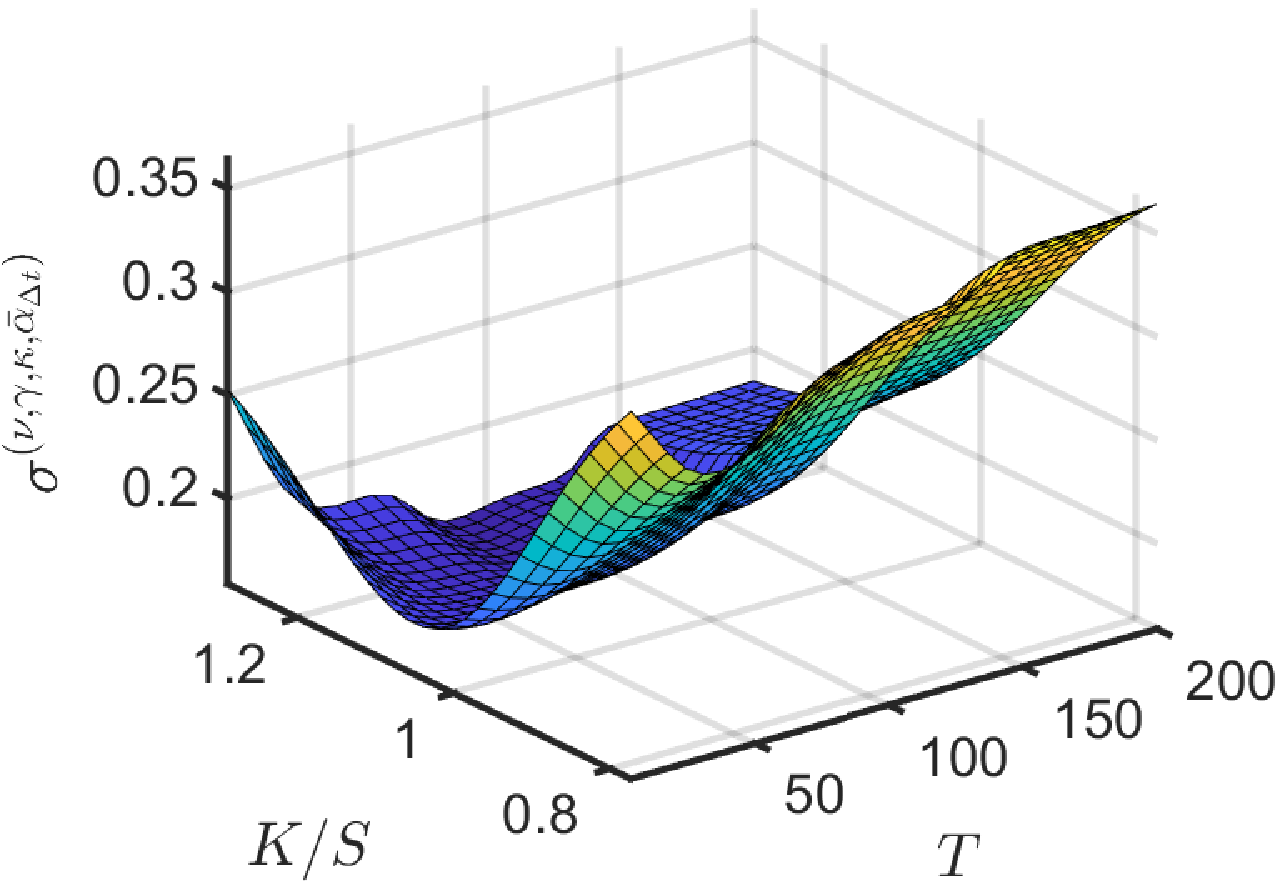}
    	\caption{}
    	\label{fig12_GJR}
    \end{subfigure}
    \begin{subfigure}[b]{0.32\textwidth} 
    	\includegraphics[width=\textwidth]{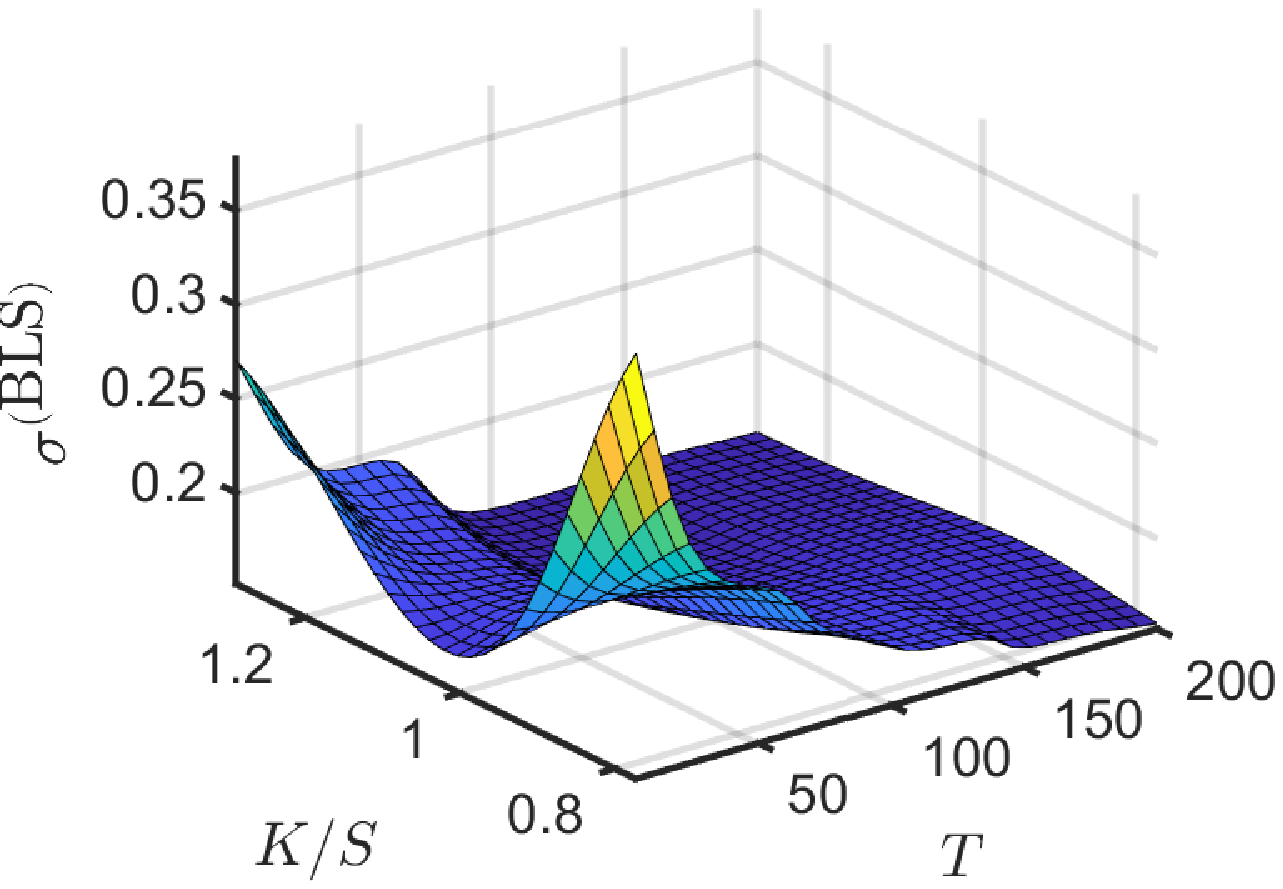}
    	\caption{}
    	\label{fig12_BLS}
    \end{subfigure}
    \begin{subfigure}[b]{0.32\textwidth}
    	\includegraphics[width=\textwidth]{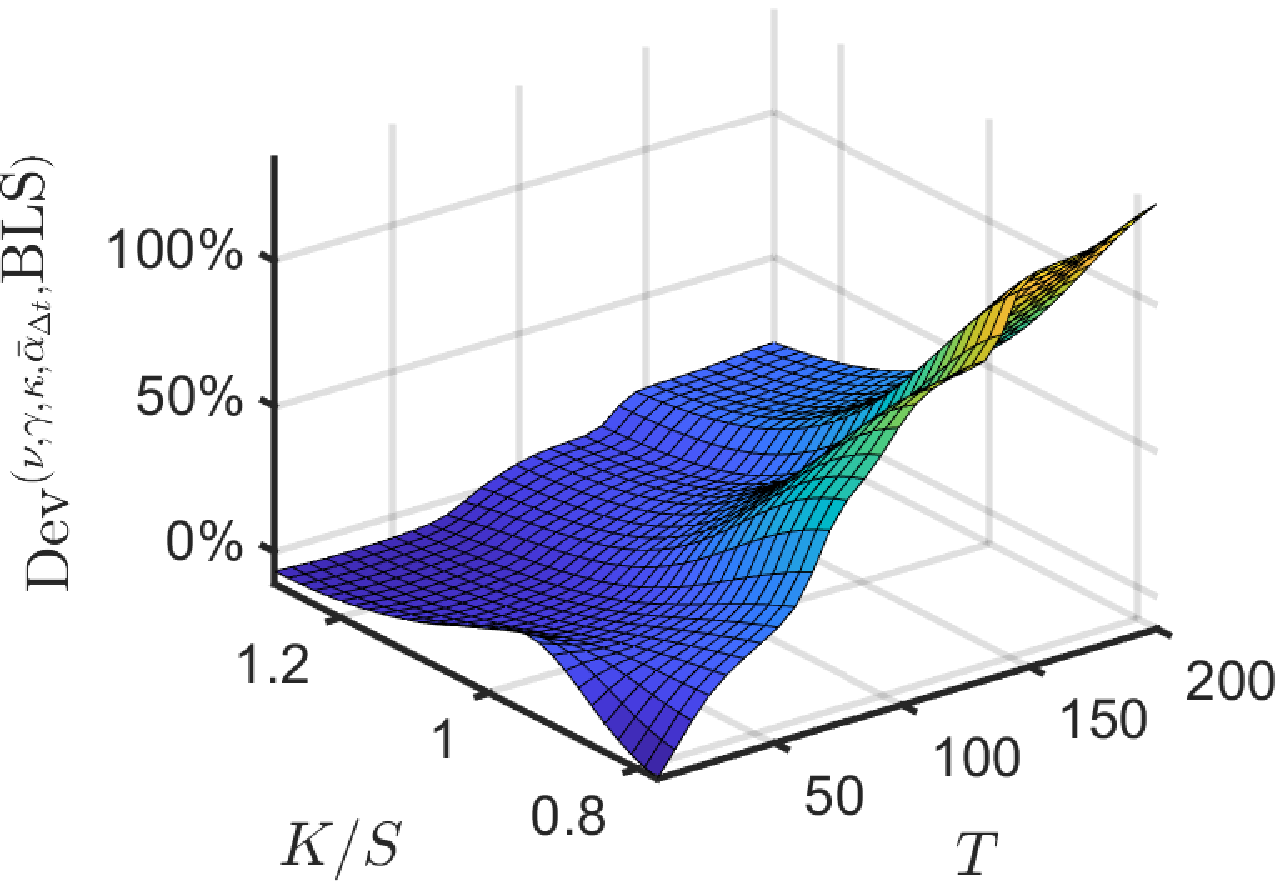}
    	\caption{}
    	\label{fig12_dev}
    \end{subfigure}
    \caption{
	(a) The implied $\sigma^{(\nu,\gamma,\kappa,\bar{\alpha}_{\Delta t})}$ surface generated
		by the path-dependent GJR pricing tree.
	(b) The implied $\sigma^{(\textrm{BLS})}$ surface using the Black-Scholes formula for implied volatility.
	(c) The percent deviation surface $\textrm{Dev}^{(\nu,\gamma,\kappa,\bar{\alpha}_{\Delta t},\textrm{BLS})}$.
	Each surface is plotted as functions of moneyness $K/S$ and time to maturity $T$ (in days).}
    \label{fig12}
\end{center}
\end{figure}
As in the numerical examples in sections \ref{sec31} and \ref{sec41},
we compute the implied volatility, $\sigma^{(\nu,\gamma,\kappa,\bar{\alpha}_{\Delta t})}$,
based on the path-dependent GJR pricing tree.
We also estimate the Black-Scholes implied volatility $\sigma^{\textrm{(BLS)}}$ surface for the same option prices,
and define the percent deviation between the two surfaces,
$\textrm{Dev}^{(\nu,\gamma,\kappa,\bar{\alpha}_{\Delta t},\textrm{BLS})}
	=100 \left(  \sigma^{(\nu,\gamma,\kappa,\bar{\alpha}_{\Delta t})} - \sigma^{\textrm{(BLS)}} \right)
	 / \sigma^{\textrm{(BLS)}}$. 
Fig. \ref{fig12} shows the  implied volatility surfaces plotted as functions of time to maturity $T$ (in days) 
and moneyness $K/S$.
For maturity times $\lesssim 50$ days, both the GJR and the Black-Scholes surfaces contain volatility smiles and
the deviation between the two surfaces is relatively small.
Beyond 50 days, the Black-Scholes implied volatility surface flattens, while the GJR surface flattens only in the
``out-of-the-money'' region, leading to large differences between the two over the ``in-the-money'' region.
\section{Conclusions}
\label{sec7}
\noindent
In the option pricing literature since 1990, academic work on continuous-time option pricing models (CPMs)
has greatly overshadowed the work on discrete-time option pricing models (DPMs).
Moreover, with very few exceptions, DPMs are used only to approximate the CPMs
(presumably to decrease the numerical complexity in generating Monte Carlo price trajectories).
Those DPMs are placed directly into the risk-neutral world and do not address an issue that is very
important for every option trader:
“Which DPM in the natural world is uniquely determined by the DPM introduced directly in the risk-neutral world?”
The reason for the “subordinated” role of the DPM’s is the  “obsessive love” of academics for CPMs,
arising from the alluring mathematical beauty of the semimartingale theory\footnote{
	See \cite{Emery2002}, \cite{Jacod1987}, and \cite{Protter2004}.}
of the Strasbourg school and the truly paramount Fundamental Theorem of Asset Pricing.\footnote{
	Delbaen and Schachermayer (1994, 1998).}
However, crucial information found in DPMs in the natural world, such as the mean-return parameter,
the probability for stock-upturn, and the distributional skewness and kurtosis, is lost in CPMs.
Most notably, in CPMs the mean-return parameter is lost due to the assumed ability of the hedger to
trade continuously in time.
DPMs with defined dynamics in the natural world do not have this issue.\footnote{
	See Kim et al. (2016,2019), Hu et al. (2020a,b).}
That observation was the main motivation for this paper.
Specifically, we have extended the classical Jarrow-Rudd pricing tree to include skewness and kurtosis in the underlying
asset’s return distribution in both the natural and in the risk-neutral world.
We have extended Merton’s option pricing tree model with hedging-transaction costs to our new
generalized Jarrow Rudd (GJR) option pricing tree model.
We have applied the Cherny-Shiryaev-Yor invariance principle and the Fama-French five-factor model to
further extend the GJR pricing tree model to cover path-dependent options.
Our numerical examples include estimation of the implied surfaces of all parameters in the GJR pricing trees.

\end{document}